\pdfoutput=1

\documentclass[11pt,twoside,a4paper,cmspaper,final,collab]{cms-tdr}
\def\svnVersion{294875}\def\cmsCernNo{2013-037}\def\cmsCernDate{\today}\def\cmsMessage{Published in the European Physical Journal C as \href{http://dx.doi.org/10.1140/epjc/s10052-015-3499-1}{\doi{10.1140/epjc/s10052-015-3499-1.}}}

\begin{document}\cmsNoteHeader{SMP-12-028}

\hyphenation{had-ron-i-za-tion}
\hyphenation{cal-or-i-me-ter}
\hyphenation{de-vices}
\RCS$Revision: 288504 $
\RCS$HeadURL: svn+ssh://svn.cern.ch/reps/tdr2/papers/SMP-12-028/trunk/SMP-12-028.tex $
\RCS$Id: SMP-12-028.tex 288504 2015-05-12 15:53:03Z alverson $
\newlength\cmsFigWidth
\ifthenelse{\boolean{cms@external}}{\setlength\cmsFigWidth{0.48\textwidth}}{\setlength\cmsFigWidth{0.7\textwidth}}
\ifthenelse{\boolean{cms@external}}{\providecommand{\cmsLeft}{top}}{\providecommand{\cmsLeft}{left}}
\ifthenelse{\boolean{cms@external}}{\providecommand{\cmsRight}{bottom}}{\providecommand{\cmsRight}{right}}
\providecommand{\alps}{\ensuremath{\alpha_S}\xspace}
\providecommand{\alpsmz}{\ensuremath{\alpha_S(M_\cPZ)}\xspace}
\providecommand{\alpsq}{\ensuremath{\alpha_S(Q)}\xspace}
\providecommand{\chisq}{\ensuremath{\chi^2}\xspace}
\providecommand{\chisqndof}{\ensuremath{\chi^2/n_\mathrm{dof}}\xspace}
\providecommand{\chipsq}{\ensuremath{\chi^2_\mathrm{p}}\xspace}
\providecommand{\chipsqndata}{\ensuremath{\chi^2_\mathrm{p}/n_\mathrm{data}}\xspace}
\providecommand{\ndata}{\ensuremath{n_{\mathrm{data}}}\xspace}
\providecommand{\ndof}{\ensuremath{n_{\mathrm{dof}}}\xspace}
\providecommand{\HERAFitter} {{\textsc{HERAFitter}}\xspace}
\providecommand{\HERWIGPP} {{\textsc{herwig++}}\xspace}
\providecommand{\NLOJETPP} {{\textsc{NLOJet++}}\xspace}
\providecommand{\fastNLO} {{\textsc{fastNLO}}\xspace}
\providecommand{\fastjet} {{\textsc{FastJet}}\xspace}
\providecommand{\GeVsq}{\ensuremath{\GeV^2}\xspace}
\providecommand{\GRV} {{\textsc{GRV}}\xspace}
\providecommand{\HOPPET} {{\textsc{HOPPET}}\xspace}
\providecommand{\POWHEGBOX} {{\textsc{powheg box}}\xspace}
\providecommand{\PYTHIAS} {{\textsc{pythia6}}\xspace}
\providecommand{\PYTHIAE} {{\textsc{pythia8}}\xspace}
\providecommand{\QCDNUM} {{\textsc{QCDNUM}}\xspace}
\providecommand{\RooUnfold} {{\textsc{RooUnfold}}\xspace}
\providecommand{\ptmax}{\ensuremath{p_{\mathrm{T,max}}}\xspace}
\providecommand{\avept}{\ensuremath{\langle p_\mathrm{T1,2}\rangle}\xspace}
\providecommand{\etaabs}{\ensuremath{\abs{\eta}}\xspace}
\providecommand{\yabs}{\ensuremath{\abs{y}}\xspace}
\providecommand{\mur}{\ensuremath{\mu_r}\xspace}
\providecommand{\muf}{\ensuremath{\mu_f}\xspace}
\providecommand{\rbthm}{\rule[-2ex]{0ex}{5ex}}
\providecommand{\rbtrr}{\rule[-0.8ex]{0ex}{3.2ex}}

\hyphenation{HERA-Fitter}
\cmsNoteHeader{SMP-12-028}
\title{Constraints on parton distribution functions and extraction of the strong coupling constant
  from the inclusive jet cross section in pp collisions at $\sqrt{s} = 7$\TeV}
\titlerunning{Parton distribution functions and $\alps$ from the inclusive jet cross section}

\date{\today}

\abstract{The inclusive jet cross section for proton-proton collisions
  at a centre-of-mass energy of 7\TeV was measured by the CMS Collaboration
  at the LHC with data corresponding to an
  integrated luminosity of 5.0\fbinv. The measurement
  covers a phase space up to 2\TeV in jet transverse momentum and 2.5
  in absolute jet rapidity. The statistical precision of these data
  leads to stringent constraints on the parton distribution functions
  of the proton. The data provide important input for the gluon
  density at high fractions of the proton momentum and for the strong
  coupling constant at large energy scales. Using predictions from
  perturbative quantum chromodynamics at next-to-leading order,
  complemented with electroweak corrections, the constraining power of
  these data is investigated and the strong coupling constant at the Z
  boson mass $M_\cPZ$ is determined to be
  $\alpha_S(M_\cPZ) = 0.1185\pm 0.0019\,(\text{exp})\,^{+0.0060}_{-0.0037}\thy$,
  which is in agreement with the world average.}

\hypersetup{%
  pdfauthor={CMS Collaboration},%
  pdftitle={Constraints on parton distribution functions and extraction of the strong coupling
    constant from the inclusive jet cross section in pp collisions at sqrt(s) = 7 TeV},%
  pdfsubject={CMS},%
  pdfkeywords={CMS, physics, QCD, PDF, jets, strong coupling constant, alpha-s}
}

\maketitle

\section{Introduction}
\label{sec:intro}

Collimated streams of particles, conventionally called jets, are
abundantly produced in highly energetic proton-proton collisions at
the LHC\@. At high transverse momenta \pt these collisions are
described by quantum chromodynamics (QCD) using perturbative
techniques (pQCD). Indispensable ingredients for QCD predictions of
cross sections in \Pp\Pp~collisions are the proton structure,
expressed in terms of parton distribution functions (PDFs), and the
strong coupling constant \alps, which is a fundamental parameter of
QCD\@. The PDFs and \alps both depend on the relevant energy scale $Q$
of the scattering process, which is identified with the jet \pt for
the reactions considered in this report. In addition, the PDFs,
defined for each type of parton, depend on the fractional momentum $x$
of the proton carried by the parton.

The large cross section for jet production at the LHC and the
unprecedented experimental precision of the jet measurements allow
stringent tests of QCD\@. In this study, the theory is confronted with
data in previously inaccessible phase space regions of $Q$ and $x$.
When jet production cross sections are combined with inclusive data
from deep-inelastic scattering (DIS), the gluon PDF for $x \gtrsim
0.01$ can be constrained and \alpsmz can be determined. In the present
analysis, this is demonstrated by means of the CMS measurement of
inclusive jet production~\cite{Chatrchyan:2012bja}. The data,
collected in 2011 and corresponding to an integrated luminosity of
5.0\fbinv, extend the accessible phase space in jet \pt up to 2\TeV,
and range up to $\yabs = 2.5$ in absolute jet rapidity.  A PDF study
using inclusive jet measurements by the ATLAS Collaboration is
described in Ref.~\cite{Aad:2013lpa}.

This paper is divided into six parts.  Section~\ref{sec:measurement}
presents an overview of the CMS detector and of the measurement,
published in Ref.~\cite{Chatrchyan:2012bja}, and proposes a modified
treatment of correlations in the experimental
uncertainties. Theoretical ingredients are introduced in
Section~\ref{sec:theory}. Section~\ref{sec:alphas} is dedicated to the
determination of \alps at the scale of the \cPZ-boson mass
$M_\cPZ$, and in Section~\ref{sec:herafitter} the influence of
the jet data on the PDFs is discussed. A summary is presented in
Section~\ref{sec:summary}.

\section{The inclusive jet cross section}
\label{sec:measurement}

\subsection{Overview of the CMS detector and of the measurement}
\label{sec:measurementoverview}

{\tolerance=1000
The central feature of the CMS detector is a superconducting solenoid
of 6\unit{m} internal diameter, providing a magnetic field of
3.8\unit{T}. Within the superconducting solenoid volume are a silicon
pixel and strip tracker, a lead tungstate crystal electromagnetic
calorimeter (ECAL), and a brass/scintillator hadron calorimeter, each
composed of a barrel and two endcap sections. Muons are measured in
gas-ionisation detectors embedded in the steel flux-return yoke
outside the solenoid. Extensive forward calorimetry (HF) complements
the coverage provided by the barrel and endcap detectors. A more
detailed description of the CMS detector, together with a definition
of the coordinate system used and the relevant kinematic variables,
can be found in Ref.~\cite{Chatrchyan:2008aa}.\par}

Jets are reconstructed with a size parameter of $R=0.7$ using the
collinear- and infrared-safe anti-\kt clustering
algorithm~\cite{Cacciari:2008gp} as implemented in the \fastjet
package~\cite{Cacciari:2011ma}. The published measurements of the
cross sections were corrected for detector effects, and include
statistical and systematic experimental uncertainties as well as
bin-to-bin correlations for each type of uncertainty.
A complete description of the measurement can be found
in Ref.~\cite{Chatrchyan:2012bja}.

The double-differential inclusive jet cross section investigated in
the following is derived from observed inclusive jet yields via
\begin{equation}
  \frac{\rd^2\sigma}{\rd\pt\,\rd y} =
  \frac{1}{\epsilon\cdot\mathcal{L}_{\text{int}}}
  \frac{N_\text{jets}}{\Delta\pt\,\left(2\cdot\Delta\yabs\right)},
\end{equation}
where $N_\text{jets}$ is the number of jets in the specific kinematic
range (bin), $\mathcal{L}_{\text{int}}$ is the integrated luminosity,
$\epsilon$ is the product of trigger and event selection efficiencies,
and $\Delta \pt$ and $\Delta\yabs$ are the bin widths in \pt and
\yabs. The factor of two reflects the folding of the distributions
around $y=0$.

\subsection{Experimental uncertainties}
\label{sec:measurementuncertainties}

The inclusive jet cross section is measured in five equally sized bins
of $\Delta\yabs = 0.5$ up to an absolute rapidity of $\yabs =
2.5$. The inner three regions roughly correspond to the barrel part of
the detector, the outer two to the endcaps.  Tracker coverage extends
up to $\yabs = 2.4$. The minimum \pt imposed on any jet is
114\GeV. The binning in jet \pt follows the jet \pt resolution of the
central detector and changes with \pt. The upper reach in \pt is given
by the available data and decreases with \yabs.

Four categories~\cite{Chatrchyan:2012bja} of experimental
uncertainties are defined: the jet energy scale (JES), the luminosity,
the corrections for detector response and resolution, and all
remaining uncorrelated effects.

The JES is the dominant source of systematic uncertainty, because a
small shift in the measured \pt translates into a large uncertainty in
the steeply falling jet \pt spectrum and hence in the cross section
for any given value of \pt. The JES uncertainty is parameterized in
terms of jet \pt and pseudorapidity $\eta = -\ln\tan(\theta/2)$ and
amounts to 1--2\%~\cite{CMS-DP-2012-006}, which translates into a
5--25\% uncertainty in the cross section. Because of its particular
importance for this analysis, more details are given in
Section~\ref{sec:measurementjec}.

The uncertainty in the integrated luminosity is
2.2\%~\cite{CMS-PAS-SMP-12-008} and translates into a normalisation
uncertainty that is fully correlated across \yabs and \pt.

The effect of the jet energy resolution (JER) is corrected for using
the D'Agostini method~\cite{D'Agostini:1994zf} as implemented in the
\RooUnfold package~\cite{Adye:2011gm}. The uncertainty due to the
unfolding comprises the effects of an imprecise knowledge of the JER,
of residual differences between data and the Monte Carlo (MC)
modelling of detector response, and of the unfolding technique
applied. The total unfolding uncertainty, which is fully correlated
across $\eta$ and \pt, is 3--4\%. Additionally, the statistical
uncertainties are propagated through the unfolding procedure, thereby
providing the correlations between the statistical uncertainties of
the unfolded measurement. A statistical covariance matrix must be used
to take this into account.

Remaining effects are collected into an uncorrelated uncertainty of
$\approx$1\%.

\subsection{Uncertainties in JES}
\label{sec:measurementjec}

The procedure to calibrate jet energies in CMS and ways to estimate
JES uncertainties are described in Ref.~\cite{Chatrchyan:2011ds}. To
use CMS data in fits of PDFs or \alpsmz, it is essential to account
for the correlations in these uncertainties among different regions of
the detector. The treatment of correlations uses 16 mutually
uncorrelated sources as in Ref.~\cite{Chatrchyan:2012bja}. Within each
source, the uncertainties are fully correlated in \pt and $\eta$. Any
change in the jet energy calibration (JEC) is described through a
linear combination of sources, where each source is assumed to have a
Gaussian probability density with a zero mean and a root-mean-square
of unity. In this way, the uncertainty correlations are encoded in a
fashion similar to that provided for PDF uncertainties using the
Hessian method~\cite{Pumplin:2001ct}. The total uncertainty is defined
through the quadratic sum of all uncertainties. The full list of
sources together with their brief descriptions can be found in
Appendix~\ref{sec:jessources}.

The JES uncertainties can be classified into four broad categories:
absolute energy scale as a function of \pt, jet flavour dependent
differences, relative calibration of JES as a function of $\eta$, and
the effects of multiple proton interactions in the same or adjacent
beam crossings (pileup). The absolute scale is a single fixed number
such that the corresponding uncertainty is fully correlated across \pt
and $\eta$.  Using photon+jet and $Z$+jet data, the JES can be
constrained directly in the jet \pt range 30--600\GeV. The response at
larger and smaller \pt is extrapolated through MC simulation.  Extra
uncertainties are assigned to this extrapolation based on the
differences between MC event generators and the single-particle
response of the detector. The absolute calibration is the most
relevant uncertainty in jet analyses at large \pt.

The categories involving jet flavour dependence and pileup effects are
important mainly at small \pt and have relatively little impact for
the phase space considered in this report.

The third category parameterizes $\eta$-dependent changes in relative
JES\@. The measurement uncertainties within different detector regions
are strongly correlated, and thus the $\eta$-dependent sources are
only provided for wide regions: barrel, endcap with upstream tracking,
endcap without upstream tracking, and the HF calorimeter. In
principle, the $\eta$-dependent effects can also have a \pt
dependence. Based on systematic studies on data and simulated events,
which indicate that the \pt and $\eta$ dependence of the uncertainties
factorise to a good approximation, this is omitted from the initial
calibration procedure. However, experiences with the calibration of
data collected in 2012 and with fits of \alpsmz reported in
Section~\ref{sec:alphas} show that this is too strong an
assumption. Applying the uncertainties and correlations in a fit of
\alpsmz to the inclusive jet data separately for each bin in \yabs
leads to results with values of \alpsmz that scatter around a central
value. Performing the same fit taking all \yabs bins together and
assuming 100\% correlation in \yabs within the JES uncertainty sources
results in a bad fit quality (high \chisq per number of degrees of
freedom $n_\mathrm{dof}$) and a value of \alpsmz that is significantly
higher than any value observed for an individual bin in
\yabs. Changing the correlation in the JES uncertainty from 0\% to
100\% produces a steep rise in \chisqndof, and influences the fitted
value of \alpsmz for correlations near 90\%, indicating an assumption
on the correlations in \yabs that is too strong. The technique of
nuisance parameters, as described in Section~\ref{sec:fitsetup},
helped in the analysis of this issue.

To implement the additional $\eta$-decorrelation induced by the
\pt-dependence in the $\eta$-dependent JEC introduced for the
calibration of 2012 data, the source from the
single-particle response JEC2, which accounts for extrapolation
uncertainties at large \pt as discussed in
Appendix~\ref{sec:jessources}, is decorrelated versus $\eta$ as
follows:
\begin{enumerate}
\item in the barrel region ($\yabs < 1.5$), the correlation of the
  single-particle response source among the three bins in \yabs is set
  to 50\%,
\item in the endcap region ($1.5 \leq \yabs < 2.5$), the correlation
  of the single-particle response source between the two bins in \yabs
  is kept at 100\%,
\item there is no correlation of the single-particle response source
  between the two detector regions of $\yabs < 1.5$ and $1.5 \leq
  \yabs < 2.5$.
\end{enumerate}
The additional freedom of \pt-dependent corrections versus $\eta$
hence leads to a modification of the previously assumed full
correlation between all $\eta$ regions to a reduced estimate of 50\%
correlation of JEC2 within the barrel region, which always contains
the tag jet of the dijet balance method~\cite{Chatrchyan:2011ds}. In
addition, the JEC2 corrections are estimated to be uncorrelated
between the barrel and endcap regions of the detector because of
respective separate \pt-dependences of these corrections.

Technically, this can be achieved by splitting the single-particle
response source into five parts (JEC2a--e), as shown in
Table~\ref{tab:cmsjets2011:nuisance}. Each of these sources is a
duplicate of the original single-particle response source that is set
to zero outside the respective ranges of $\yabs < 1.5$, $1.5 \leq
\yabs < 2.5$, $\yabs < 0.5$, $0.5 \leq \yabs < 1.0$, and $1.0 \leq
\yabs < 1.5$, such that the original full correlation of
\begin{equation}
  \mathrm{corr}_\mathrm{JEC2,old} =
  \begin{pmatrix}
    1 & 1 & 1 & 1 & 1\\
    1 & 1 & 1 & 1 & 1\\
    1 & 1 & 1 & 1 & 1\\
    1 & 1 & 1 & 1 & 1\\
    1 & 1 & 1 & 1 & 1\\
  \end{pmatrix}
\end{equation}
is replaced by the partially uncorrelated version of
\begin{equation}
  \mathrm{corr}_\mathrm{JEC2,new} =
  \begin{pmatrix}
    1   & 0.5 & 0.5 & 0 & 0\\
    0.5 & 1   & 0.5 & 0 & 0\\
    0.5 & 0.5 & 1   & 0 & 0\\
    0   & 0   & 0   & 1 & 1\\
    0   & 0   & 0   & 1 & 1\\
  \end{pmatrix},
  \label{eqn:decorr}
\end{equation}
which is more accurate as justified by studies based on
2012 data.
For the proper normalisation of the five new correlated sources,
normalisation factors of $1/\sqrt{2}$ (JEC2a, JEC2c--JEC2f) and $1$
(JEC2b) must be applied. With these factors, the sum of the five
sources reproduces the original uncertainty for each \yabs, while the
additional freedom gives the estimated level of correlation among the
\yabs regions.

All results presented in this paper are based on this improved
treatment of the correlation of JES uncertainties. While some
decorrelation of these uncertainties versus $\eta$ is important for
the fits of \alpsmz described in Section~\ref{sec:alphas}, the exact
size of the estimated decorrelation is not. Varying the assumptions
according to Eq.~(\ref{eqn:decorr}) from 50\% to 20 or 80\% in the
barrel region, from 100 to 80\% in the endcap region, or from 0 to
20\% between the barrel and endcap regions leads to changes in the
fitted value of \alpsmz that are negligible with respect to other
experimental uncertainties.

\section{Theoretical ingredients}
\label{sec:theory}

The theoretical predictions for the inclusive jet cross section
comprise a next-to-leading order (NLO) pQCD calculation with
electroweak corrections
(EW)~\cite{Butterworth:2014efa,Dittmaier:2012kx}. They are
complemented by a nonperturbative (NP) factor that corrects for
multiple-parton interactions (MPI) and hadronization (HAD)
effects. Parton shower (PS) corrections, derived from NLO predictions
with matched parton showers, are tested in an additional study in
Section~\ref{section:results_a_s}, but are not applied to the main
result.

\subsection{Fixed-order prediction in perturbative QCD}
\label{sec:fixedorder}

The same NLO prediction as in Ref.~\cite{Chatrchyan:2012bja} is used,
\ie the calculations are based on the parton-level program \NLOJETPP
version~4.1.3~\cite{Nagy:2001fj,Nagy:2003tz} and are performed within
the \fastNLO framework version~2.1~\cite{Britzger:2012bs}. The
renormalization and factorisation scales, \mur and \muf respectively,
are identified with the individual jet \pt. The number of active
(massless) flavours $N_f$ in \NLOJETPP has been set to five.

Five sets of PDFs are available for a series of values of \alpsmz,
which is a requisite for a determination of \alpsmz from data. For an
overview, these PDF sets are listed in Table~\ref{tab:pdfsets}
together with the respective references. The ABM11 PDF set employs a
fixed-flavour number scheme with five active flavours, while the other
PDF sets use a variable-flavour number scheme with a maximum of five
flavours, $N_{f,\mathrm{max}} = 5$, except for NNPDF2.1 which has
$N_{f,\mathrm{max}} = 6$. All sets exist at next-to-leading and
next-to-next-to-leading evolution order. The PDF uncertainties are
provided at $68.3\%$ confidence level (CL) except for CT10, which
provides uncertainties at $90\%$ CL\@. For a uniform treatment of all
PDFs, the CT10 uncertainties are downscaled by a factor of
$\sqrt{2}\erf^{-1}{(0.9)} \approx 1.645$.

\begin{table*}[tbp]
  \centering
  \topcaption{The PDF sets used in comparisons to the data together with
    the evolution order (Evol.),
    the corresponding number of active flavours $N_f$, the assumed masses
    $M_\cPqt$ and $M_\cPZ$ of the top quark and the \cPZ\
    boson, respectively, the default values of
    \alpsmz, and the range in \alpsmz variation
    available for fits. For CT10 the updated versions of 2012 are taken.}
  \label{tab:pdfsets}
\begin{tabular}{lllcclcc}
    Base set & Refs. & Evol.\ & $N_f$ & $M_\cPqt$ (\GeVns{}) &
    $M_\cPZ$ (\GeVns{}) &\alpsmz & \alpsmz range\rbthm\\
    \hline
    ABM11     & \cite{Alekhin:2012ig} & NLO  &       5  & 180 & 91.174 & 0.1180 & 0.110--0.130\rbtrr\\
    ABM11     & \cite{Alekhin:2012ig} & NNLO &       5  & 180 & 91.174 & 0.1134 & 0.104--0.120\rbtrr\\
    CT10      & \cite{Lai:2010vv}     & NLO  & ${\leq}5$ & 172 & 91.188 & 0.1180 & 0.112--0.127\rbtrr\\
    CT10      & \cite{Lai:2010vv}     & NNLO & ${\leq}5$ & 172 & 91.188 & 0.1180 & 0.110--0.130\rbtrr\\
    HERAPDF1.5& \cite{Aaron:2009aa}   & NLO  & ${\leq}5$ & 180 & 91.187 & 0.1176 & 0.114--0.122\rbtrr\\
    HERAPDF1.5& \cite{Aaron:2009aa}   & NNLO & ${\leq}5$ & 180 & 91.187 & 0.1176 & 0.114--0.122\rbtrr\\
    MSTW2008  & \cite{Martin:2009iq,Martin:2009bu} & NLO  & ${\leq}5$ & $10^{10}$ & 91.1876 & 0.1202 & 0.110--0.130\rbtrr\\
    MSTW2008  & \cite{Martin:2009iq,Martin:2009bu} & NNLO & ${\leq}5$ & $10^{10}$ & 91.1876 & 0.1171 & 0.107--0.127\rbtrr\\
    NNPDF2.1  & \cite{Ball:2011mu}    & NLO  & ${\leq}6$ & 175 & 91.2 & 0.1190 & 0.114--0.124\rbtrr\\
    NNPDF2.1  & \cite{Ball:2011mu}    & NNLO & ${\leq}6$ & 175 & 91.2 & 0.1190 & 0.114--0.124\rbtrr\\
  \end{tabular}
\end{table*}

The electroweak corrections to the hard-scattering cross section have
been computed with the CT10-NLO PDF set for a fixed number of five
flavours and with the \pt of the leading jet, \ptmax, as scale choice
for \mur and \muf instead of the \pt of each jet. At high jet \pt and
central rapidity, where the electroweak effects become sizeable, NLO
calculations with either of the two scale settings differ by less than
one percent. Given the small impact of the electroweak corrections on
the final results in Sections~\ref{sec:alphas}
and~\ref{sec:herafitter}, no uncertainty on their size has been
assigned.

\subsection{Theoretical prediction from MC simulations including
  parton showers and nonperturbative effects}

The most precise theoretical predictions for jet measurements are
usually achieved in fixed-order pQCD, but are available at parton
level only. Data that have been corrected for detector effects,
however, refer to measurable particles, \ie to colour-neutral
particles with mean decay lengths such that $c\tau>10\unit{mm}$.  Two
complications arise when comparing fixed-order perturbation theory to
these measurements: emissions of additional partons close in phase
space, which are not sufficiently accounted for in low-order
approximations, and effects that cannot be treated by perturbative
methods. The first problem is addressed by the parton shower
concept~\cite{Marchesini:1987cf,Knowles:1988hu,Knowles:1988vs} within
pQCD, where multiple parton radiation close in phase space is taken
into account through an all-orders approximation of the dominant terms
including coherence effects. Avoiding double counting, these parton
showers are combined with leading-order (LO) calculations in MC event
generators, such as \PYTHIA~\cite{Sjostrand:2006za} and
\HERWIGPP~\cite{Bahr:2008pv}.

The second issue concerns NP corrections, which comprise supplementary
parton-parton scatters within the same colliding protons, \ie MPI,
and the hadronization process including particle decays.  The
MPI~\cite{Sjostrand:1987su,Bahr:2008dy} model for additional
soft-particle production, which is detected as part of the underlying
event, is implemented in \PYTHIA as well as \HERWIGPP.  Hadronization
describes the transition phase from coloured partons to colour-neutral
particles, where perturbative methods are no longer applicable. Two
models for hadronization are in common use, the Lund string
fragmentation~\cite{Andersson:1983ia,Andersson:1983jt,Sjostrand:1984iu}
that is used in \PYTHIA, and the cluster
fragmentation~\cite{Webber:1983if} that has been adopted by \HERWIGPP.

Beyond LO combining fixed-order predictions with parton showers, MPI,
and hadronization models is much more complicated.  Potential double
counting of terms in the perturbative expansion and the PS has to be
avoided.  In recent years programs have become available for dijet
production at NLO that can be matched to PS MC event generators. In
the following, one such program, the \POWHEG
package~\cite{Frixione:2007vw,Alioli:2010xd} will be used for
comparisons with dijet events~\cite{Alioli:2010xa} to the LO MC event
generators.

\subsection{NP corrections from \texorpdfstring{\PYTHIAS}{PYTHIA6} and
  \texorpdfstring{\HERWIGPP}{HERWIG++}}\label{sec:assumptions}

For the comparison of theoretical predictions to the measurement
reported in Ref.~\cite{Chatrchyan:2012bja}, the NP correction was
derived as usual~\cite{Campbell:2006wx} from the average prediction of
two LO MC event generators and more specifically from \PYTHIA
version~6.4.22 tune~Z2 and \HERWIGPP version~2.4.2 with the default
tune of version~2.3. Tune Z2 is identical to tune Z1 described
in~\cite{Field:2010bc} except that Z2 employs the
CTEQ6L1~\cite{Pumplin:2002vw} PDF set, while Z1 uses the
CTEQ5L~\cite{Lai:1999wy} PDF set. The NP correction factor can be
defined for each bin in \pt and \yabs as
\begin{equation}
  C _\mathrm{LO}^{\text{NP}} = \frac{\sigma_{\mathrm{LO+PS+HAD+MPI}}}{\sigma_{\mathrm{LO+PS}}}\,
  \label{e:C_LO_NP}
\end{equation}
where $\sigma$ represents the inclusive jet cross section and the
subscripts ``LO+PS+HAD+MPI'' and ``LO+PS'' indicate which steps of a
general MC event generation procedure have been run, see also
Refs.~\cite{Campbell:2006wx,Buckley:2011ms}. The central value is
calculated by taking the average of the two predictions from \PYTHIAS
and \HERWIGPP.

In applying these factors as corrections for NP effects to NLO theory
predictions, it is assumed that the NP corrections are universal,
\ie they are similar for LO and NLO\@.

\subsection{NP and PS corrections from \texorpdfstring{\POWHEG +
    \PYTHIAS}{POWHEG + PYTHIA6}}
\label{sec:powhegpythia}

Alternative corrections are derived, which use the \POWHEGBOX
revision~197 with the CT10-NLO PDF set
for the hard subprocess at NLO plus the leading
emission~\cite{Nason:2004rx} complemented with the matched showering,
MPI, and hadronization from \PYTHIAS version~6.4.26. The NLO event
generation within the \POWHEG framework, and the showering and
hadronization process performed by \PYTHIAS are done in independent
steps.

\begin{figure}[btp]
  \centering
  \includegraphics[width=\cmsFigWidth]{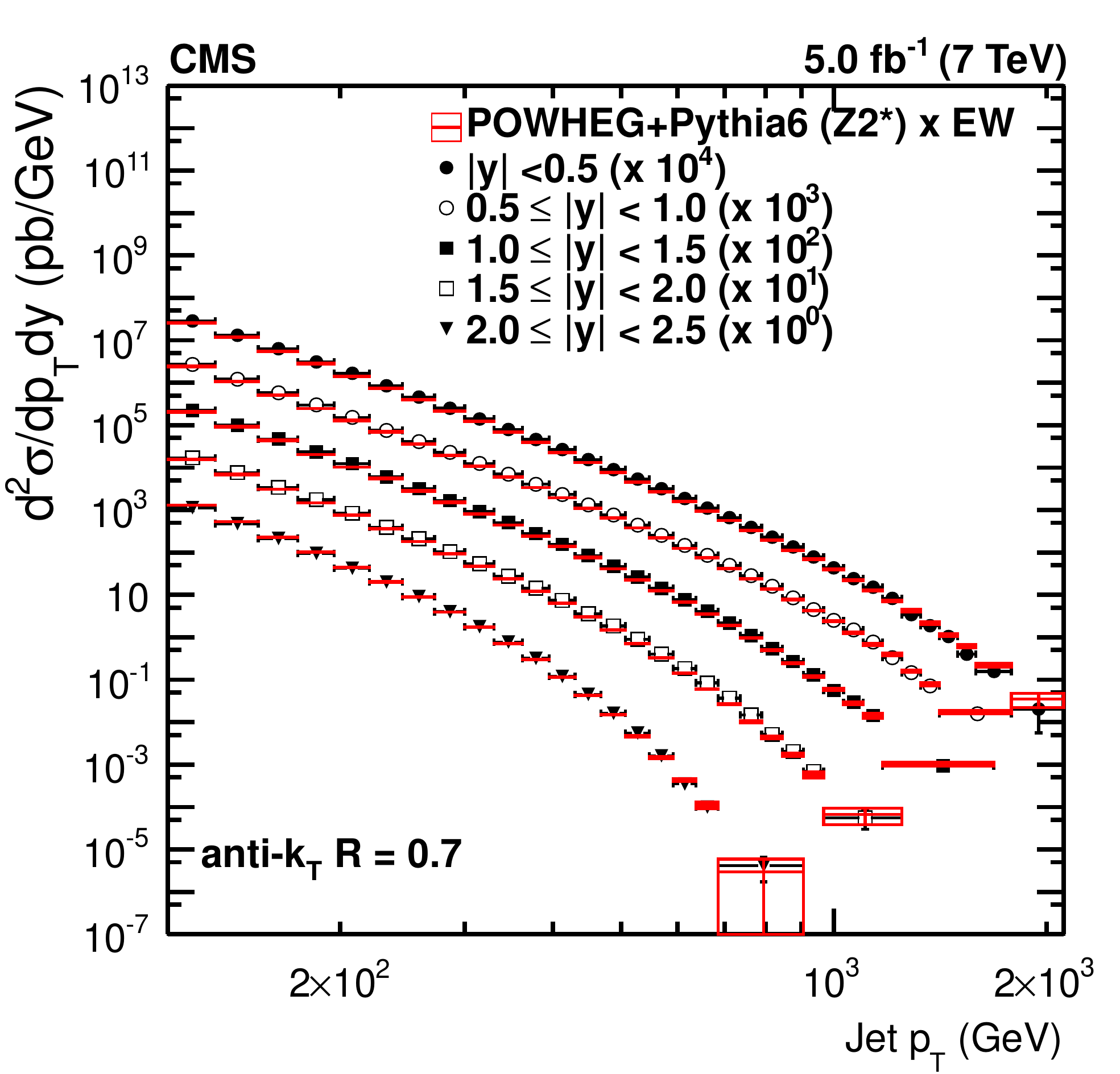}
  \caption{Measured inclusive jet cross section from
    Ref.~\cite{Chatrchyan:2012bja} compared to the prediction by
    \POWHEG + \PYTHIAS tune Z2* at particle level complemented with
    electroweak corrections. The boxes indicate the statistical
    uncertainty of the calculation.}
  \label{fig:DataTheory_comp4a}
\end{figure}

For illustration, Fig.~\ref{fig:DataTheory_comp4a} shows the
comparison of the inclusive jet data with the \POWHEG + \PYTHIAS tune
Z2* particle-level prediction complemented with electroweak
corrections. The tune Z2* is derived from the earlier tune Z2, where
the \PYTHIAS parameters PARP(82) and PARP(90) that control the
energy dependence of the MPI are retuned, yielding
1.921 and 0.227, respectively. The error boxes indicate statistical
uncertainties. Ratio plots of this comparison for each separate region
in \yabs can be found in Appendix~\ref{theory_data}.

The corrections to NLO parton-level calculations that are derived this
way consist of truly nonperturbative contributions, which are
optionally complemented with parton shower effects. They are
investigated separately in the following two sections. A previous
investigation can be found in Ref.~\cite{Dooling:2012uw}.

\subsubsection{NP corrections from \texorpdfstring{\POWHEG +
    \PYTHIAS}{POWHEG + PYTHIA6}}
\label{sec:npcorrection}

The NP corrections using a NLO prediction with a matched PS event
generator can be defined analogously as in Eq.~(\ref{e:C_LO_NP}):
\begin{equation}
  C_\mathrm{NLO}^{\text{NP}} = \frac{\sigma_{\text{NLO+PS+HAD+MPI}}}{\sigma_{\text{NLO+PS}}},
  \label{e:C_NLO_NP}
\end{equation}
\ie the numerator of this NP correction is defined by the inclusive
cross section, where parton showers, hadronization, and multiparton
interactions are turned on, while the inclusive cross section in the
denominator does not include hadronization and multiparton
interactions. A NLO calculation can then be corrected for NP effects
as
\begin{equation}
  \frac{\rd^2 \sigma_\mathrm{theo}}{\rd\pt\, \rd{}y} =
  \frac{\rd^2 \sigma_\mathrm{NLO}}{\rd\pt\, \rd{}y} \cdot C_\mathrm{NLO}^\mathrm{NP}.
\end{equation}

{\tolerance=900
In contrast to the LO MC event generation with \PYTHIAS, the
parameters of the NP and PS models, however, have not been retuned to
data for the use with NLO+PS predictions by \POWHEG. Therefore two
different underlying event tunes of \PYTHIAS for LO+PS predictions,
P11~\cite{Skands:2010ak} and Z2*, are used. In both cases a
parameterization using a functional form of $a_0 + a_1 / \pt^{a_2}$ is
employed to smoothen statistical fluctuations. For $\pt > 100\GeV$ the
difference in the NP correction factor between the two tunes is very
small such that their average is taken as
$C_\mathrm{NLO}^{\text{NP}}$.\par}

Since procedures to estimate uncertainties inherent to the NLO+PS
matching procedure are not yet well established and proper tunes to
data for \POWHEG + \PYTHIAS are lacking, the centre of the envelope
given by the three curves from \PYTHIAS, \HERWIGPP, and the \POWHEG +
\PYTHIAS average of tunes Z2* and P11 is adopted as the final NP
correction for the central results in Sections~\ref{sec:alphas}
and~\ref{sec:herafitter}. Half the spread among these three
predictions defines the uncertainty.

\begin{figure*}[phtb]
  \centering
    \includegraphics[width=0.48\textwidth]{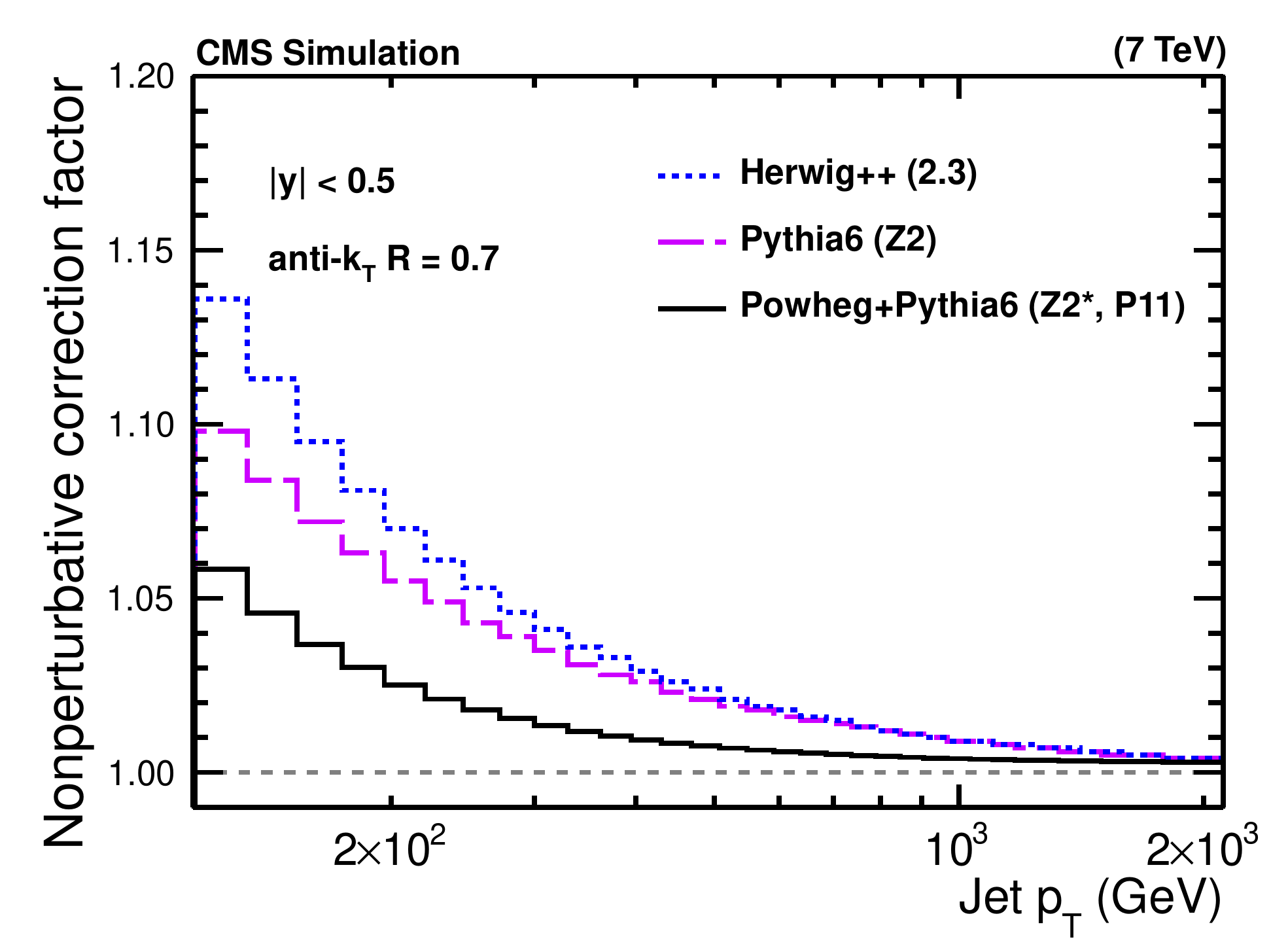}
    \includegraphics[width=0.48\textwidth]{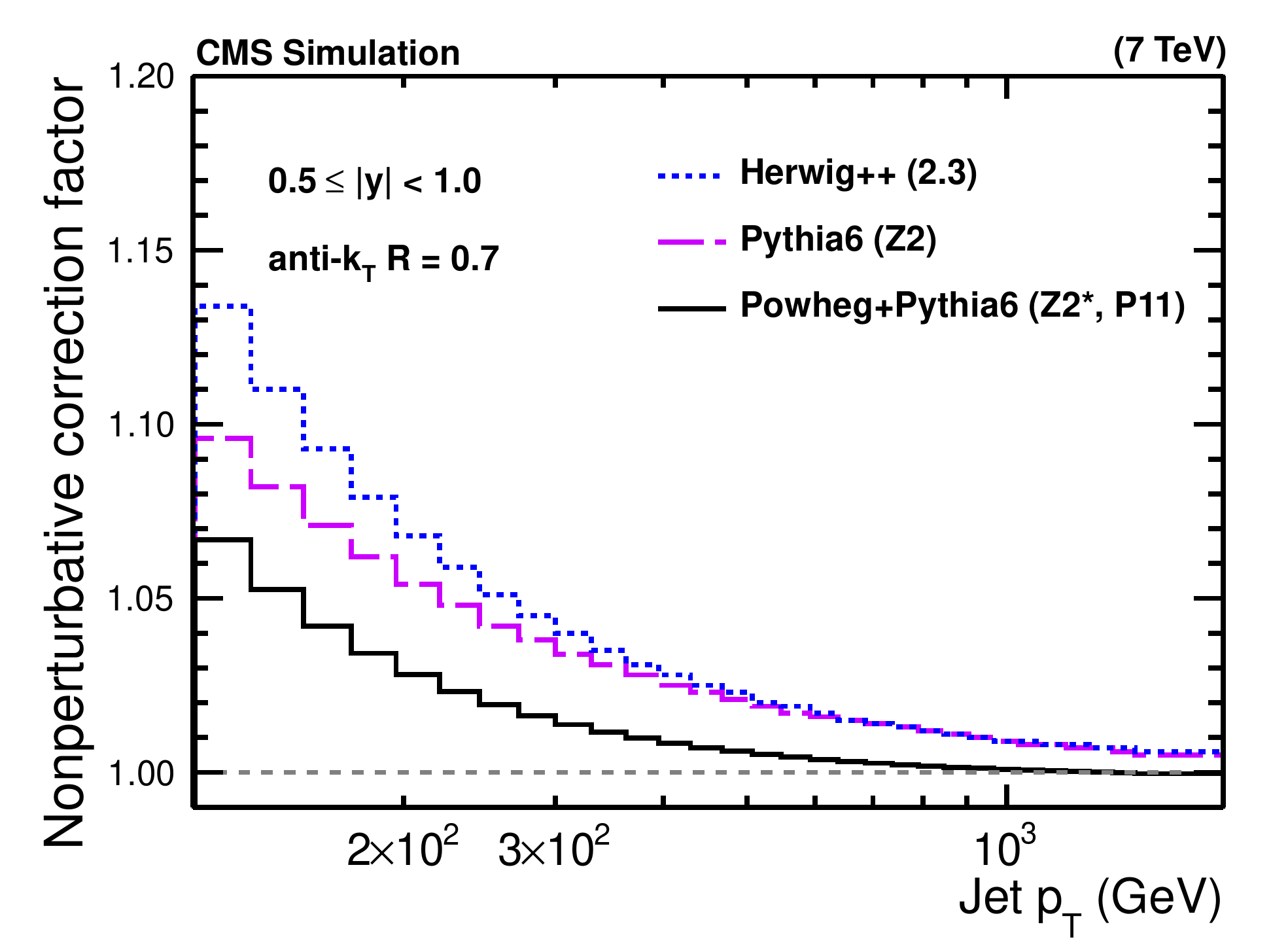}
    \includegraphics[width=0.48\textwidth]{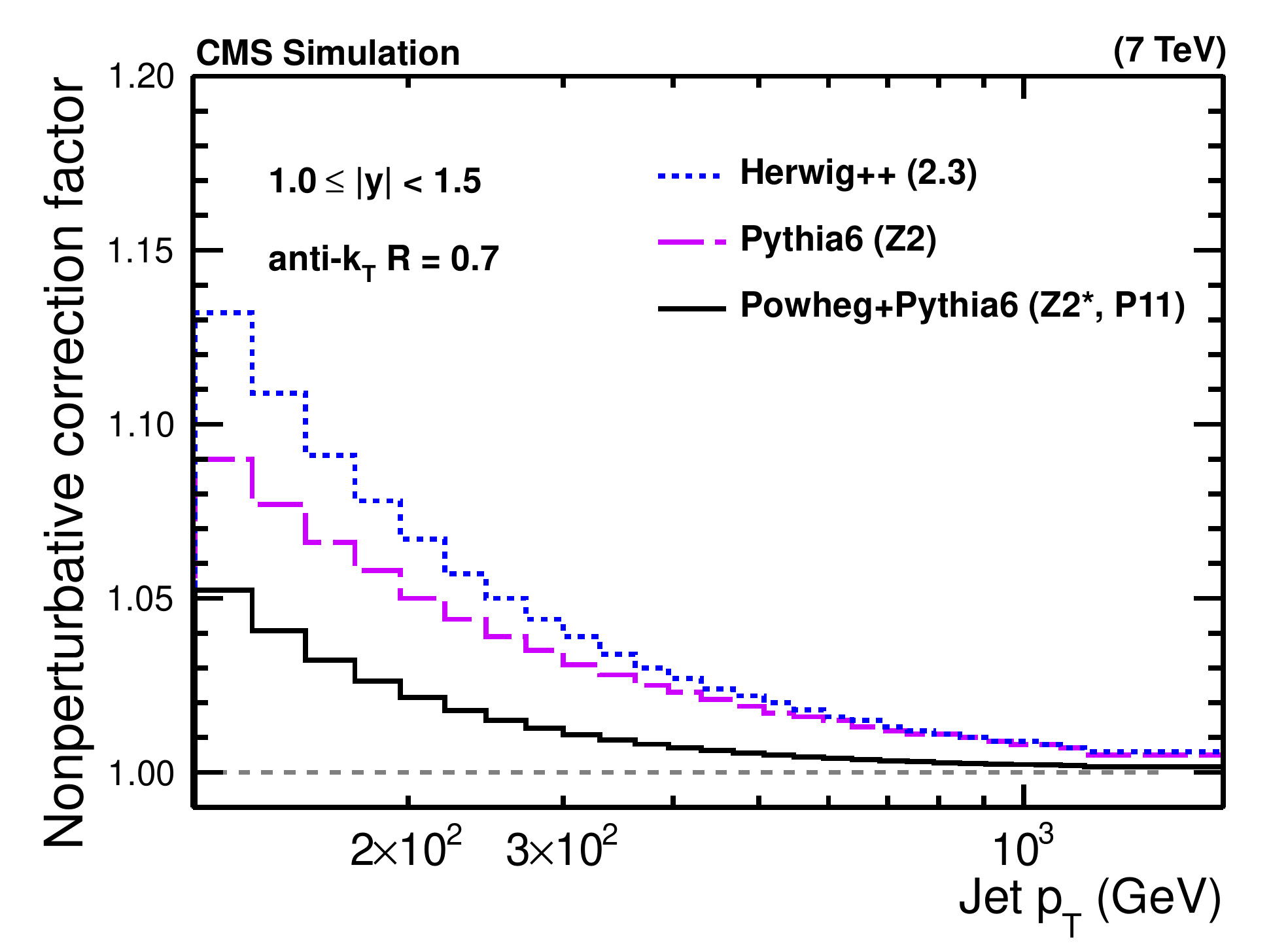}
    \includegraphics[width=0.48\textwidth]{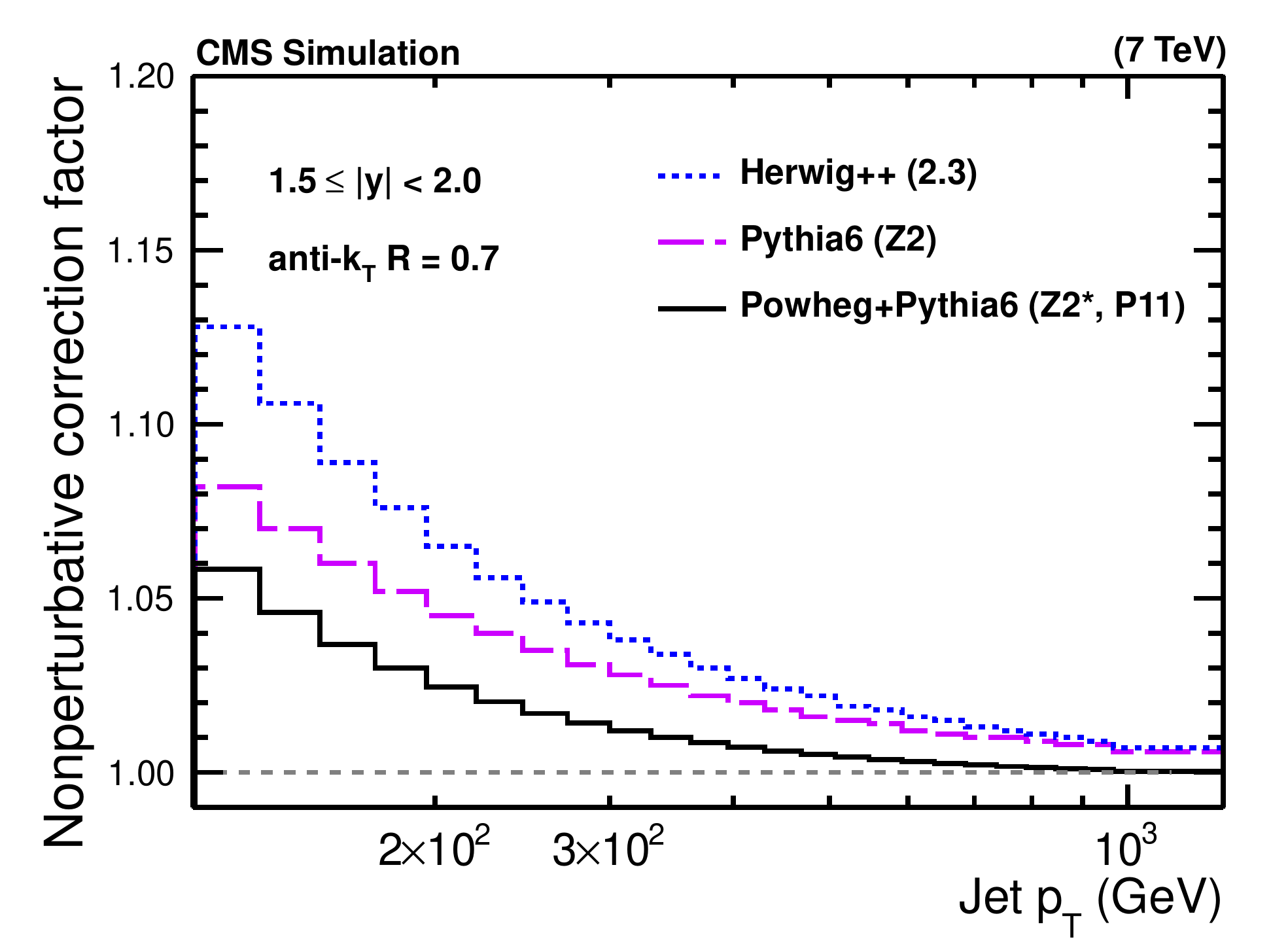}
    \includegraphics[width=0.48\textwidth]{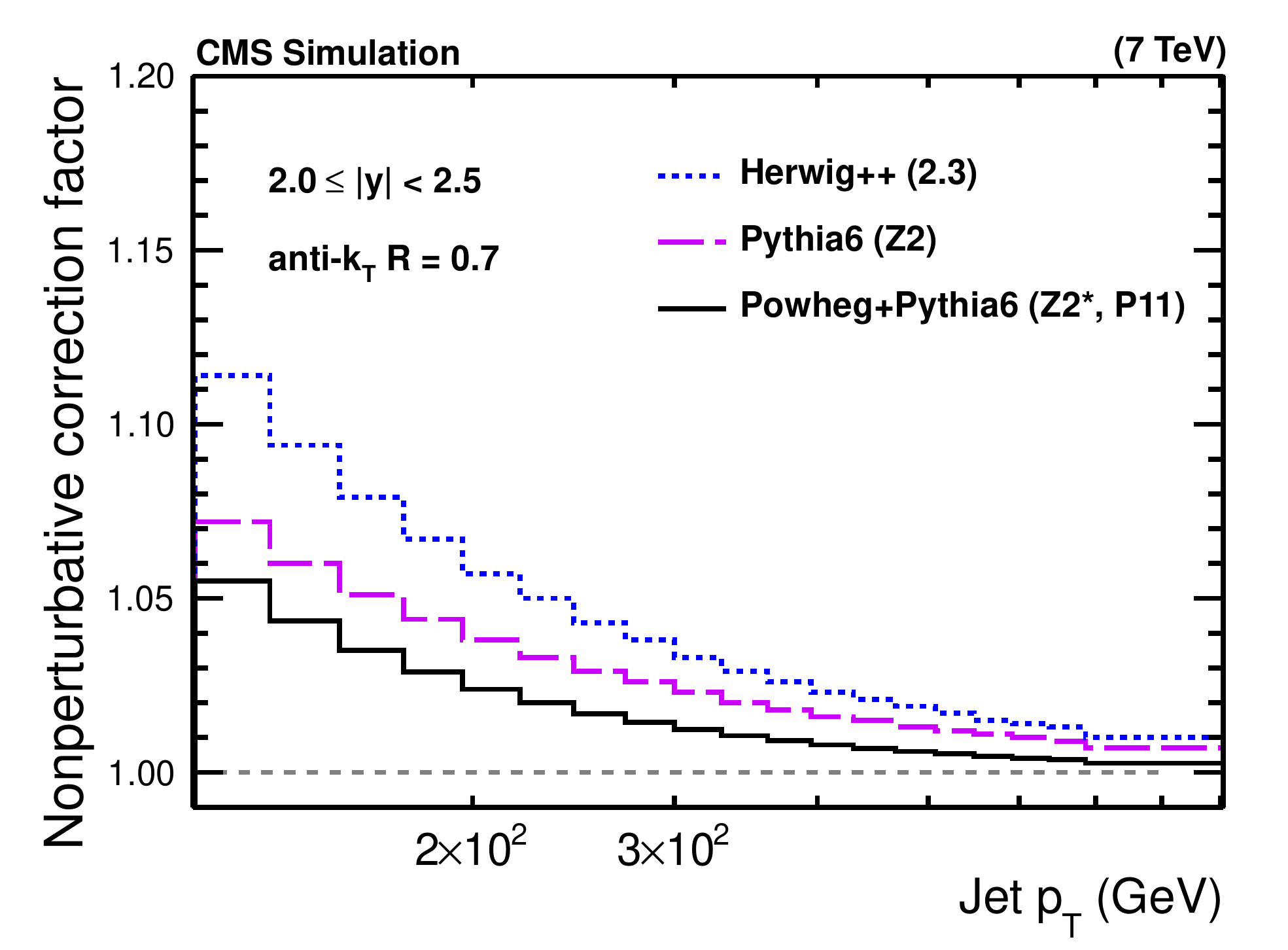}
    \caption{NP corrections for the five regions in \yabs as derived
      in Ref.~\cite{Chatrchyan:2012bja}, using \PYTHIAS tune Z2 and
      \HERWIGPP with the default tune of version~2.3, in comparison to
      corrections obtained from \POWHEG using \PYTHIAS for showering
      with the two underlying event tunes P11 and Z2*.}
    \label{fig:np_corrections_powheg_pythia}
\end{figure*}

The NP correction, as defined for \POWHEG + \PYTHIAS, is shown in
Fig.~\ref{fig:np_corrections_powheg_pythia} together with the original
factors from \PYTHIAS and \HERWIGPP, as a function of the jet \pt for
five ranges in absolute rapidity \yabs of size 0.5 up to $\yabs =
2.5$. The factors derived from both, LO+PS and NLO+PS MC event
generators, are observed to decrease with increasing jet \pt and to
approach unity at large \pt. Within modelling uncertainties, the
assumption of universal NP corrections that are similar for LO+PS and
NLO+PS MC event generation holds approximately above a jet \pt of a
few hundred \GeV.

\subsubsection{PS corrections from \texorpdfstring{\POWHEG +
    \PYTHIAS}{POWHEG + PYTHIA6}}
\label{sec:pscorrection}

Similarly to the NP correction of Eq.~(\ref{e:C_NLO_NP}), a PS
correction factor can be defined as the ratio of the differential
cross section including PS effects divided by the NLO prediction, as
given by \POWHEG, \ie including the leading emission:
\begin{equation}
  C _\mathrm{NLO}^{\text{PS}} =
  \frac{\sigma_{\text{NLO+PS}}}{\sigma_{\text{NLO}}}.
  \label{e:C_NLO_PS}
\end{equation}

The combined correction for NP and PS effects can then be written as
\begin{equation}
  \frac{\rd^2 \sigma_\mathrm{theo}}{\rd\pt\, \rd{}y} =
  \frac{\rd^2 \sigma_\mathrm{NLO}}{\rd\pt\, \rd{}y} \cdot C_\mathrm{NLO}^\mathrm{NP}
  \cdot C_\mathrm{NLO}^\mathrm{PS}.
\end{equation}

\begin{figure*}[phtb]
  \centering
    \includegraphics[width=0.48\textwidth]{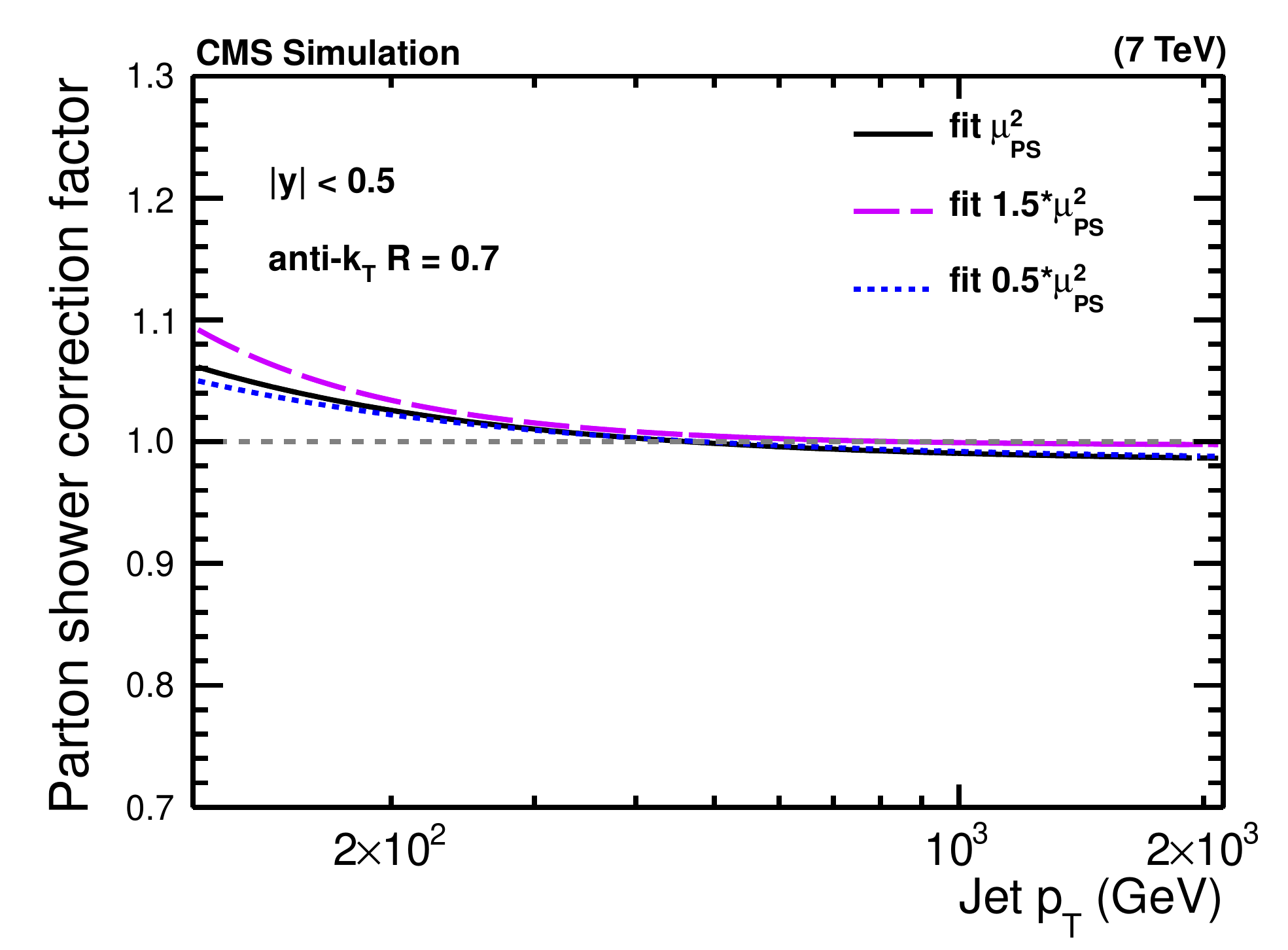}
    \includegraphics[width=0.48\textwidth]{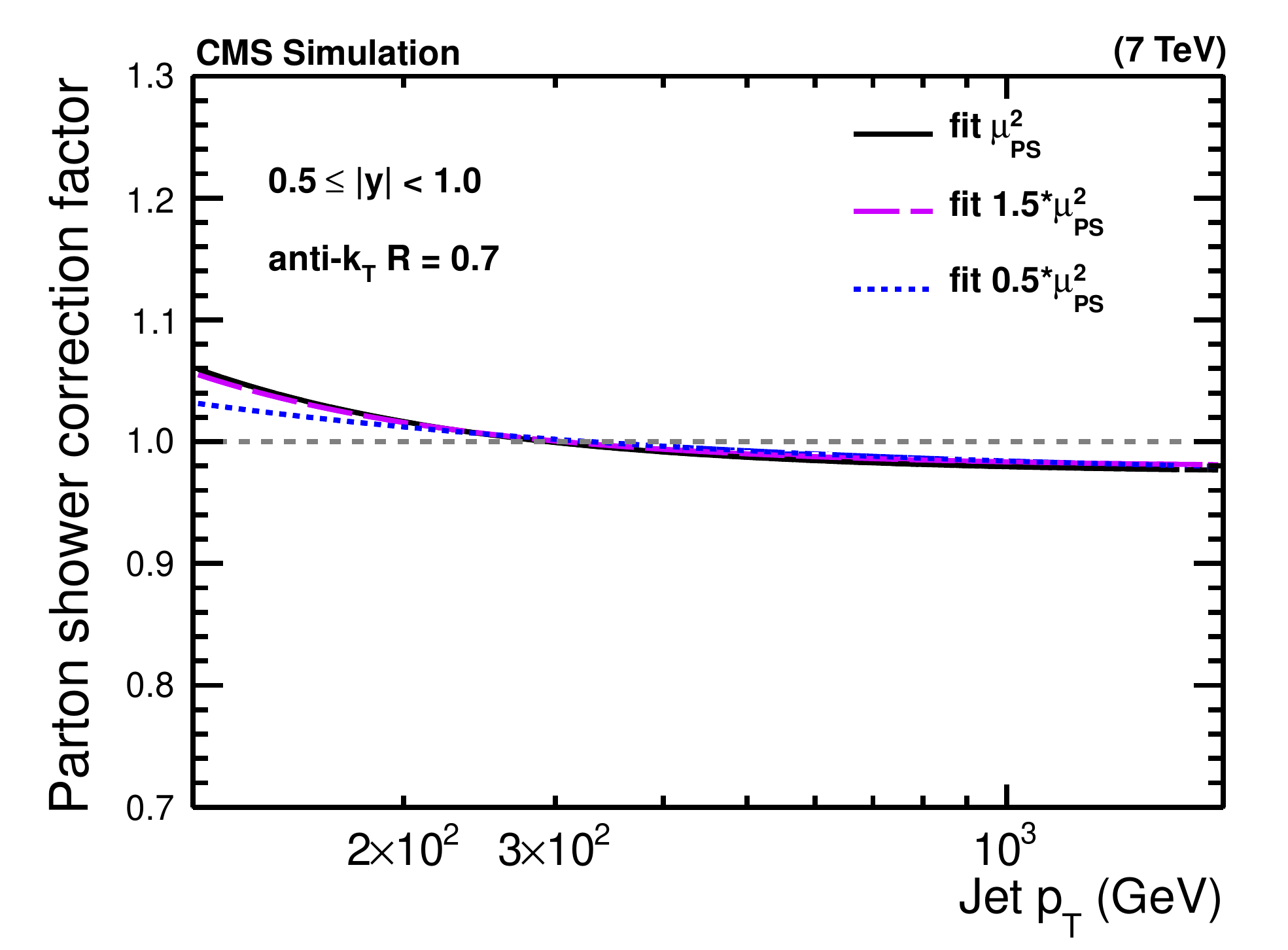}
    \includegraphics[width=0.48\textwidth]{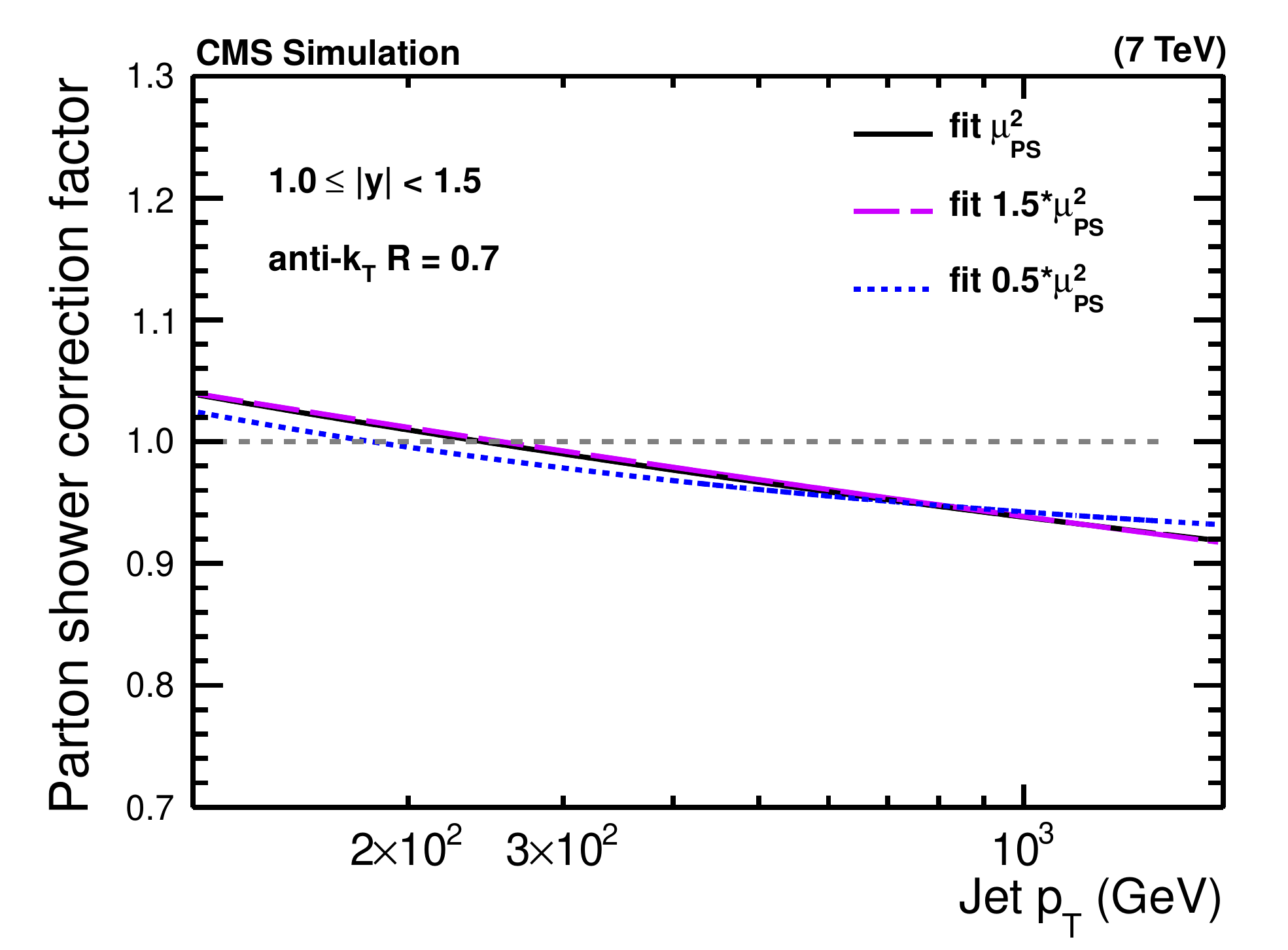}
    \includegraphics[width=0.48\textwidth]{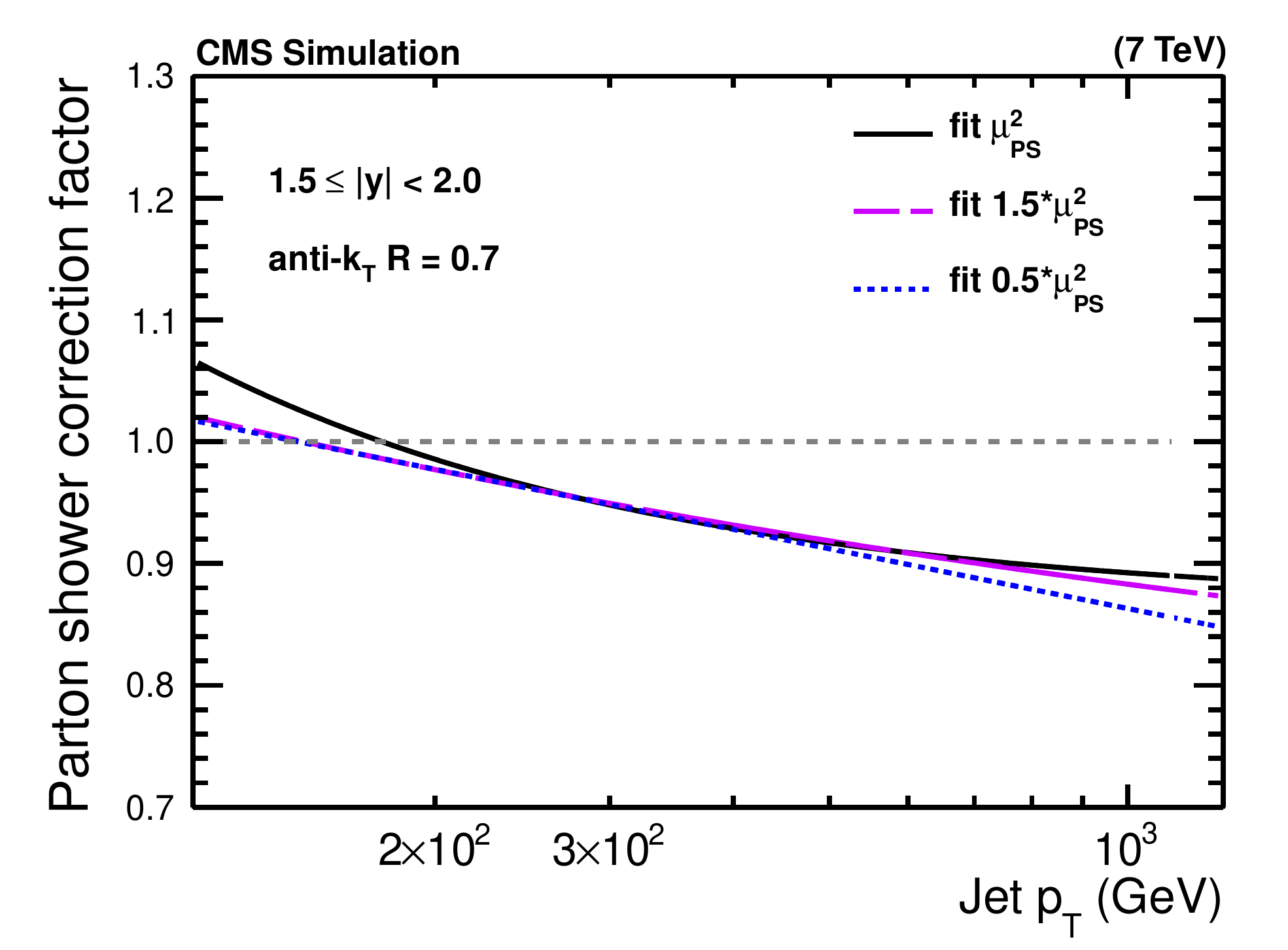}
    \includegraphics[width=0.48\textwidth]{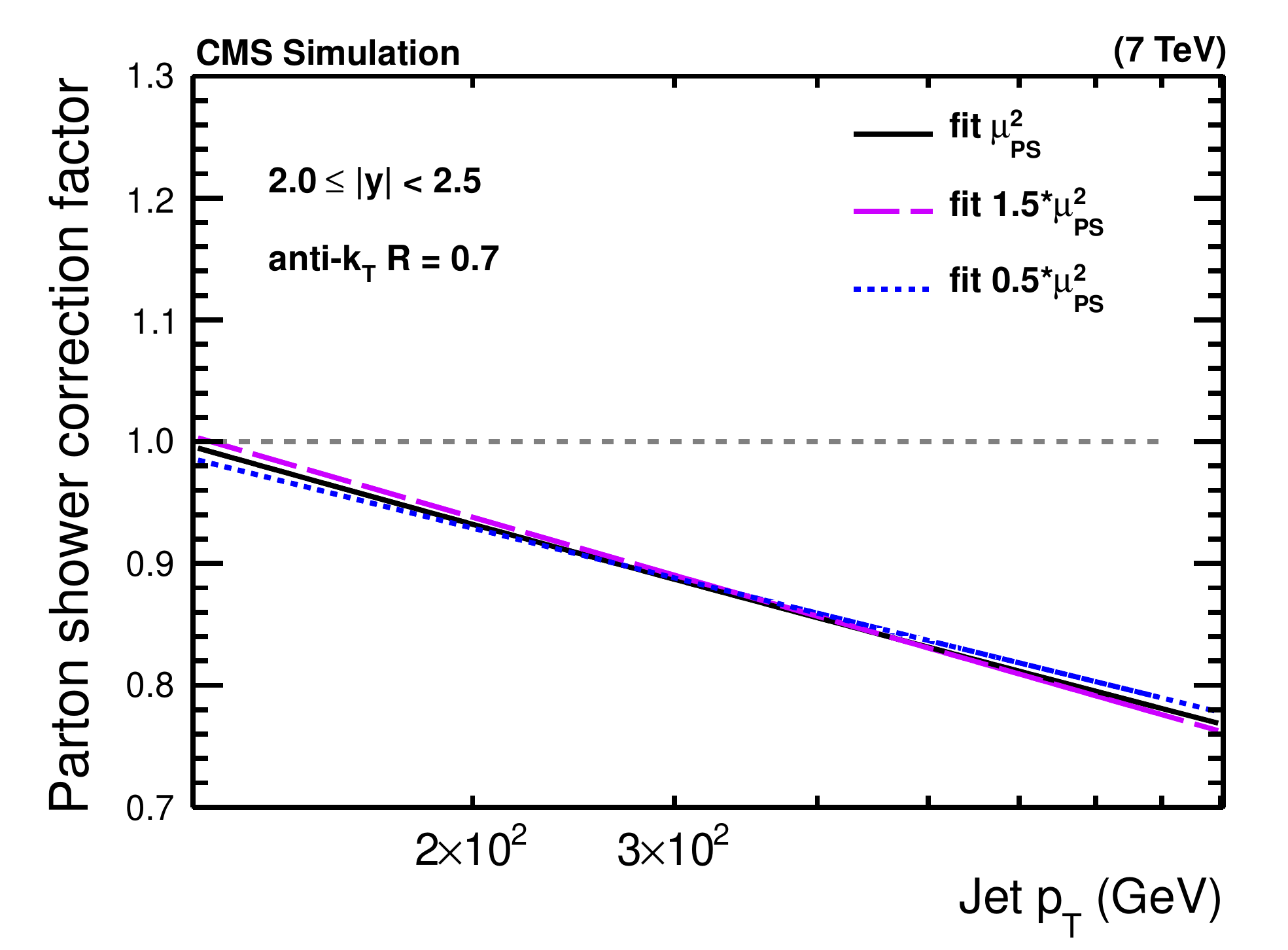}
    \caption{PS corrections for the five regions in \yabs obtained
      from \POWHEG using \PYTHIAS for showering for different upper
      scale limits of the parton shower evolution in \PYTHIAS tune
      Z2*\@. The curves parameterize the correction factors as a
      function of the jet \pt.}
    \label{fig:ps_corrections_powheg_pythia}
\end{figure*}

The PS corrections derived with \POWHEG + \PYTHIAS are presented in
Fig.~\ref{fig:ps_corrections_powheg_pythia}. They
are significant at large \pt, particularly at high
rapidity, where the factors approach $-20$\%. However, the
combination of \POWHEG + \PYTHIAS has never been tuned
to data and the Z2* tune strictly is only valid for a LO+PS tune
with \PYTHIAS, but not with showers matched to \POWHEG. Moreover,
\POWHEG employs the CT10-NLO PDF, while the Z2* tune requires the
CTEQ6L1-LO PDF to be used for the showering part. Therefore, such PS
corrections can be considered as only an illustrative test, as
reported in Section~\ref{section:results_a_s}.

The maximum parton virtuality allowed in the parton shower evolution,
$\mu_\mathrm{PS}^2$, is varied by factors of 0.5 and 1.5 by changing
the corresponding parameter PARP(67) in \PYTHIAS from its default
value of 4 to 2 and 6, respectively.  The resulting changes in the PS
factors are shown in Fig.~\ref{fig:ps_corrections_powheg_pythia}. The
\POWHEG + \PYTHIAS PS factors employed in an illustrative test later
are determined as the average of the predictions from the two extreme
scale limits. Again, a parameterization using a functional form of
$a_0 + a_1 / \pt^{a_2}$ is employed to smoothen statistical
fluctuations.

\begin{figure}[btp]
  \centering
    \includegraphics[width=0.48\textwidth]{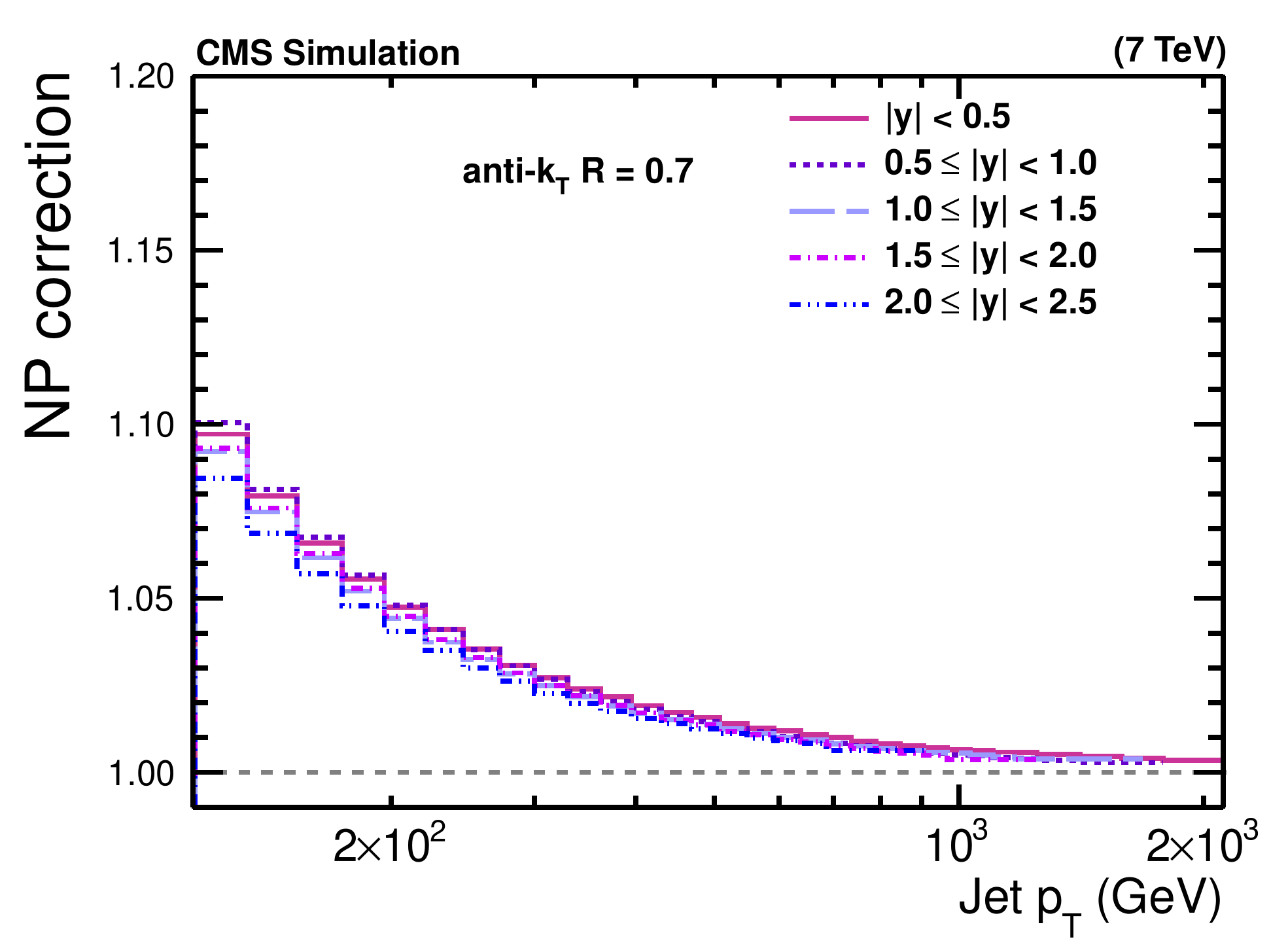}
    \includegraphics[width=0.48\textwidth]{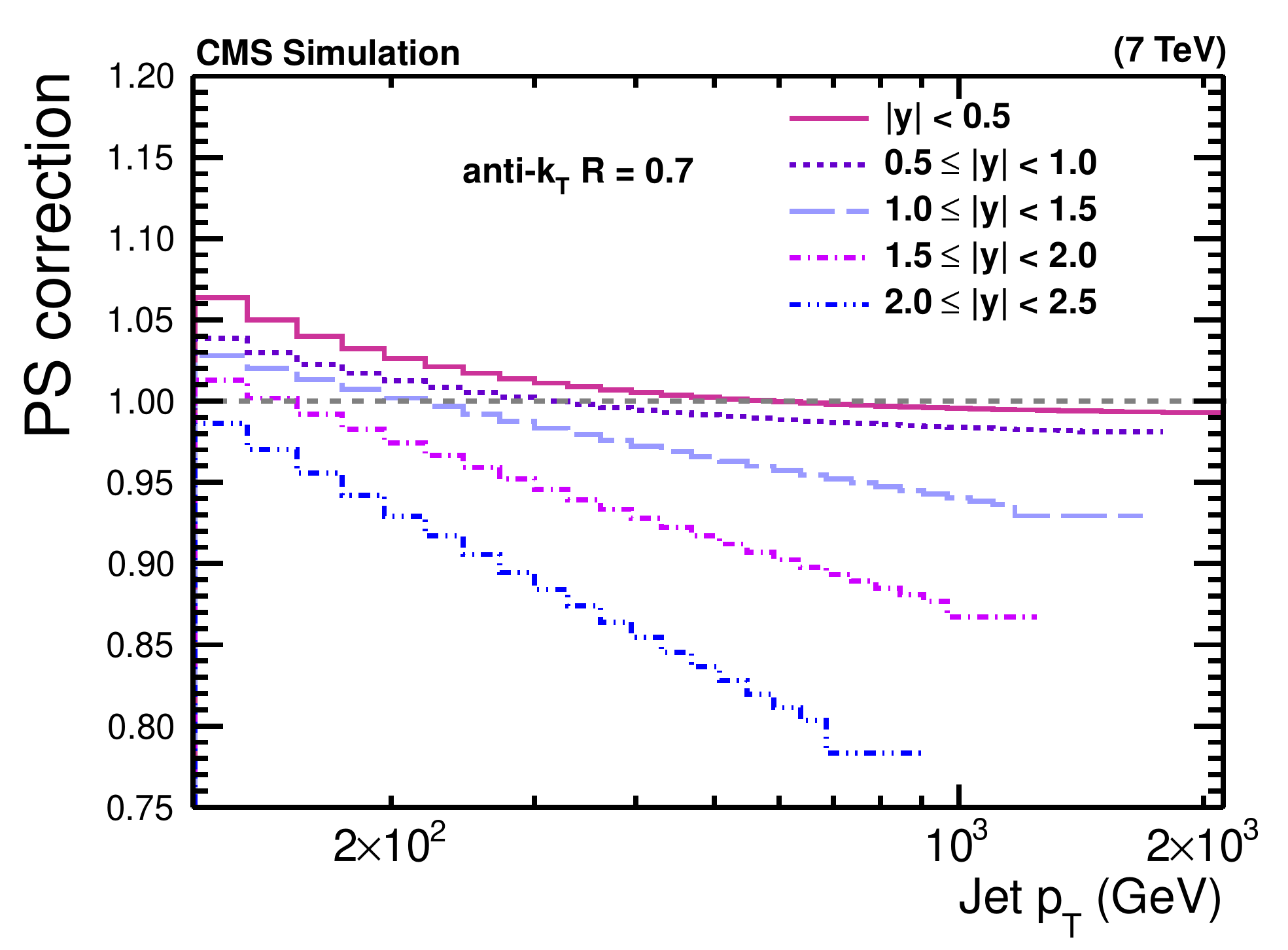}
    \includegraphics[width=0.48\textwidth]{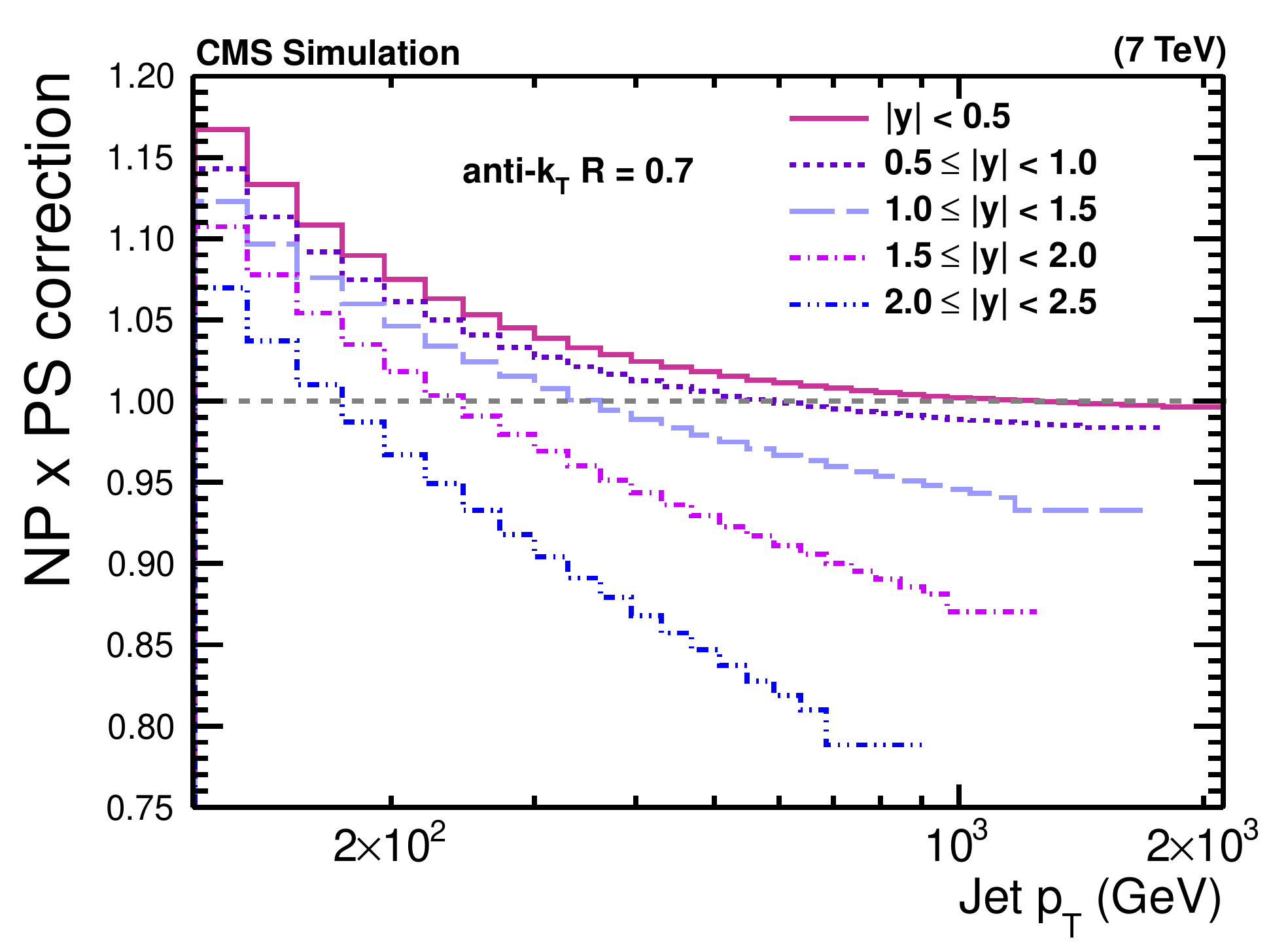}
    \caption{NP correction (top left) obtained from the envelope of
      the predictions of \PYTHIAS tune Z2, \HERWIGPP tune 2.3, and
      \POWHEG + \PYTHIAS with the tunes P11 and Z2*, PS correction
      (top right) obtained from the average of the predictions of
      \POWHEG + \PYTHIAS tune Z2* with scale factor variation, and
      combined correction (bottom), defined as the product of the NP
      and PS correction, for the five regions in \yabs.}
    \label{fig:total_corrections_powheg_pythia}
\end{figure}

Finally, Fig.~\ref{fig:total_corrections_powheg_pythia} presents an
overview of the NP, PS, and combined corrections for all five ranges
in \yabs.

\section{Determination of the strong coupling constant}
\label{sec:alphas}

The measurement of the inclusive jet cross
section~\cite{Chatrchyan:2012bja}, as described in
Section~\ref{sec:measurement}, can be used to determine \alpsmz, where
the proton structure in the form of PDFs is taken as a
prerequisite. The necessary theoretical ingredients are specified in
Section~\ref{sec:theory}.  The choice of PDF sets is restricted to
global sets that fit data from different experiments, so that only the
most precisely known gluon distributions are employed. Combined fits
of \alpsmz and the gluon content of the proton are investigated in
Section~\ref{sec:combinedfits}.

In the following, the sensitivity of the inclusive jet cross section
to \alpsmz is demonstrated. Subsequently, the fitting procedure is
given in detail before presenting the outcome of the various fits of
\alpsmz.

\subsection{Sensitivity of the inclusive jet cross section to
  \texorpdfstring{\alpsmz}{alpha-S(M(Z))}\label{section:sens_as}}

\begin{figure*}[pt]
  \centering
  \includegraphics[width=0.47\textwidth]{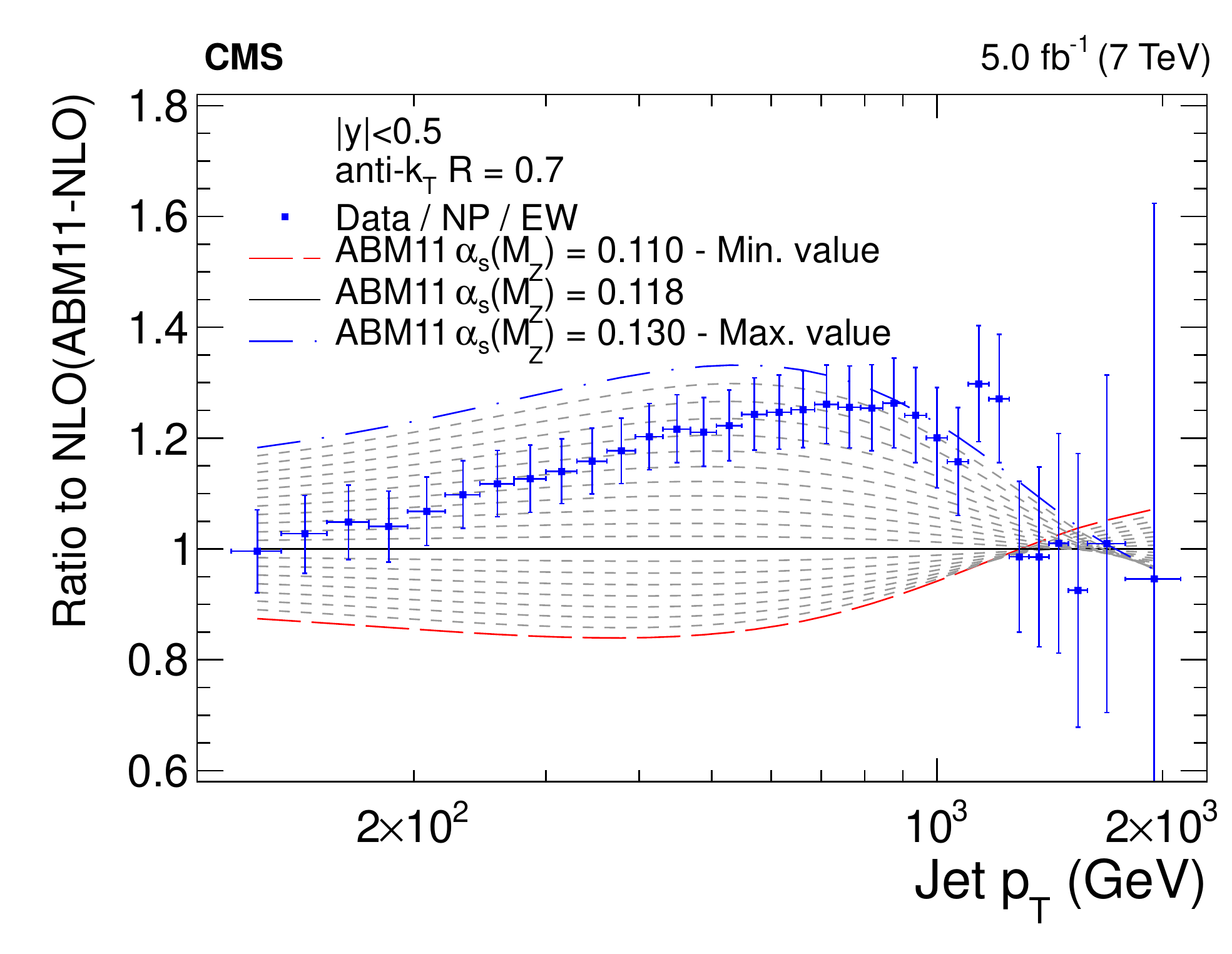}\hfill%
  \includegraphics[width=0.47\textwidth]{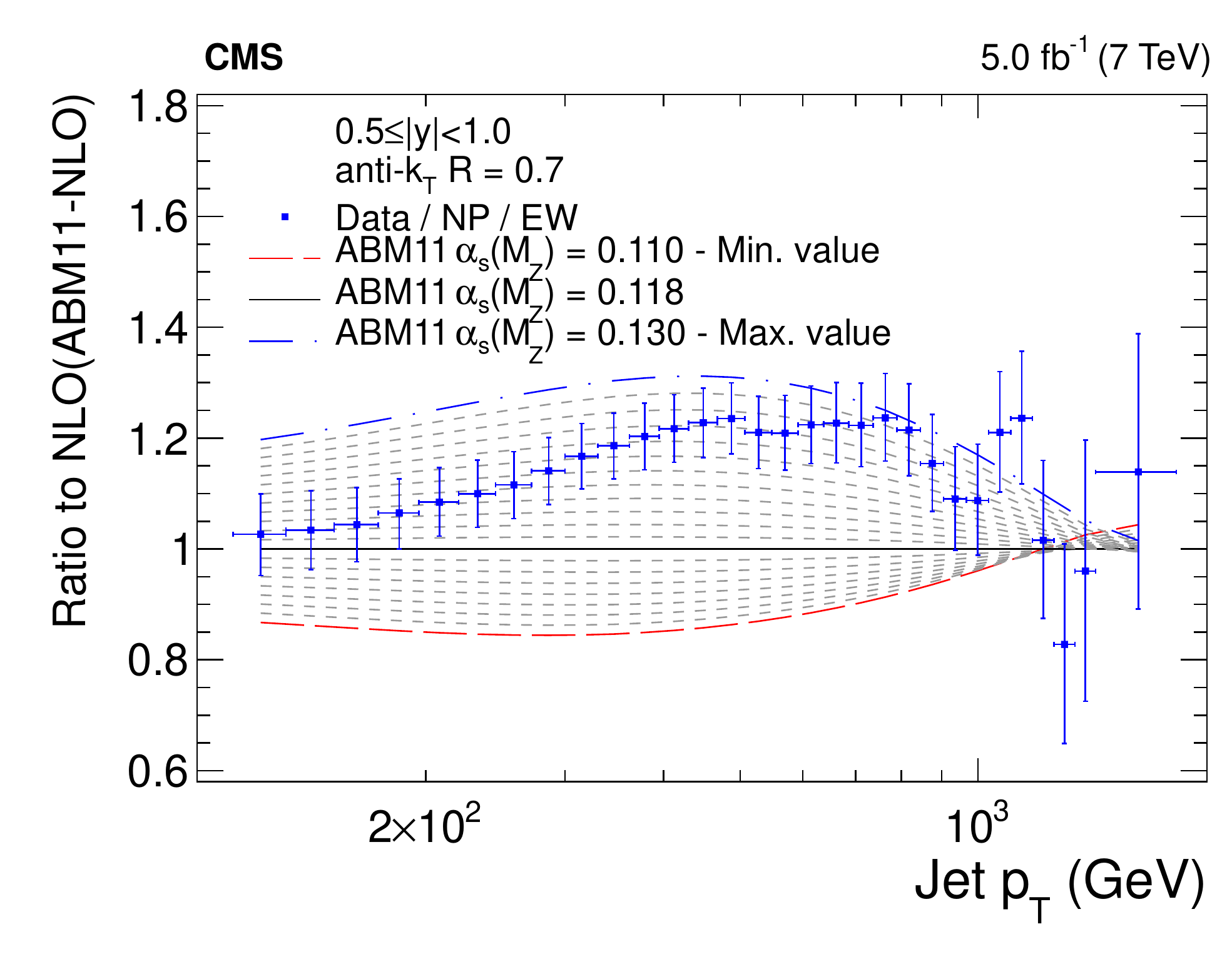}
  \includegraphics[width=0.47\textwidth]{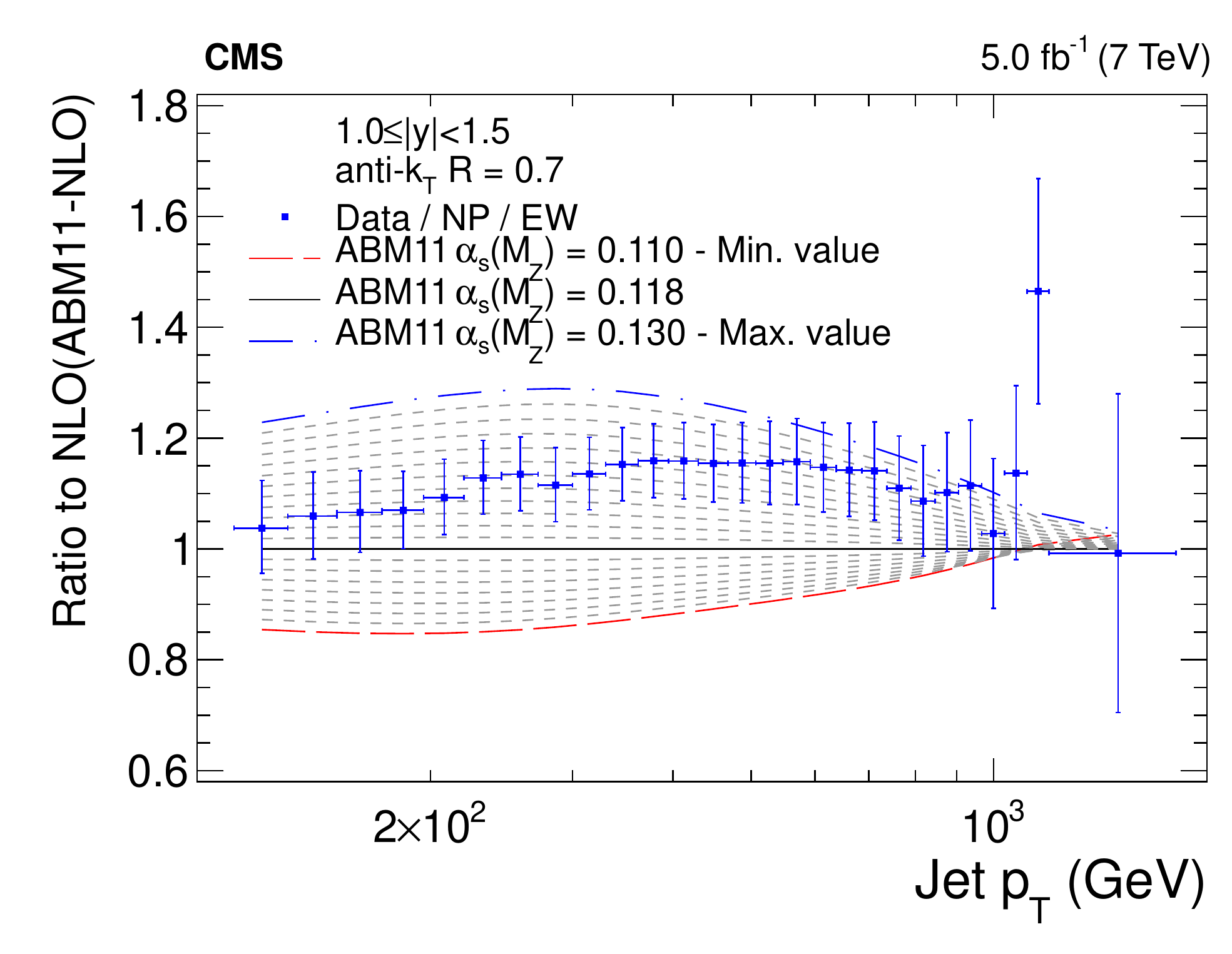}\hfill%
  \includegraphics[width=0.47\textwidth]{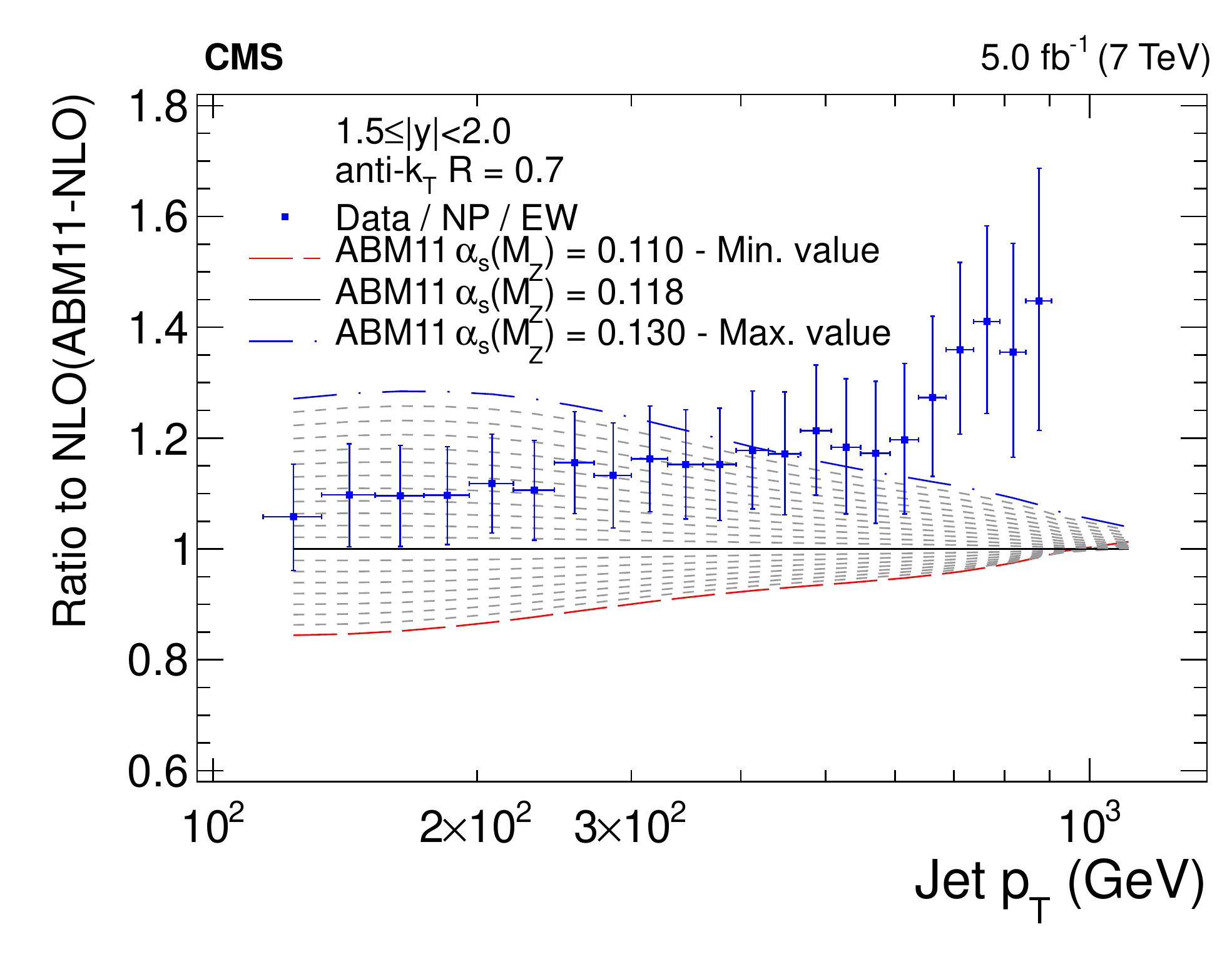}
  \includegraphics[width=0.47\textwidth]{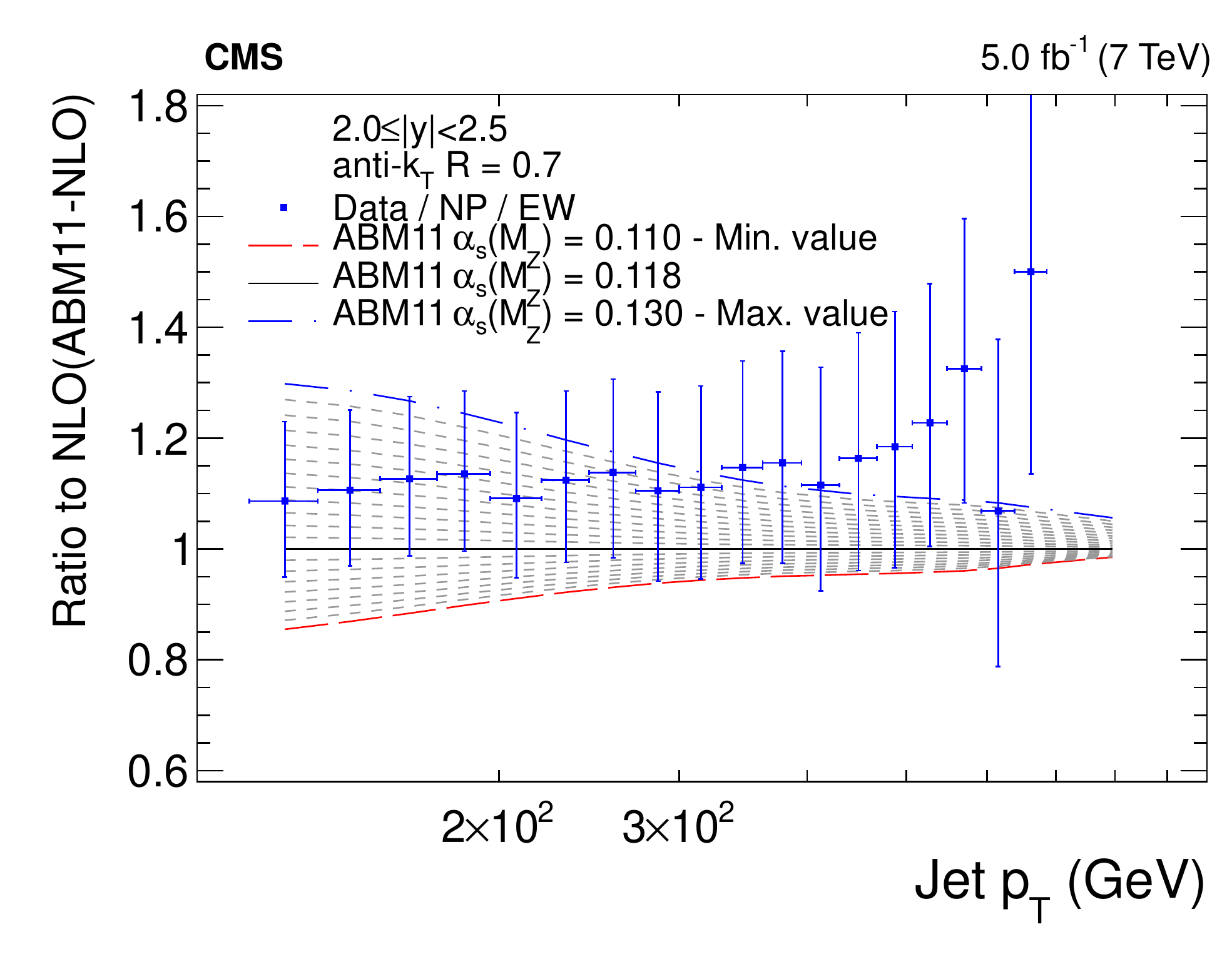}
  \caption{Ratio of the inclusive jet cross section to theoretical
    predictions using the ABM11-NLO PDF set for the five rapidity
    bins, where the \alpsmz value is varied in the range
    0.110--0.130 in steps of 0.001. The error bars correspond to
    the total uncertainty.}
  \label{fig:DataTheory_as_ABM11nlo}
\end{figure*}

\begin{figure*}[pt]
  \centering
  \includegraphics[width=0.47\textwidth]{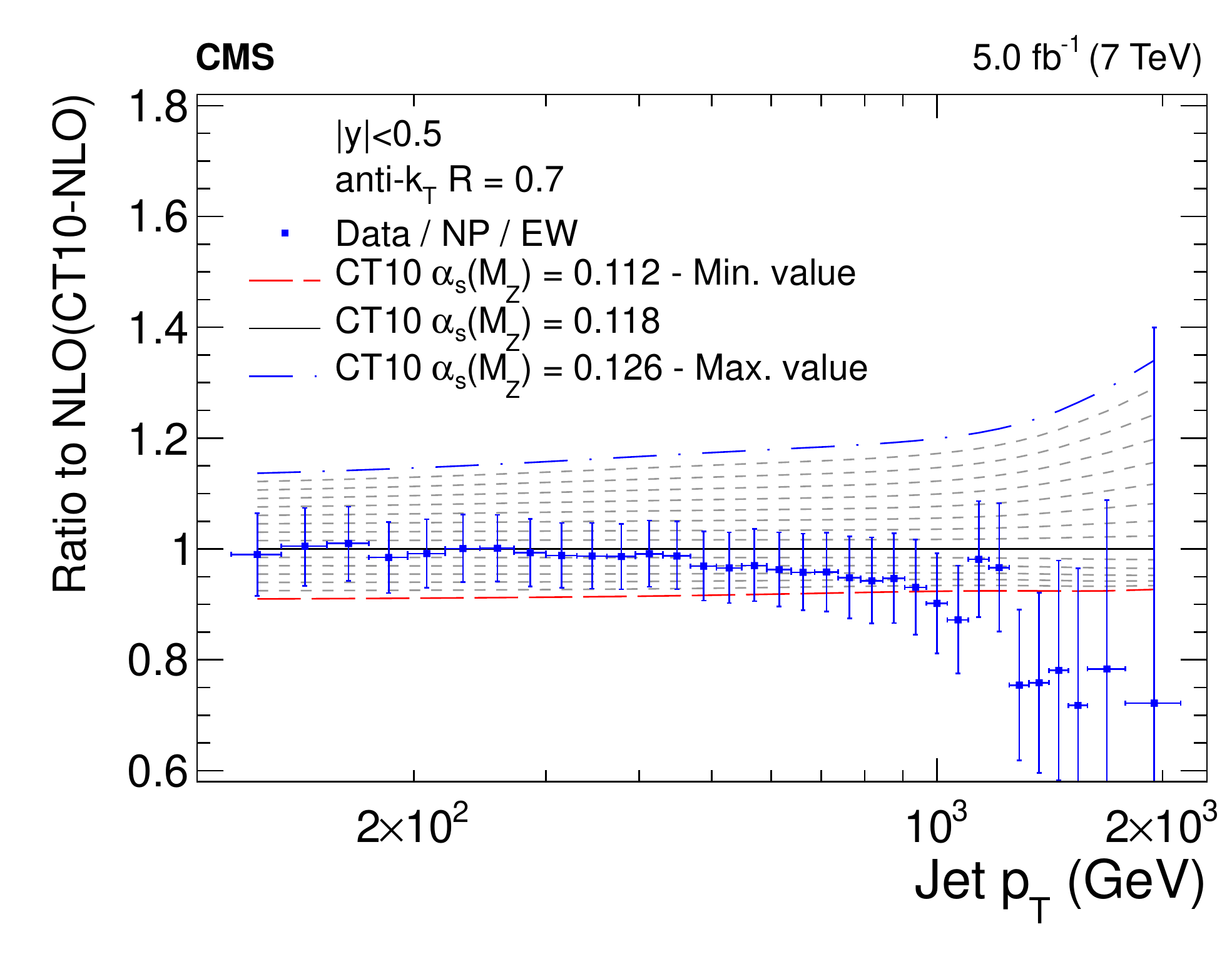}\hfill%
  \includegraphics[width=0.47\textwidth]{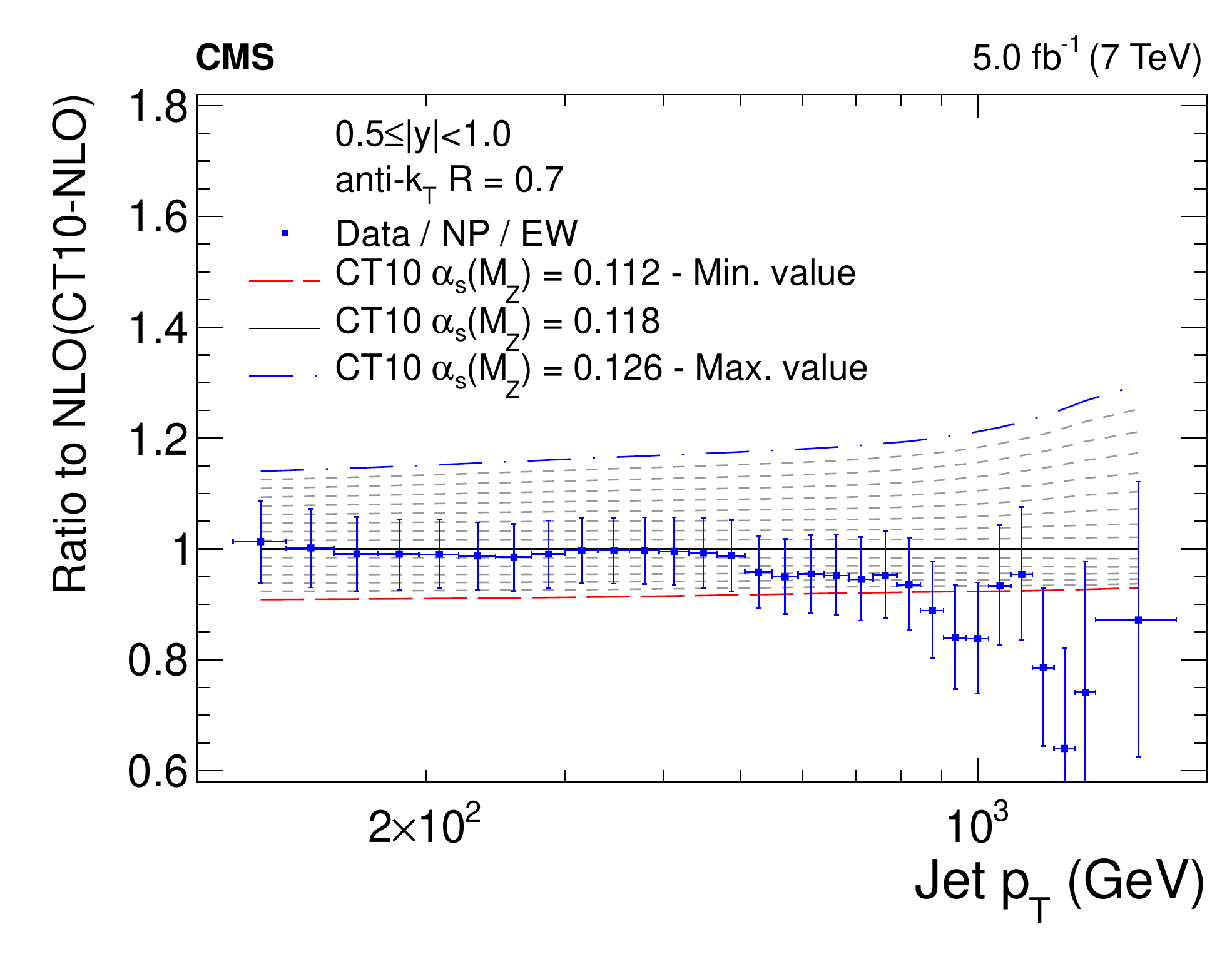}
  \includegraphics[width=0.47\textwidth]{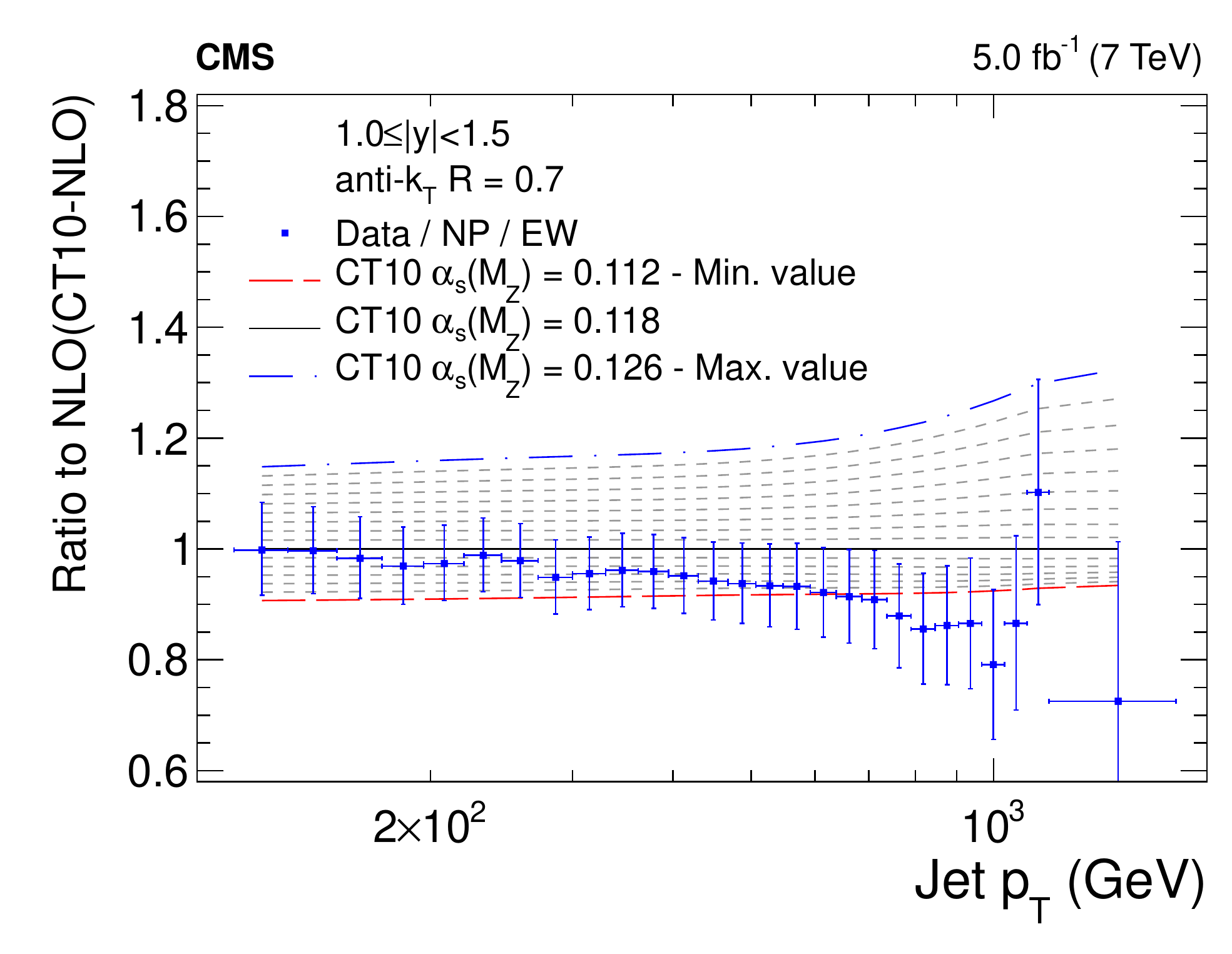}\hfill%
  \includegraphics[width=0.47\textwidth]{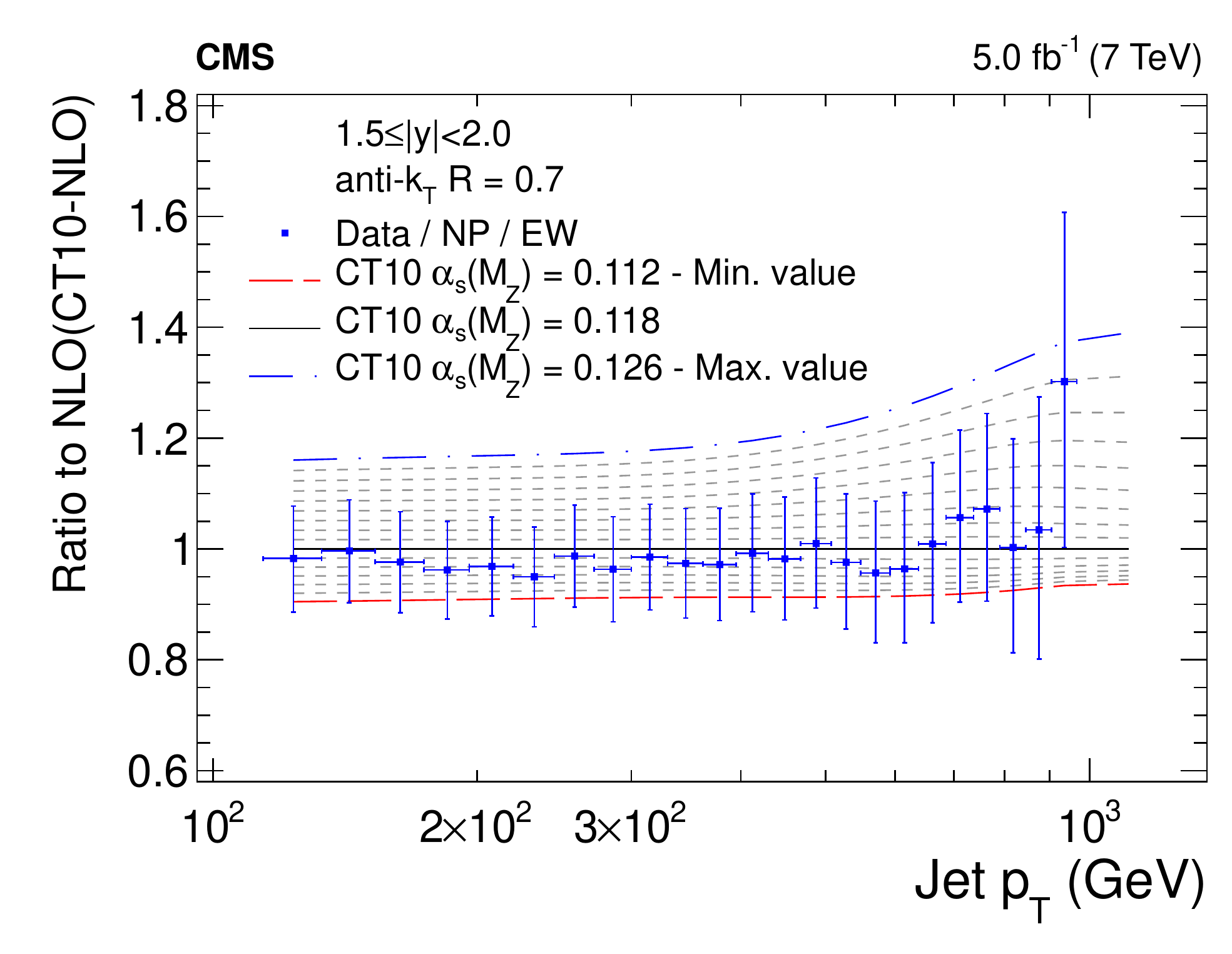}
  \includegraphics[width=0.47\textwidth]{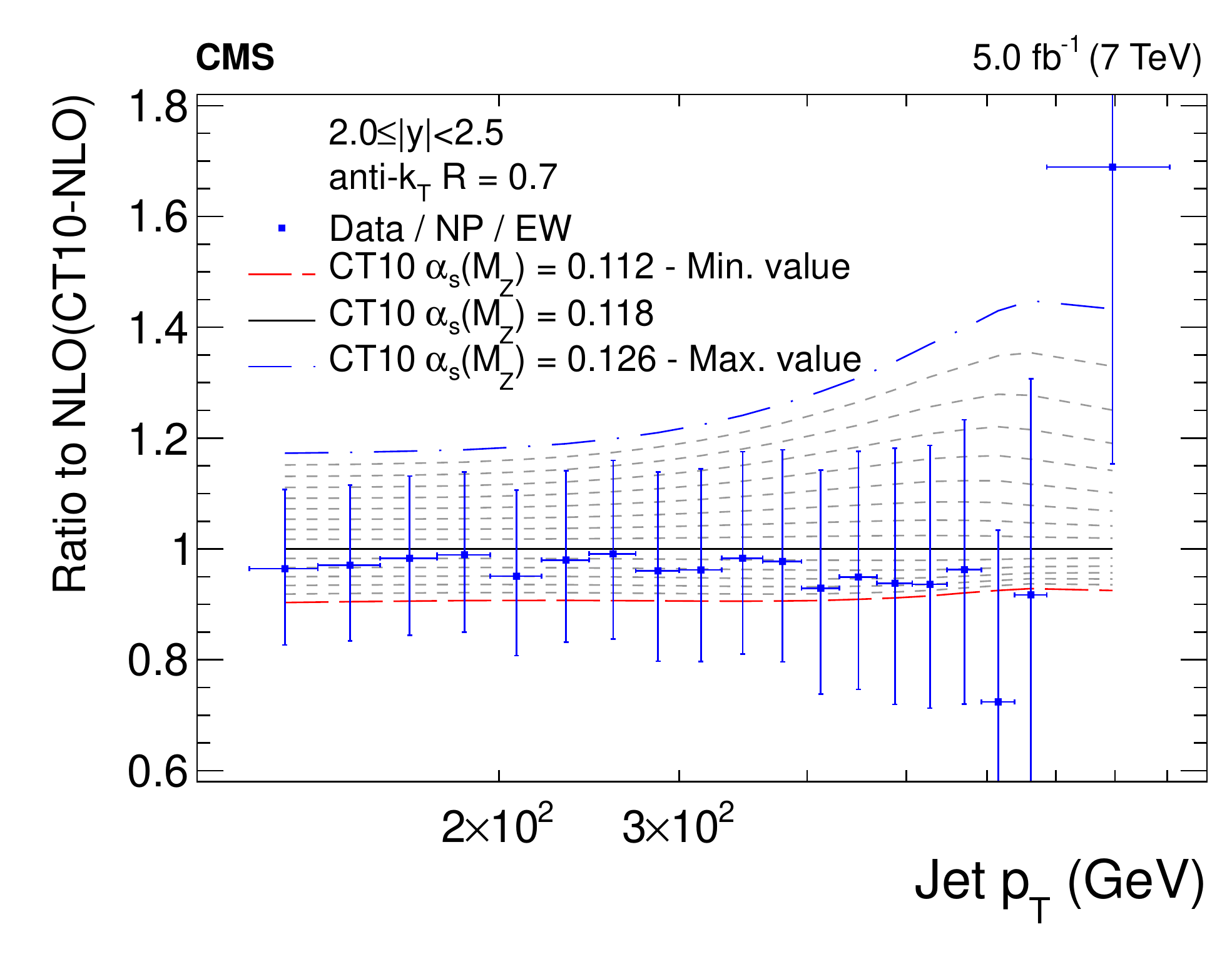}
  \caption{Ratio of the inclusive jet cross section to theoretical
    predictions using the CT10-NLO PDF set for the five rapidity bins,
    where the \alpsmz value is varied in the range $0.112$--$0.126$ in
    steps of 0.001. The error bars correspond to the total
    uncertainty.}
  \label{fig:DataTheory_as_CT10nlo}
\end{figure*}

\begin{figure*}[pt]
  \centering
  \includegraphics[width=0.47\textwidth]{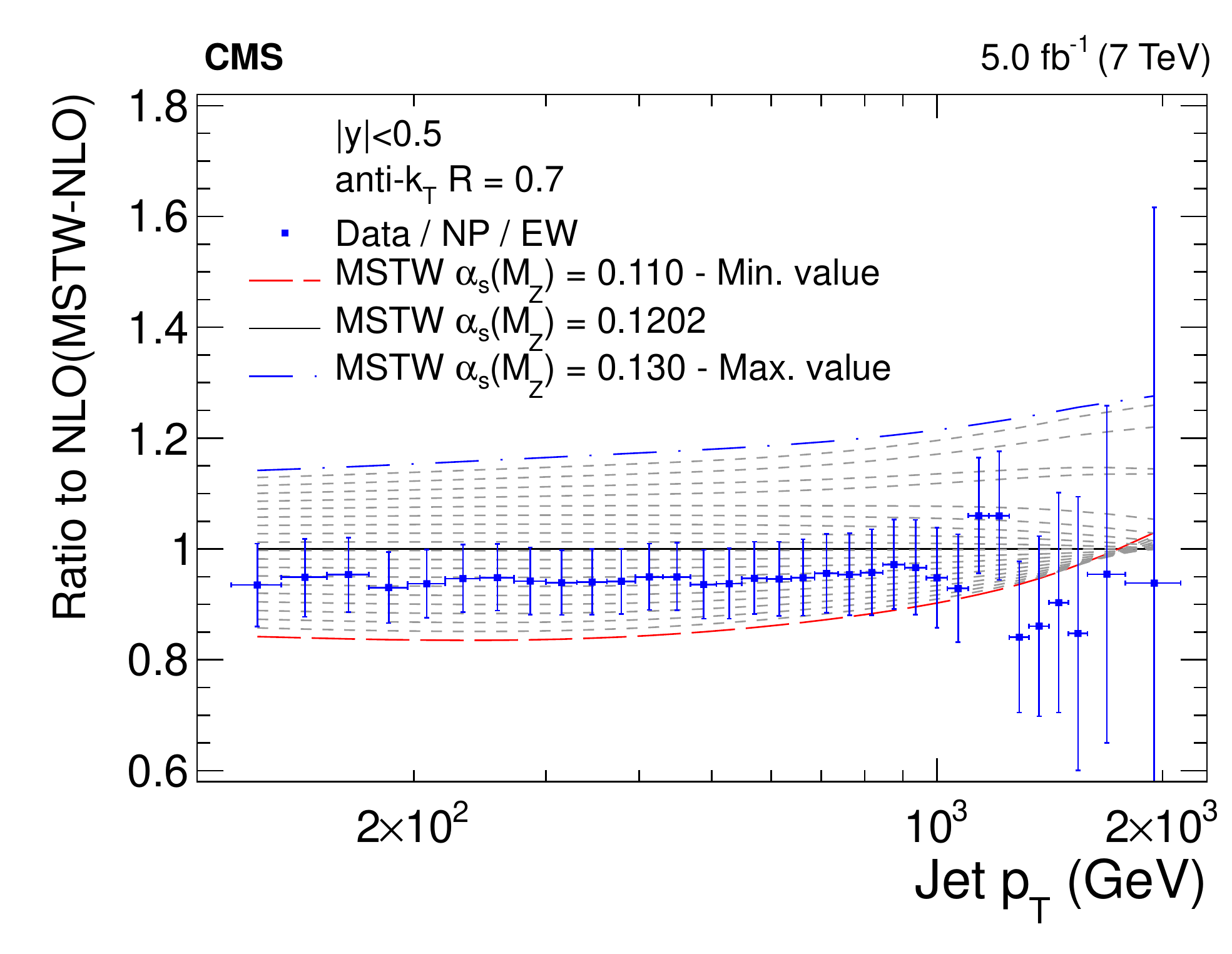}\hfill%
  \includegraphics[width=0.47\textwidth]{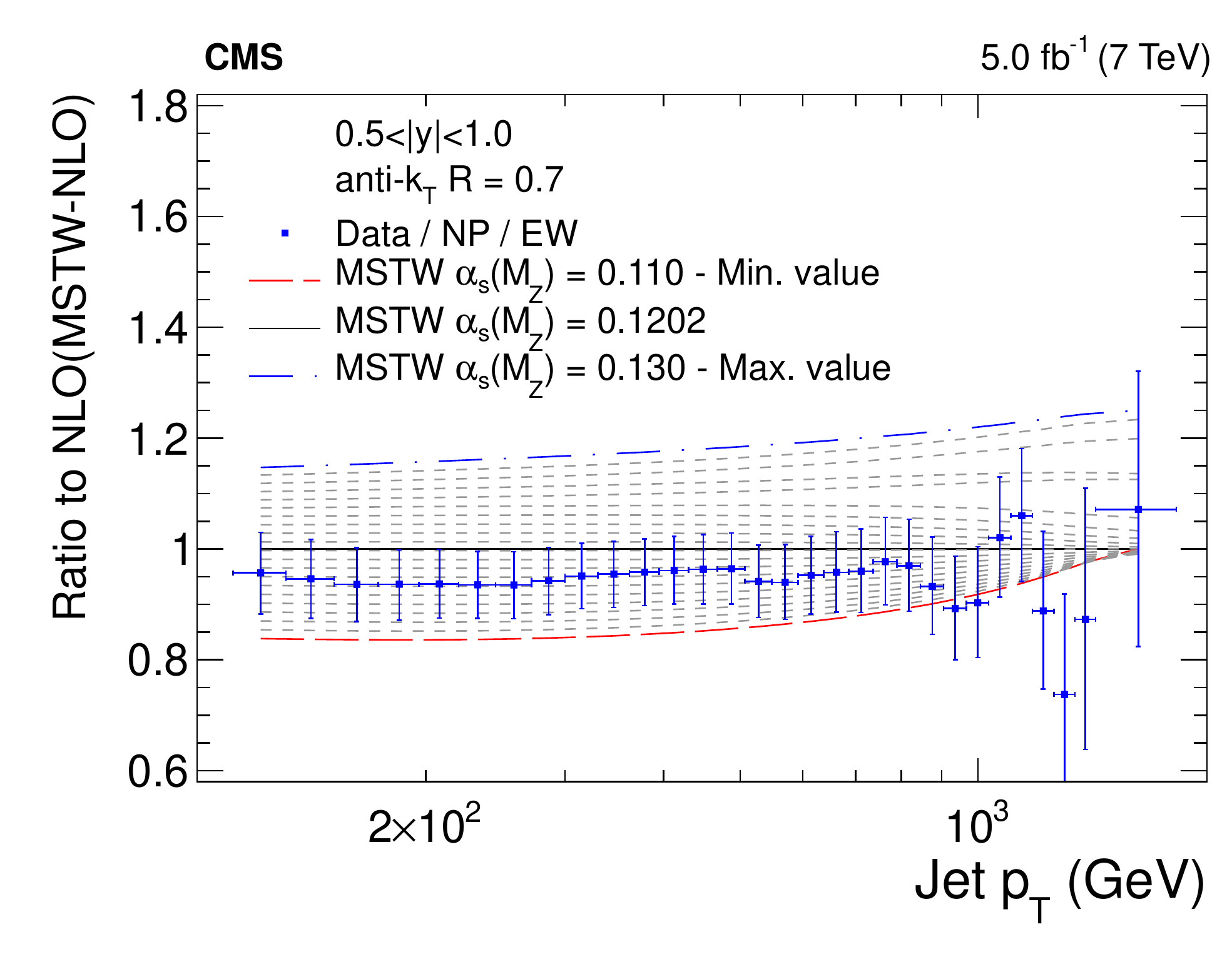}
  \includegraphics[width=0.47\textwidth]{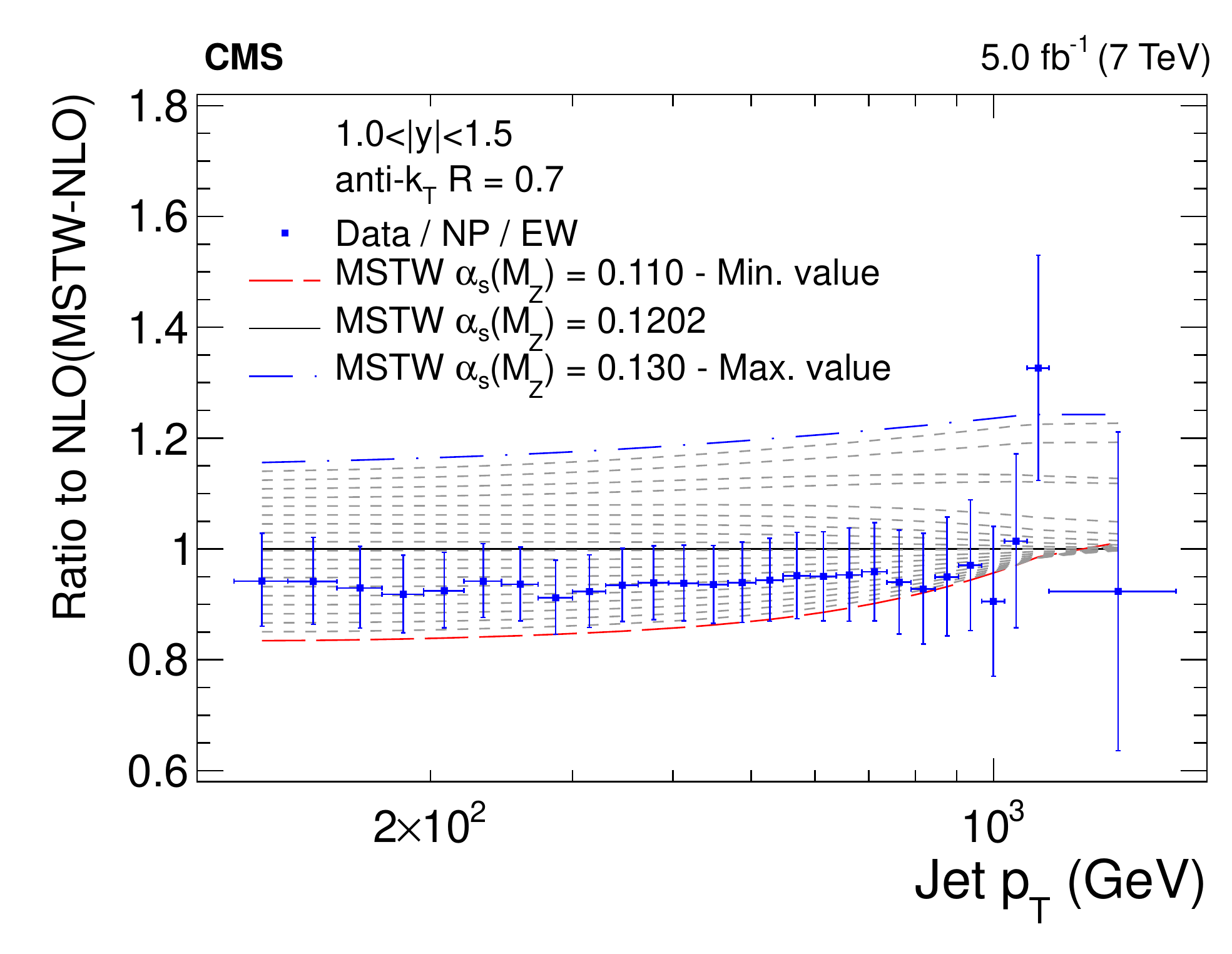}\hfill%
  \includegraphics[width=0.47\textwidth]{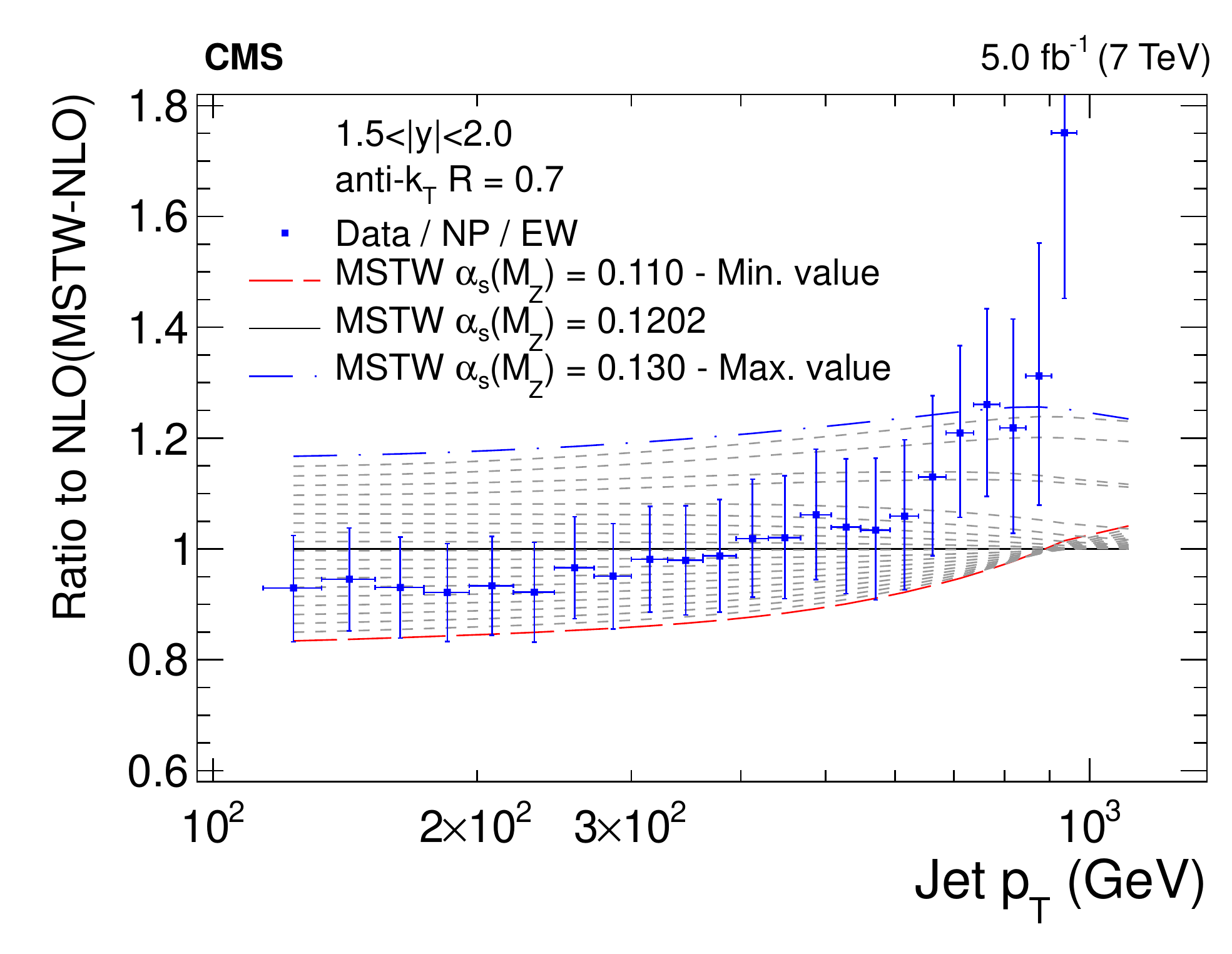}
  \includegraphics[width=0.47\textwidth]{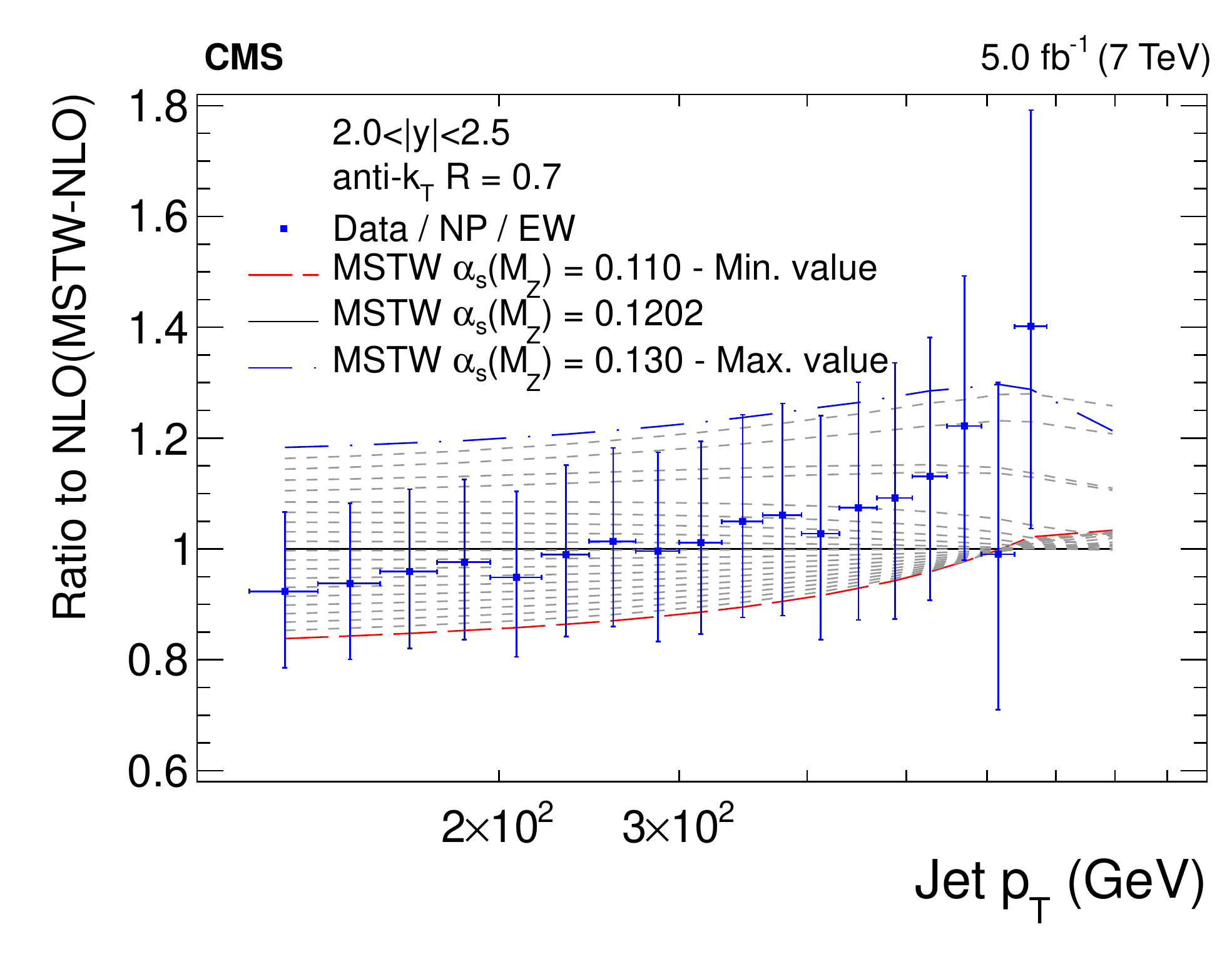}
  \caption{Ratio of the inclusive jet cross section to theoretical
    predictions using the MSTW2008-NLO PDF set for the five rapidity
    bins, where the \alpsmz value is varied in the range
    $0.110$--$0.130$ in steps of 0.001. The error bars correspond to
    the total uncertainty.}
  \label{fig:DataTheory_as_MSTW2008nlo}
\end{figure*}

\begin{figure*}[pt]
  \centering
  \includegraphics[width=0.47\textwidth]{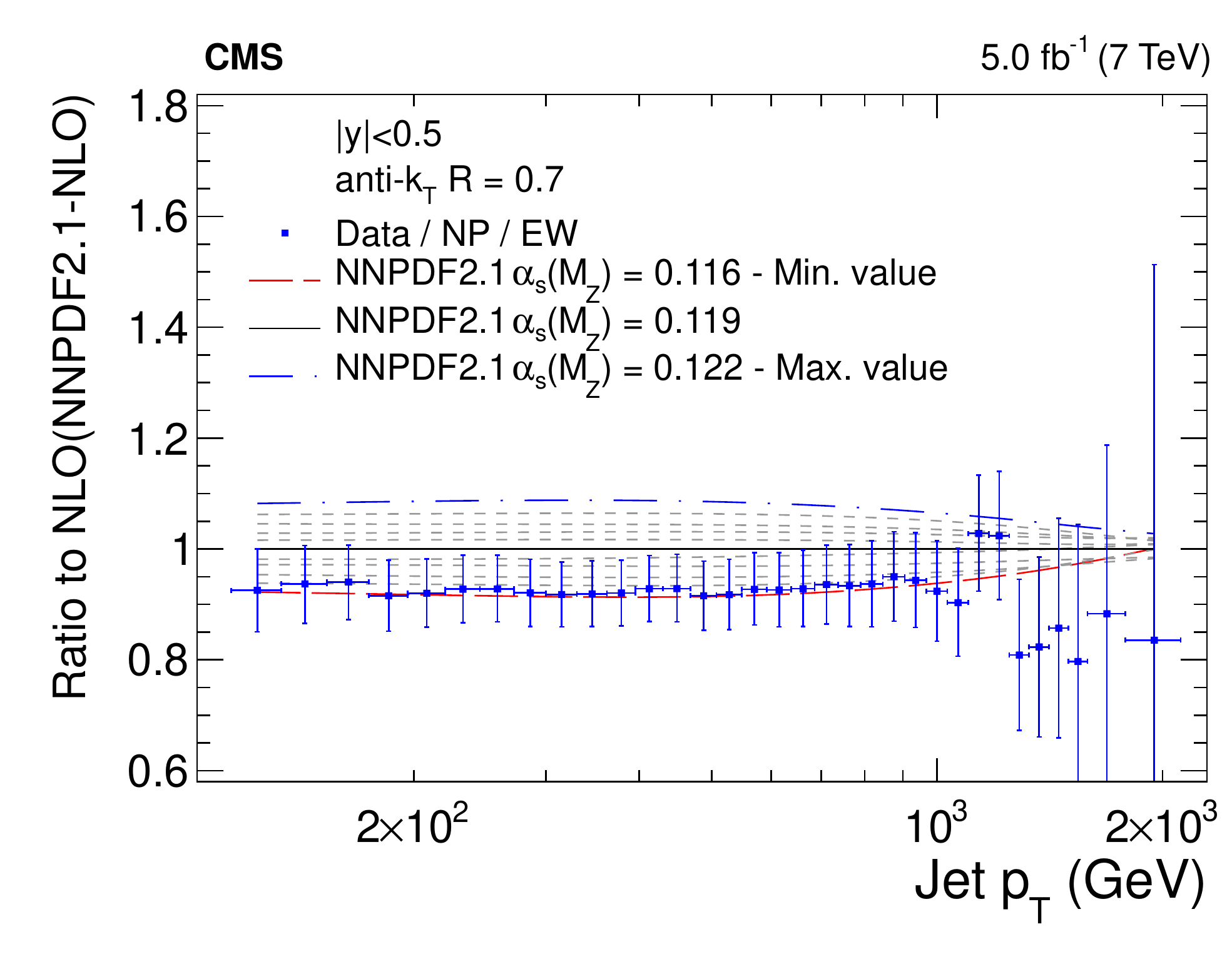}\hfill%
  \includegraphics[width=0.47\textwidth]{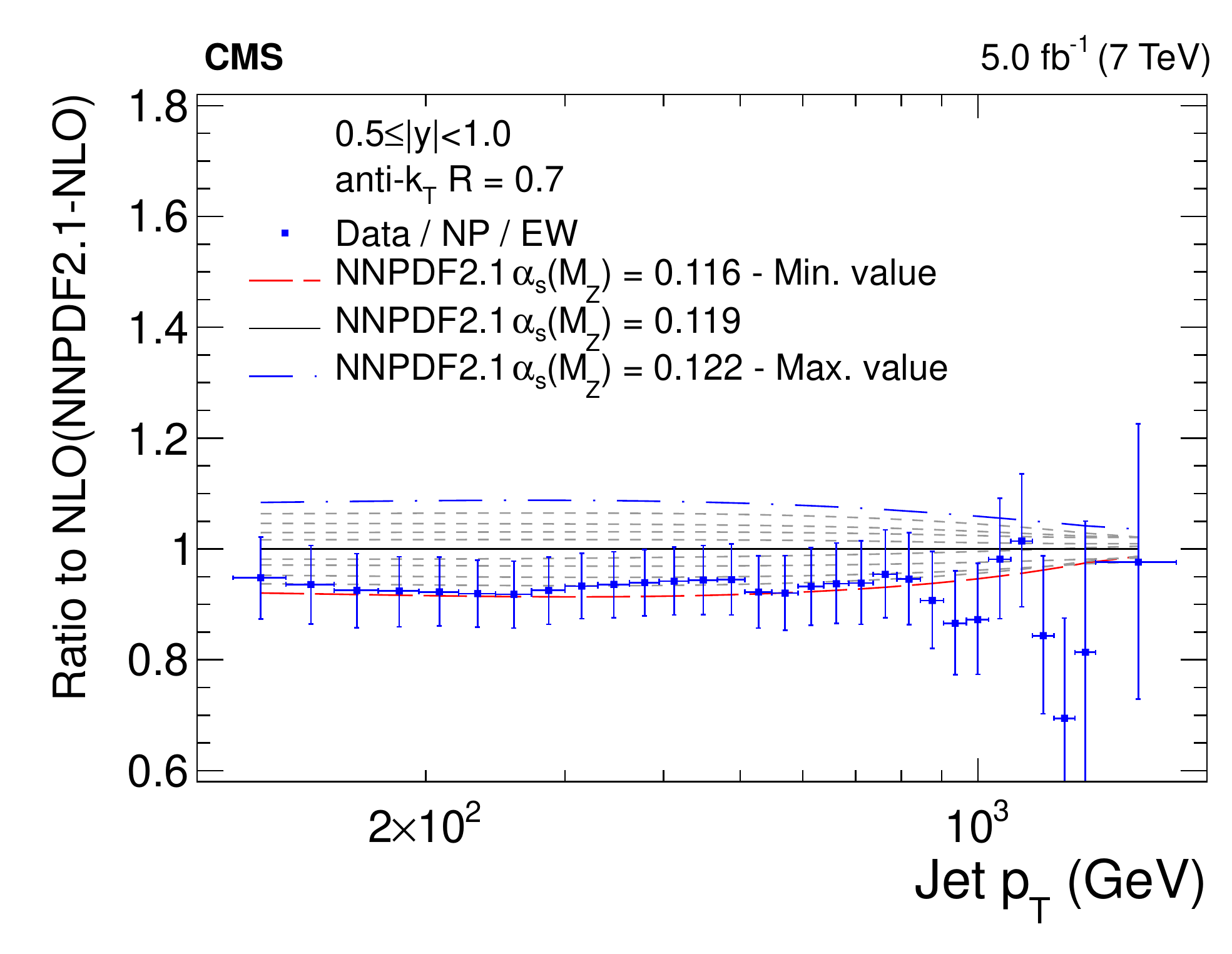}
  \includegraphics[width=0.47\textwidth]{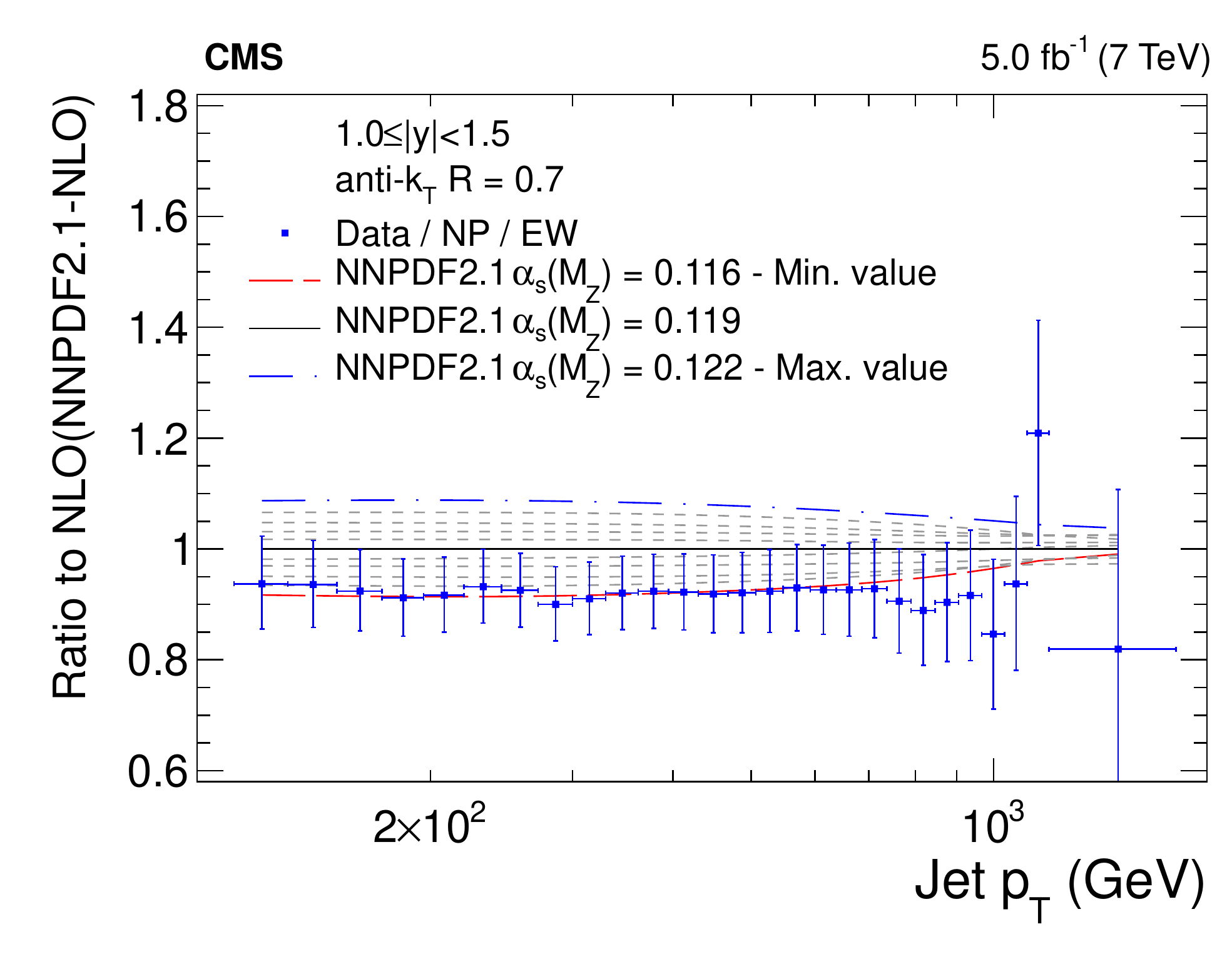}\hfill%
  \includegraphics[width=0.47\textwidth]{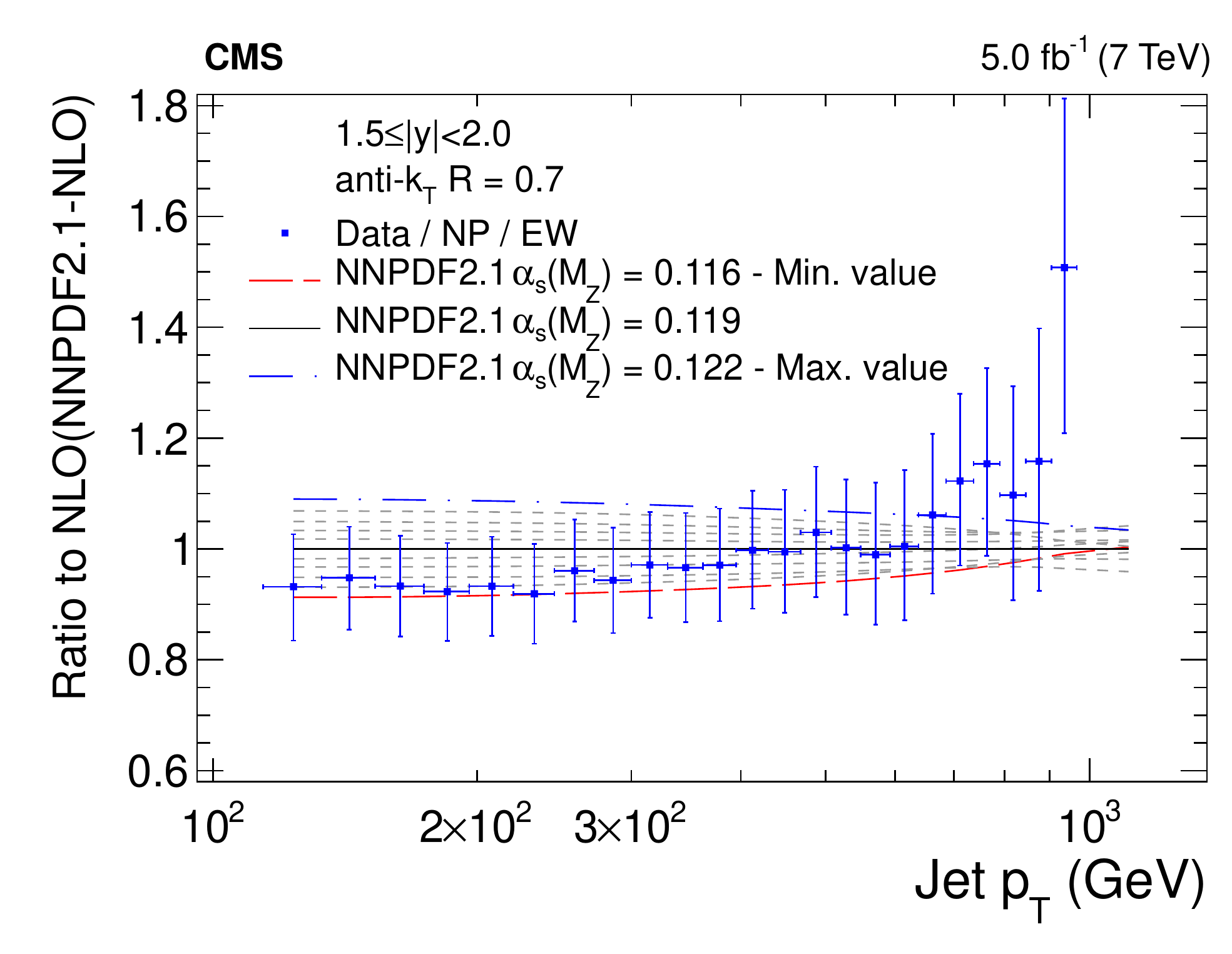}
  \includegraphics[width=0.47\textwidth]{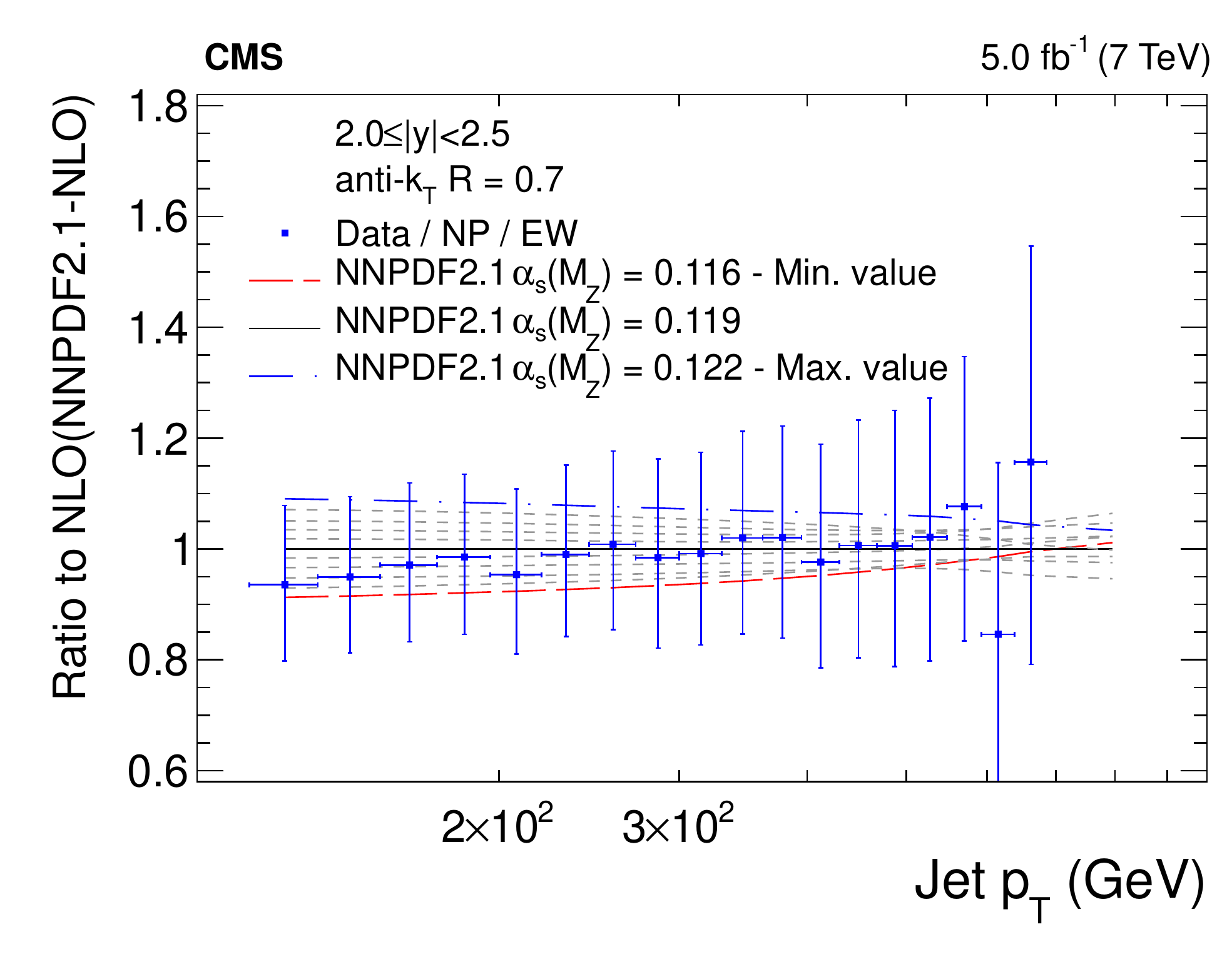}
  \caption{Ratio of the inclusive jet cross section to theoretical
    predictions using the NNPDF2.1-NLO PDF set for the five rapidity
    bins, where the \alpsmz value is varied in the range
    0.116--0.122 in steps of 0.001. The error bars correspond to
    the total uncertainty.}
  \label{fig:DataTheory_as_NNPDF21nlo}
\end{figure*}

Figures~\ref{fig:DataTheory_as_ABM11nlo}--\ref{fig:DataTheory_as_NNPDF21nlo}
present the ratio of data to the theoretical predictions for all
variations in \alpsmz available for the PDF sets ABM11, CT10,
MSTW2008, and NNPDF2.1 at next-to-leading evolution order, as
specified in Table~\ref{tab:pdfsets}. Except for the ABM11 PDF set,
which leads to QCD predictions significantly different in shape to the
measurement, all PDF sets give satisfactory theoretical descriptions
of the data and a strong sensitivity to \alpsmz is
demonstrated. Because of the discrepancies, ABM11 is excluded from
further investigations. The CT10-NLO PDF set is chosen for the main
result on \alpsmz, because the value of \alpsmz preferred by the CMS
jet data is rather close to the default value of this PDF
set. As crosschecks fits are performed with the NNPDF2.1-NLO and
MSTW2008-NLO sets. The CT10-NNLO, NNPDF2.1-NNLO, and MSTW2008-NNLO PDF
sets are employed for comparison.

\subsection{The fitting procedure\label{section:fit_proc}}

The value of \alpsmz is determined by minimising the \chisq between
the $N$ measurements $D_i$ and the theoretical predictions $T_i$. The
\chisq is defined as
\begin{equation}
  \chi^2 = \sum_{ij}^N \left(D_i - T_i\right) \mathrm{C}_{ij}^{-1}
  \left(D_j - T_j\right),
  \label{chi2_square}
\end{equation}
where the covariance matrix $C_{ij}$ is composed of the following
terms:
\ifthenelse{\boolean{cms@external}}{
\begin{multline}
  \label{eqn:C_matrix}
  C = \cov_\text{stat} + \cov_\text{uncor} +
  \left(\sum_\text{sources}\cov_\mathrm{JES}\right) +
  \cov_\text{unfolding} \\+ \cov_\text{lumi} +
  \cov_\mathrm{PDF},
\end{multline}
}{
\begin{equation}
  \label{eqn:C_matrix}
  C = \cov_\text{stat} + \cov_\text{uncor} +
  \left(\sum_\text{sources}\cov_\mathrm{JES}\right) +
  \cov_\text{unfolding} + \cov_\text{lumi} +
  \cov_\mathrm{PDF},
\end{equation}
}
and the terms in the sum represent
\begin{enumerate}
\item{$\cov_\text{stat}$: statistical uncertainty including
    correlations induced through unfolding};
\item{$\cov_\text{uncor}$: uncorrelated systematic
    uncertainty summing up small residual effects such as trigger and
    identification inefficiencies, time dependence of the jet \pt
    resolution, or the uncertainty on the trigger prescale factor};
\item{$\cov_\mathrm{JES\,sources}$: systematic uncertainty for
    each JES uncertainty source};
\item{$\cov_\text{unfolding}$: systematic uncertainty of the
    unfolding};
\item{$\cov_\text{lumi}$: luminosity uncertainty}; and
\item{$\cov_\mathrm{PDF}$: PDF uncertainty}.
\end{enumerate}

All JES, unfolding, and luminosity uncertainties are treated as
100\% correlated across the \pt and \yabs bins, with the exception of
the single-particle response JES source as described in
Section~\ref{sec:measurementjec}. The JES, unfolding, and luminosity
uncertainties are treated as multiplicative to avoid the statistical
bias that arises when estimating uncertainties from
data~\cite{Lyons:1989gh,D'Agostini:2003nk,Ball:2009qv}.

The derivation of PDF uncertainties follows prescriptions for each
individual PDF set. The CT10 and MSTW PDF sets both employ the
eigenvector method with upward and downward variations for each
eigenvector. As required by the use of covariance matrices, symmetric
PDF uncertainties are computed following Ref.~\cite{Pumplin:2002vw}.
The NNPDF2.1 PDF set uses the MC pseudo-experiments instead of the
eigenvector method in order to provide PDF uncertainties. A hundred
so-called replicas, whose averaged predictions give the central
result, are evaluated following the prescription in
Ref.~\cite{Ball:2010de} to derive the PDF uncertainty for NNPDF\@.

As described in Section~\ref{sec:npcorrection}, the NP correction is
defined as the centre of the envelope given by \PYTHIAS, \HERWIGPP,
and the \POWHEG + \PYTHIAS average of tunes Z2* and P11\@. Half the
spread among these three numbers is taken as the uncertainty. This is
the default NP correction used in this analysis. Alternatively, the PS
correction factor, defined in Section~\ref{sec:pscorrection}, is
applied in addition as an illustrative test to complement the main results.

The uncertainty in \alpsmz due to the NP uncertainties is evaluated by
looking for maximal offsets from a default fit. The theoretical
prediction $T$ is varied by the NP uncertainty $\Delta\mathrm{NP}$ as
$T\cdot\mathrm{NP} \to T\cdot\left(\mathrm{NP} \pm
  \Delta\mathrm{NP}\right)$. The fitting procedure is repeated for
these variations, and the deviation from the central \alpsmz values is
considered as the uncertainty in \alpsmz.

{\tolerance=900
Finally the uncertainty due to the renormalization and factorisation
scales is evaluated by applying the same method as for the NP
corrections: \mur and \muf are varied from the default choice of
$\mur=\muf=\pt$ between $\pt/2$ and $2\pt$ in the following six
combinations: $(\mur/\pt,\muf/\pt) = (1/2,1/2)$, $(1/2,1)$, $(1,1/2)$,
$(1,2)$, $(2,1)$, and $(2,2)$. The \chisq minimisation with respect to
\alpsmz is repeated in each case. The contribution from the \mur and
\muf scale variations to the uncertainty is evaluated by considering
the maximal upwards and downwards deviation of \alpsmz from the
central result.\par}

\subsection{The results on \texorpdfstring{\alpsmz}{alpha-s(MZ)}
  \label{section:results_a_s}}

\begin{figure}[tbp]
  \centering
    \includegraphics[width=\cmsFigWidth]{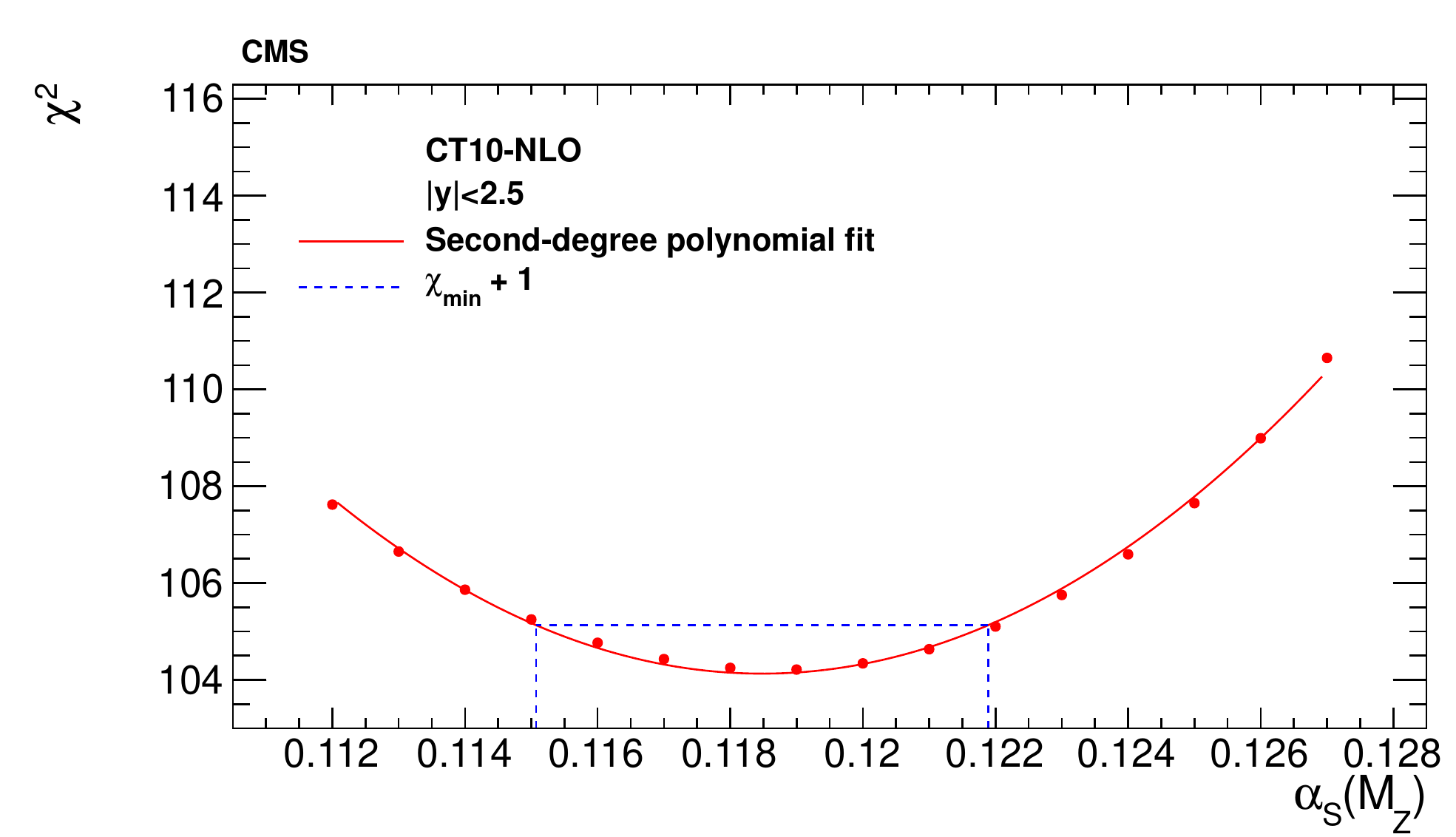}
    \caption{The \chisq minimisation with respect to \alpsmz using the
      CT10-NLO PDF set and data from all rapidity bins. The
      experimental uncertainty is obtained from the \alpsmz values for
      which \chisq is increased by one with respect to the minimum
      value, indicated by the dashed line. The curve corresponds to a
      second-degree polynomial fit through the available \chisq
      points.}
    \label{fig:chi2_points}
\end{figure}

The values of \alpsmz obtained with the CT10-NLO PDF set are listed in
Table~\ref{tbl:CT10_nlo_as_results} together with the experimental,
PDF, NP, and scale uncertainties for each bin in rapidity and for a
simultaneous fit of all rapidity bins. To disentangle the
uncertainties of experimental origin from those of the PDFs,
additional fits without the latter uncertainty source are performed.
An example for the evaluation of the uncertainties in a $\chi^{2}$ fit
is shown in Fig.~\ref{fig:chi2_points}. The NP and scale uncertainties
are determined via separate fits, as explained above.

For the two outer rapidity bins ($1.5<\yabs<2.0$ and $2.0<\yabs<2.5$)
the series in values of \alpsmz of the CT10-NLO PDF set does not reach
to sufficiently low values of \alpsmz. As a consequence the shape of
the \chisq curve at minimum up to $\chisq+1$ can not be determined
completely. To avoid extrapolations based on a polynomial fit to the
available points, the alternative \alps evolution code of the \HOPPET
package~\cite{Salam:2008qg} is employed. This is the same evolution
code as chosen for the creation of the CT10 PDF set.  Replacing the
original \alps evolution in CT10 by \HOPPET, \alpsmz can be set freely
and in particular different from the default value used in a PDF set,
but at the expense of losing the correlation between the value of
\alpsmz and the fitted PDFs. Downwards or upwards deviations from the
lowest and highest values of \alpsmz, respectively, provided in a PDF
series are accepted for uncertainty evaluations up to a limit of
$\abs{\Delta\alpsmz} = 0.003$. Applying this method for comparisons,
within the available range of \alpsmz values, an additional
uncertainty is estimated to be negligible.

For comparison the CT10-NNLO PDF set is used for the determination of
\alpsmz. These results are presented in
Table~\ref{tbl:CT10_nnlo_as_results}.

\begin{table*}[tbp]
  \centering
  \topcaption{Determination of \alpsmz in bins of rapidity using the CT10-NLO PDF set.
    The last row presents the result of a simultaneous fit in all rapidity bins.
    \label{tbl:CT10_nlo_as_results}}
\begin{tabular}{lcr@{$\,\pm\,$}lc}
    \multirow{2}{*}{$\yabs$ range} & No.\ of data & \multicolumn{2}{c}{\multirow{2}{*}{\alpsmz}} & \multirow{2}{*}{\chisqndof} \\
    & points  &\multicolumn{2}{c}{}  & \\\hline
   $\yabs<0.5$        & 33 & 0.1189 & $ 0.0024\,(\text{exp}) \pm 0.0030\,(\mathrm{PDF})$ & $16.2/32$ \\
    &     & & $0.0008\,(\mathrm{NP}) ^{+0.0045}_{-0.0027}\,(\text{scale})$      & \\\hline
    $0.5\leq\yabs<1.0$ & 30 & 0.1182 & $ 0.0024\,(\text{exp}) \pm 0.0029\,(\mathrm{PDF})$ & $25.4/29$ \\
    &     & & $0.0008\,(\mathrm{NP}) ^{+0.0050}_{-0.0025}\,(\text{scale})$      & \\\hline
    $1.0\leq\yabs<1.5$ & 27 & 0.1165 & $ 0.0027\,(\text{exp}) \pm 0.0024\,(\mathrm{PDF})$ & $9.5/26$  \\
    &     & & $0.0008\,(\mathrm{NP}) ^{+0.0043}_{-0.0020}\,(\text{scale})$      & \\\hline
    $1.5\leq\yabs<2.0$ & 24 & 0.1146 & $ 0.0035\,(\text{exp}) \pm 0.0031\,(\mathrm{PDF})$ & $20.2/23$ \\
    &     & & $0.0013\,(\mathrm{NP})  ^{+0.0037}_{-0.0020}\,(\text{scale})$     & \\\hline
    $2.0\leq\yabs<2.5$ & 19 & 0.1161 & $ 0.0045\,(\text{exp}) \pm 0.0054\,(\mathrm{PDF})$ & $12.6/18$ \\
    &     & & $0.0015\,(\mathrm{NP})  ^{+0.0034}_{-0.0032}\,(\text{scale})$     & \\\hline
    $\yabs<2.5$        & 133& 0.1185 & $ 0.0019\,(\text{exp}) \pm 0.0028\,(\mathrm{PDF})$ & $104.1/132$ \\
    &     & & $0.0004\,(\mathrm{NP}) ^{+0.0053}_{-0.0024}\,(\text{scale})$      & \\
  \end{tabular}
\end{table*}

\begin{table*}[tbp]
  \centering
  \topcaption{Determination of \alpsmz in bins of rapidity using the CT10-NNLO PDF set.
    The last row presents the result of a simultaneous fit in all rapidity bins.
    \label{tbl:CT10_nnlo_as_results}}
\begin{tabular}{lcr@{$\,\pm\,$}lc}
    \multirow{2}{*}{$\yabs$ range} & No.\ of data & \multicolumn{2}{c}{\multirow{2}{*}{\alpsmz}} & \multirow{2}{*}{\chisqndof} \\
    & points  &\multicolumn{2}{c}{}  & \\\hline
    $\yabs<0.5$        & 33 & 0.1180 & $0.0017\,(\text{exp}) \pm 0.0027\,(\mathrm{PDF})$ & $15.4/32$ \\
    &     & & $0.0006\,(\mathrm{NP})^{+0.0031}_{-0.0026}\,(\text{scale})$       & \\\hline
    $0.5\leq\yabs<1.0$ & 30 & 0.1176 & $0.0016\,(\text{exp}) \pm 0.0026\,(\mathrm{PDF})$ & $23.9/29$ \\
    &     & &$ 0.0006\,(\mathrm{NP}) ^{+0.0033}_{-0.0023}\,(\text{scale})$      & \\\hline
    $1.0\leq\yabs<1.5$ & 27 & 0.1169 &$ 0.0019\,(\text{exp}) \pm 0.0024\,(\mathrm{PDF})$ & $10.5/26$ \\
    &     & &$ 0.0006\,(\mathrm{NP}) ^{+0.0033}_{-0.0019}\,(\text{scale})$      & \\\hline
    $1.5\leq\yabs<2.0$ & 24 & 0.1133 &$ 0.0023\,(\text{exp}) \pm 0.0028\,(\mathrm{PDF})$ & $22.3/23$ \\
    &     & &$ 0.0010\,(\mathrm{NP}) ^{+0.0039}_{-0.0029}\,(\text{scale})$      & \\\hline
    $2.0\leq\yabs<2.5$ & 19 & 0.1172 &$ 0.0044\,(\text{exp}) \pm 0.0039\,(\mathrm{PDF})$ & $13.8/18$ \\
    &     & &$ 0.0015\,(\mathrm{NP}) ^{+0.0049}_{-0.0060}\,(\text{scale})$      & \\\hline
    $\yabs<2.5$        & 133& 0.1170 &$ 0.0012\,(\text{exp}) \pm 0.0024\,(\mathrm{PDF})$ & $105.7/132$ \\
    &     & &$ 0.0004\,(\mathrm{NP}) ^{+0.0044}_{-0.0030}\,(\text{scale})$      & \\
  \end{tabular}
\end{table*}

The final result using all rapidity bins and the CT10-NLO PDF set is
(last row of Table~\ref{tbl:CT10_nlo_as_results})
\ifthenelse{\boolean{cms@external}}{
\begin{equation}
  \label{eqn:analytic_result}
  \begin{split}
    \alpsmz  = &0.1185 \pm 0.0019\,\text{(exp)} \\
               &\pm 0.0028\,(\mathrm{PDF}) \pm 0.0004\,(\mathrm{NP})^{+0.0053}_{-0.0024}\,(\text{scale})\\
     = &0.1185 \pm 0.0034\,\text{(all except scale)}^{+0.0053}_{-0.0024}\,(\text{scale}) \\
     = &0.1185^{+0.0063}_{-0.0042},
  \end{split}
\end{equation}
}{
\begin{equation}
  \label{eqn:analytic_result}
  \begin{split}
    \alpsmz & = 0.1185 \pm 0.0019\,\text{(exp)} \pm
    0.0028\,(\mathrm{PDF})
    \pm 0.0004\,(\mathrm{NP})^{+0.0053}_{-0.0024}\,(\text{scale})\\
    & = 0.1185 \pm 0.0034\,\text{(all except scale)}^{+0.0053}_{-0.0024}\,(\text{scale}) =
    0.1185^{+0.0063}_{-0.0042},
  \end{split}
\end{equation}
}

where experimental, PDF, NP, and scale uncertainties have been added
quadratically to give the total uncertainty. The result is in
agreement with the world average value of $\alpsmz = 0.1185 \pm
0.0006$~\cite{Agashe:2014kda}, with the Tevatron results
\cite{Affolder:2001hn,Abazov:2009nc,Abazov:2012lua}, and recent
results obtained with LHC data
\cite{Malaescu:2012ts,Chatrchyan:2013txa,CMS-PAPERS-TOP-12-022}.  The
determination of \alpsmz, which is based on the CT10-NLO PDF set, is
also in agreement with the result obtained using the NNPDF2.1-NLO and
MSTW2008-NLO sets, as shown in
Table~\ref{tbl:ALL_nlo_nnlo_as_results}. For comparison this table
also shows the results using the CT10, MSTW2008, and NNPDF2.1 PDF sets
at NNLO\@. The \alpsmz values are in agreement among the different NLO
PDF sets within the uncertainties.

\begin{table*}[tbp]
  \centering
  \topcaption{Determination of \alpsmz using the CT10 and MSTW2008 PDF sets at NLO
           and the CT10, NNPDF2.1, MSTW2008 PDF sets at NNLO\@.
           The results are obtained by a simultaneous fit to all rapidity bins.
           \label{tbl:ALL_nlo_nnlo_as_results}}
\begin{tabular}{lr@{$\,\pm\,$}lc}
    PDF set        & \multicolumn{2}{c}{\alpsmz}     & \chisqndof   \\\hline
    CT10-NLO       & 0.1185 &$ 0.0019\,(\text{exp}) \pm 0.0028\,(\mathrm{PDF})$ & $104.1/132$\\
    & &$0.0004\,(\mathrm{NP})^{+0.0053}_{-0.0024}\,(\text{scale})$   \\\hline
    NNPDF2.1-NLO & 0.1150 &$ 0.0015\,(\text{exp}) \pm  0.0024\,(\mathrm{PDF})$  & $103.5/132$\\
    & &$0.0003\,(\mathrm{NP})^{+0.0025}_{-0.0025}\,(\text{scale})$   \\\hline
    MSTW2008-NLO   & 0.1159 &$ 0.0012\,(\text{exp}) \pm 0.0014\,(\mathrm{PDF})$ & $107.9/132$\\
    & &$0.0001\,(\mathrm{NP})^{+0.0024}_{-0.0030}\,(\text{scale})$   \\\hline
    CT10-NNLO      & 0.1170 &$ 0.0012\,(\text{exp}) \pm 0.0024\,(\mathrm{PDF})$ & $105.7/132$\\
    & &$0.0004\,(\mathrm{NP})^{+0.0044}_{-0.0030}\,(\text{scale})$   \\\hline
    NNPDF2.1-NNLO  & 0.1175 &$ 0.0012\,(\text{exp}) \pm 0.0019\,(\mathrm{PDF})$ & $103.0/132$\\
    & &$0.0001\,(\mathrm{NP})^{+0.0018}_{-0.0020}\,(\text{scale})$   \\\hline
    MSTW2008-NNLO  & 0.1136 &$ 0.0010\,(\text{exp}) \pm 0.0011\,(\mathrm{PDF})$ & $108.8/132$\\
    & &$0.0001\,(\mathrm{NP})^{+0.0019}_{-0.0024}\,(\text{scale})$   \\
  \end{tabular}
\end{table*}

Applying the PS correction factor to the NLO theory prediction in
addition to the NP correction as discussed in
Section~\ref{sec:pscorrection}, the fit using all rapidity bins and
the CT10-NLO PDF set yields $\alpsmz = 0.1204 \pm
0.0018\,(\text{exp})$. This value is in agreement with our main
result of Eq.~(\ref{eqn:analytic_result}), which is obtained using
only the NP correction factor.

To investigate the running of the strong coupling, the fitted region
is split into six bins of \pt and the fitting procedure is repeated in
each of these bins. The six extractions of \alpsmz are reported in
Table~\ref{tbl:as_values}. The \alpsmz values are evolved to the
corresponding energy scale $Q$ using the two-loop solution to the
renormalization group equation (RGE) within \HOPPET. The value of $Q$
is calculated as a cross section weighted average in each fit
region. These average scale values $Q$, derived again with the
\fastNLO framework, are identical within about 1\GeV for different
PDFs. To emphasise that theoretical uncertainties limit the achievable
precision, Tables~\ref{tbl:as_unc} and~\ref{tbl:as_q_unc} present for
the six bins in \pt the total uncertainty as well as the experimental,
PDF, NP, and scale components, where the six experimental
uncertainties are all correlated.

\begin{table*}[tbp]
  \centering
  \topcaption{Determination of \alps in separate bins of jet \pt using
    the CT10-NLO PDF set.\label{tbl:as_values}}
  \begin{tabular}{cccccc}
    \pt range        & $Q$      & \multirow{2}{*}{\alpsmz}       & \multirow{2}{*}{\alpsq}        & No.\ of data & \multirow{2}{*}{$\chisqndof$} \\
    (\GeVns{})         & (\GeVns{}) &                                &                                & points       & \\\hline
    114--196  \rbtrr & 136      & $0.1172\,_{-0.0043}^{+0.0058}$ & $0.1106\,^{+0.0052}_{-0.0038}$ & 20\rbtrr     & $6.2/19$\\
    196--300  \rbtrr & 226      & $0.1180\,_{-0.0046}^{+0.0063}$ & $0.1038\,^{+0.0048}_{-0.0035}$ & 20\rbtrr     & $7.6/19$\\
    300--468  \rbtrr & 345      & $0.1194\,_{-0.0049}^{+0.0064}$ & $0.0993\,^{+0.0044}_{-0.0034}$ & 25\rbtrr     & $8.1/24$\\
    468--638  \rbtrr & 521      & $0.1187\,_{-0.0051}^{+0.0067}$ & $0.0940\,^{+0.0041}_{-0.0032}$ & 20\rbtrr     & $10.6/19$\\
    638--905  \rbtrr & 711      & $0.1192\,_{-0.0056}^{+0.0074}$ & $0.0909\,^{+0.0042}_{-0.0033}$ & 22\rbtrr     & $11.2/21$\\
    905--2116 \rbtrr & 1007     & $0.1176\,_{-0.0065}^{+0.0111}$ & $0.0866\,^{+0.0057}_{-0.0036}$ & 26\rbtrr     & $33.6/25$\\
  \end{tabular}
\end{table*}

\begin{table*}[tbp]
  \centering
    \topcaption{Uncertainty composition for \alpsmz from the
      determination of \alpsq in bins of \pt using the CT10-NLO PDF
      set.\label{tbl:as_unc}}
    \begin{tabular}{ccccccc}
      \pt range        & $Q$      & \multirow{2}{*}{\alpsmz}& \multirow{2}{*}{exp.} & \multirow{2}{*}{PDF} & \multirow{2}{*}{NP} & \multirow{2}{*}{scale}\\
      (\GeVns)         & (\GeVns{}) &                         &                       &                      &                     &  \\\hline
      114--196  \rbtrr & 136      & 0.1172  & $\pm{0.0031}$ & $\pm{0.0018}$ & $\pm{0.0007}$ & $_{-0.0022}^{+0.0045}$\\
      196--300  \rbtrr & 226      & 0.1180  & $\pm{0.0034}$ & $\pm{0.0019}$ & $\pm{0.0011}$ & $_{-0.0025}^{+0.0048}$\\
      300--468  \rbtrr & 345      & 0.1194  & $\pm{0.0032}$ & $\pm{0.0023}$ & $\pm{0.0010}$ & $_{-0.0027}^{+0.0049}$\\
      468--638  \rbtrr & 521      & 0.1187  & $\pm{0.0029}$ & $\pm{0.0031}$ & $\pm{0.0006}$ & $_{-0.0027}^{+0.0052}$\\
      638--905  \rbtrr & 711      & 0.1192  & $\pm{0.0034}$ & $\pm{0.0032}$ & $\pm{0.0005}$ & $_{-0.0030}^{+0.0057}$\\
      905--2116 \rbtrr & 1007     & 0.1176  & $\pm{0.0047}$ & $\pm{0.0040}$ & $\pm{0.0002}$ & $_{-0.0020}^{+0.0092}$\\
    \end{tabular}

\end{table*}

\begin{table*}[tbp]
  \centering
    \topcaption{Uncertainty composition for \alpsq in bins of \pt using
      the CT10-NLO PDF set.\label{tbl:as_q_unc}}
    \begin{tabular}{ccccccc}
      \pt range        & $Q$      & \multirow{2}{*}{\alpsq}& \multirow{2}{*}{exp.} & \multirow{2}{*}{PDF} & \multirow{2}{*}{NP} & \multirow{2}{*}{scale}\\
      (\GeVns)         & (\GeVns{}) &         &               &               &               &  \\\hline
      114--196  \rbtrr & 136    & 0.1106  & $\pm{0.0028}$ & $\pm{0.0016}$   & $\pm{0.0006}$ &  $_{-0.0020}^{+0.0040}$\\
      196--300  \rbtrr & 226    & 0.1038  & $\pm{0.0026}$ & $\pm{0.0015}$   & $\pm{0.0008}$ &  $_{-0.0019}^{+0.0037}$\\
      300--468  \rbtrr & 345    & 0.0993  & $\pm{0.0022}$ & $\pm{0.0016}$   & $\pm{0.0007}$ &  $_{-0.0019}^{+0.0033}$\\
      468--638  \rbtrr & 521    & 0.0940  & $\pm{0.0018}$ & $\pm{0.0019}$   & $\pm{0.0004}$ &  $_{-0.0017}^{+0.0032}$\\
      638--905  \rbtrr & 711    & 0.0909  & $\pm{0.0019}$ & $\pm{0.0018}$   & $\pm{0.0003}$ &  $_{-0.0017}^{+0.0032}$\\
      905--2116 \rbtrr & 1007   & 0.0866  & $\pm{0.0025}$ & $\pm{0.0021}$   & $\pm{0.0001}$ &  $_{-0.0011}^{+0.0048}$\\
    \end{tabular}

\end{table*}

\begin{figure}[tbp]
  \centering
    \includegraphics[width=\cmsFigWidth]{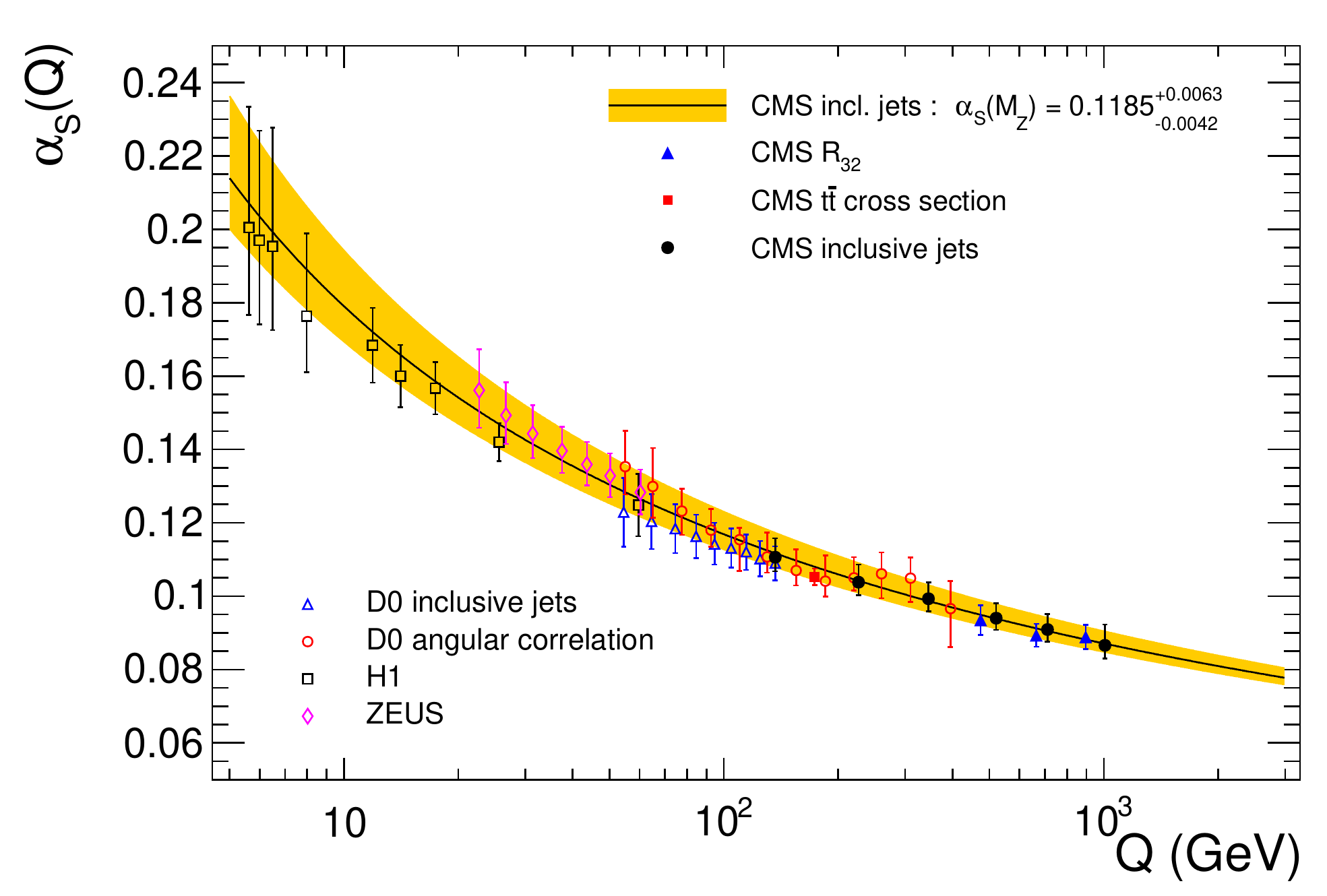}
    \caption{The strong coupling \alpsq (full line) and its total
      uncertainty (band) as determined in this analysis using a
      two-loop solution to the RGE as a function of the momentum
      transfer $Q=\pt$. The extractions of \alpsq in six separate
      ranges of $Q$ as presented in Table~\ref{tbl:as_values} are
      shown together with results from the
      H1~\cite{Aaron:2010ac,Andreev:2014wwa},
      ZEUS~\cite{Abramowicz:2012jz}, and
      D0~\cite{Abazov:2009nc,Abazov:2012lua} experiments at the HERA
      and Tevatron colliders. Other recent CMS
      measurements~\cite{Chatrchyan:2013txa,CMS-PAPERS-TOP-12-022} are
      displayed as well. The uncertainties represented by error
      bars are subject to correlations.}
    \label{fig:as_running}
\end{figure}

Figure~\ref{fig:as_running} presents the running of the strong
coupling \alpsq and its total uncertainty as determined in this
analysis. The extractions of \alpsq in six separate ranges of $Q$, as
presented in Table~\ref{tbl:as_values}, are also shown. In the same
figure the values of \alps at lower scales determined by the
H1~\cite{Aaron:2009vs,Aaron:2010ac,Andreev:2014wwa},
ZEUS~\cite{Abramowicz:2012jz}, and
D0~\cite{Abazov:2009nc,Abazov:2012lua} collaborations are shown for
comparison. Recent CMS
measurements~\cite{Chatrchyan:2013txa,CMS-PAPERS-TOP-12-022}, which
are in agreement with the \alpsmz determination of this study, are
displayed as well. The results on \alps reported here are consistent
with the energy dependence predicted by the RGE\@.

\section{Study of PDF constraints with
  \texorpdfstring{\HERAFitter}{HERAFitter}}
\label{sec:herafitter}

The PDFs of the proton are an essential ingredient for precision
studies in hadron-induced reactions. They are derived from
experimental data involving collider and fixed-target experiments. The
DIS data from the HERA-I \Pe\Pp~collider cover most of the kinematic
phase space needed for a reliable PDF extraction. The \Pp\Pp~inclusive
jet cross section contains additional information that can constrain
the PDFs, in particular the gluon, in the region of high fractions $x$
of the proton momentum.

The \HERAFitter project~\cite{Alekhin:2014rma,HERAFitter:2013hf} is an
open-source framework designed among other things to fit PDFs to data.
It has a modular structure, encompassing a variety of theoretical
predictions for different processes and phenomenological approaches
for determining the parameters of the PDFs. In this study, the
recently updated \HERAFitter version~1.1.1 is employed to estimate the
impact of the CMS inclusive jet data on the PDFs and their
uncertainties. Theory is used at NLO for both processes, i.e.\ up to
order $\alps^2$ for DIS and up to order $\alps^3$ for inclusive jet
production in \Pp\Pp~collisions.

\subsection{Correlation between inclusive jet production and the PDFs}
\label{sec:pdf_sensitivity}

\begin{figure*}[pt]
  \centering
  \includegraphics[width=0.45\textwidth]{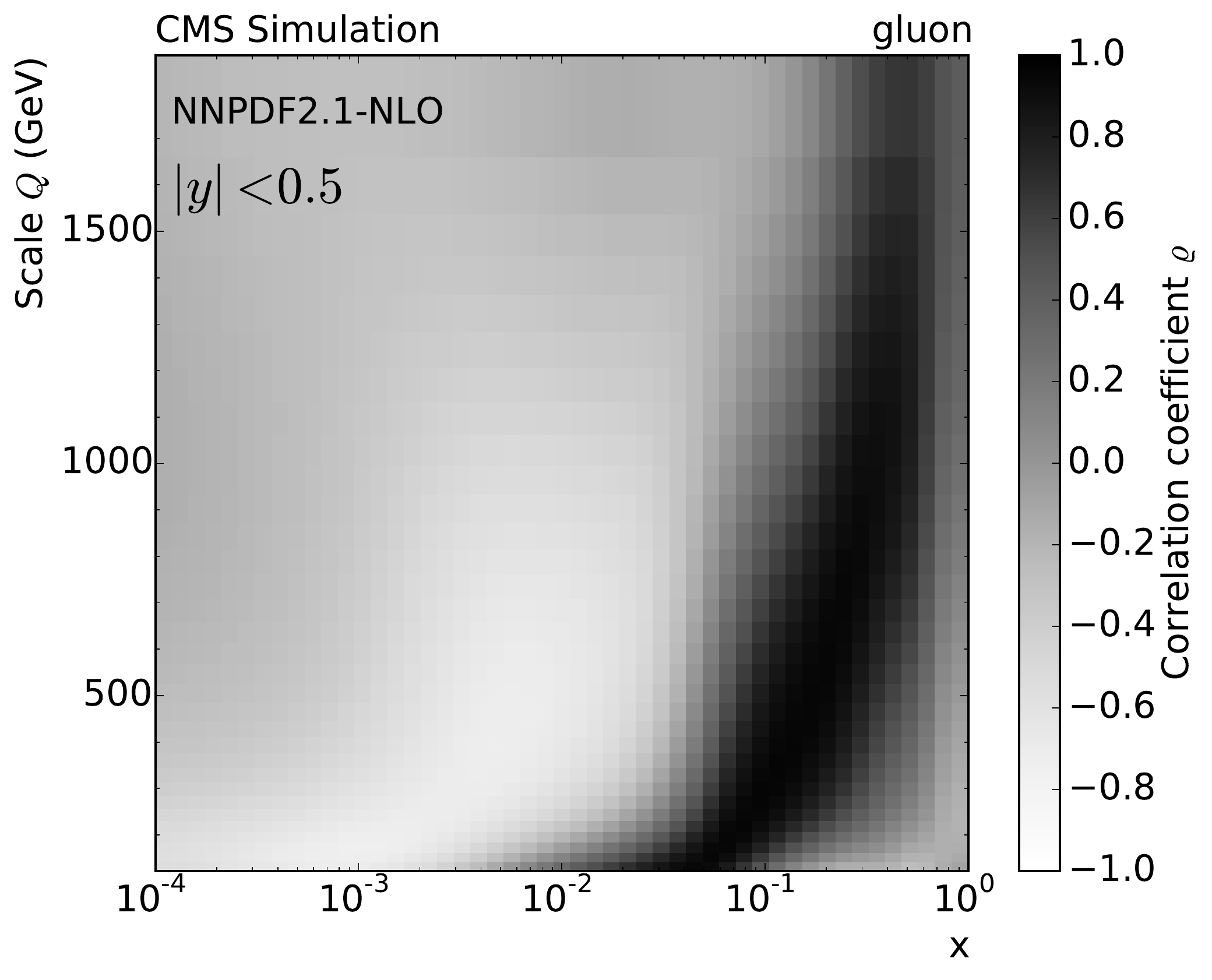}\hfill%
  \includegraphics[width=0.45\textwidth]{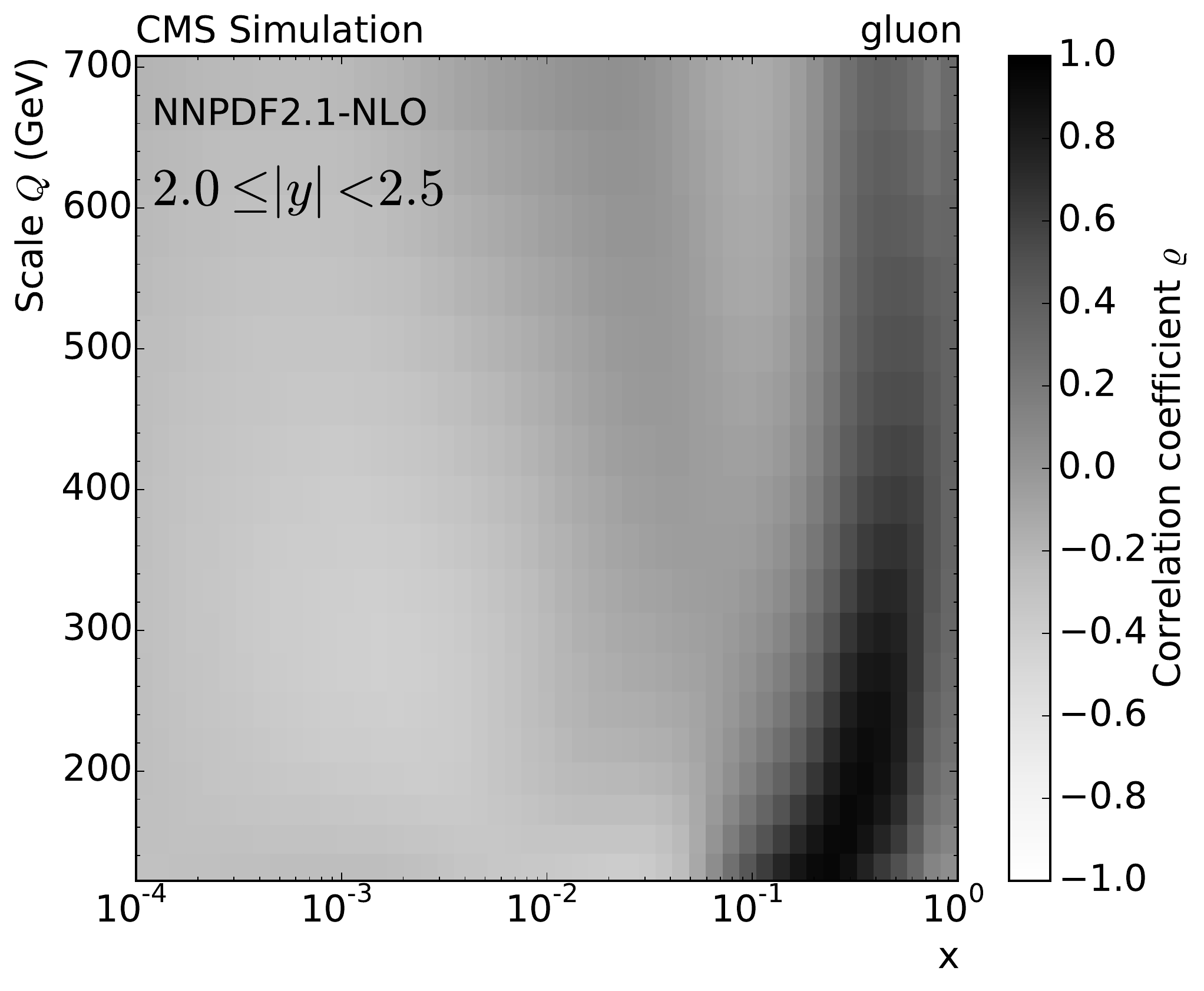}
  \includegraphics[width=0.45\textwidth]{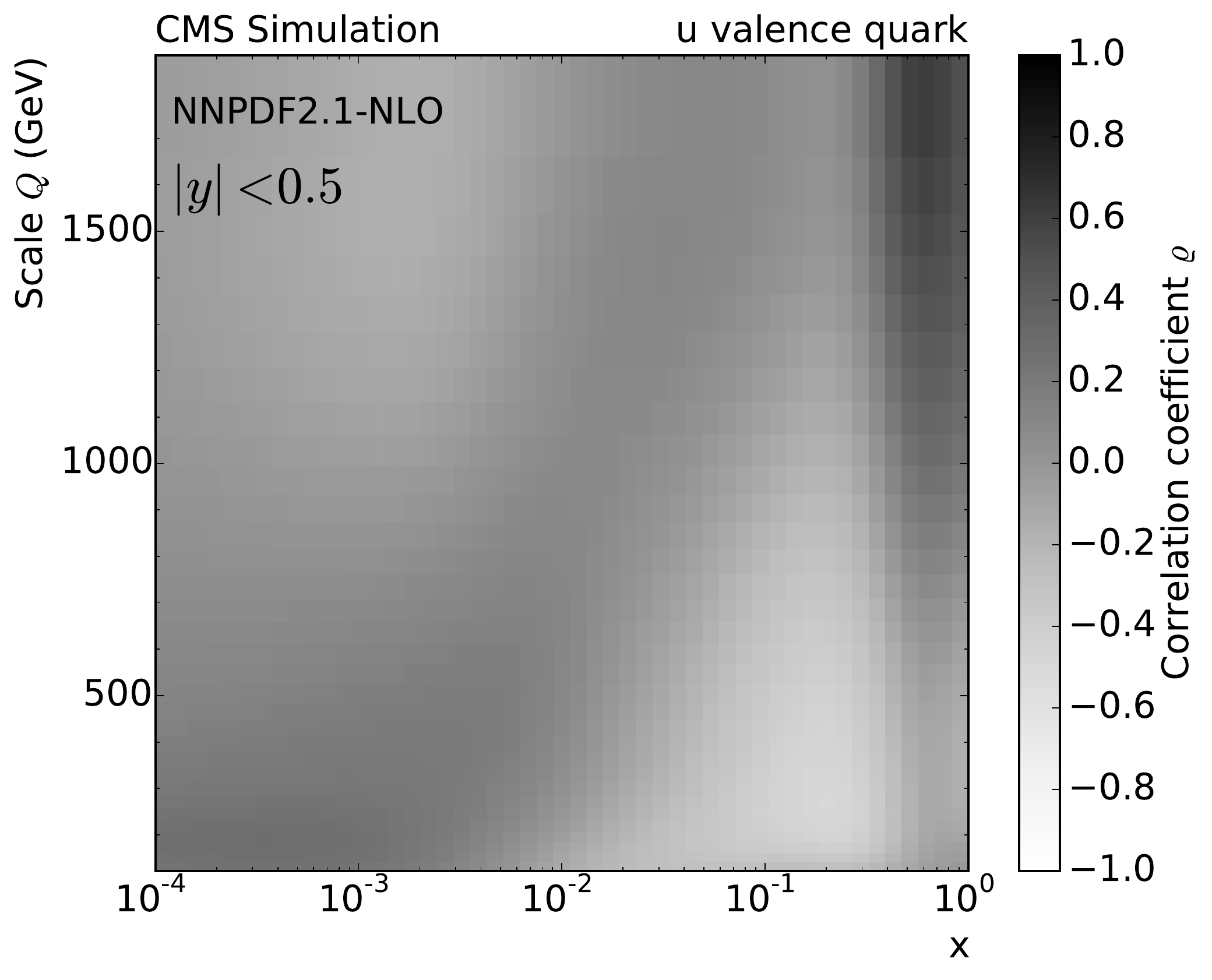}\hfill%
  \includegraphics[width=0.45\textwidth]{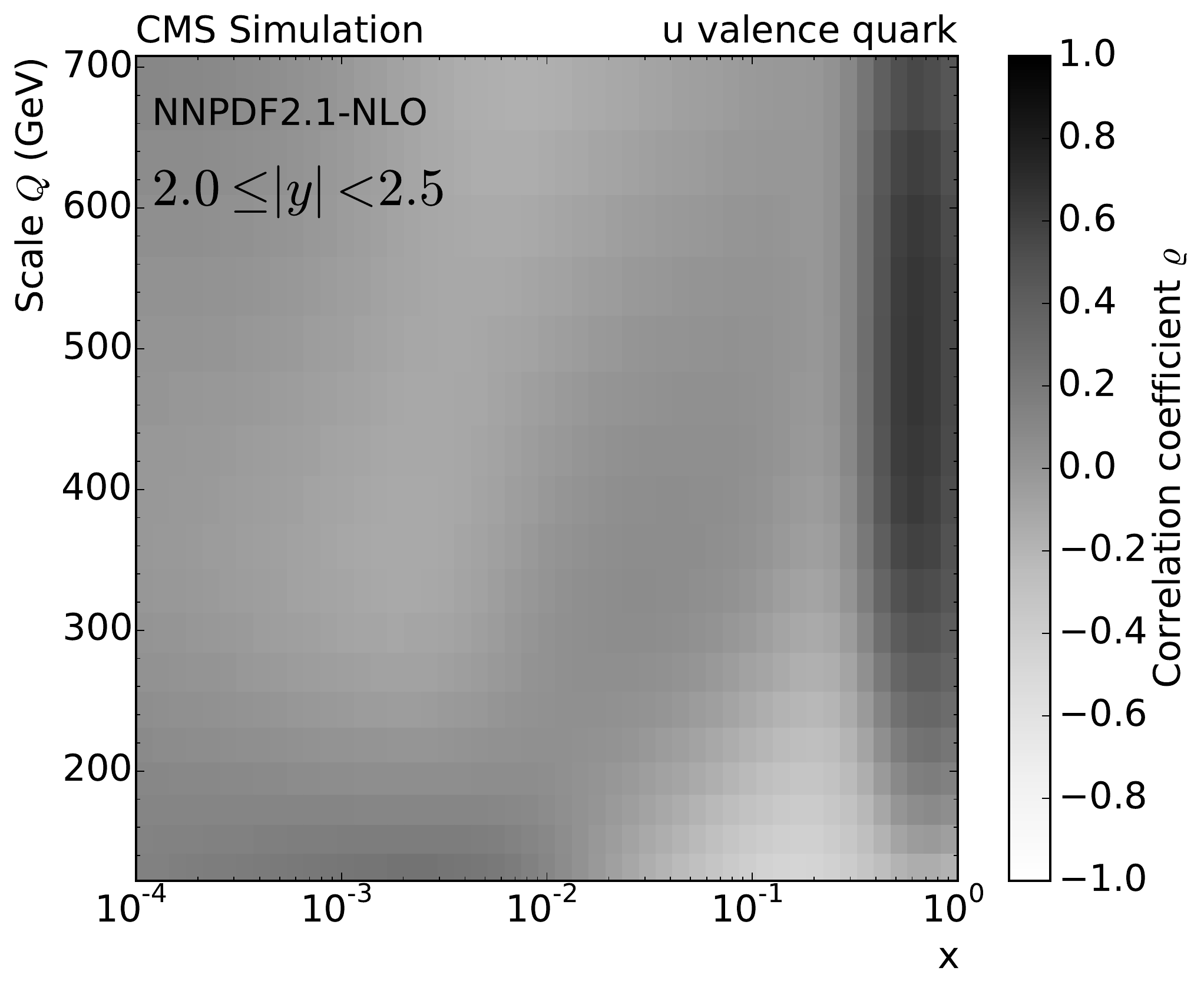}
  \includegraphics[width=0.45\textwidth]{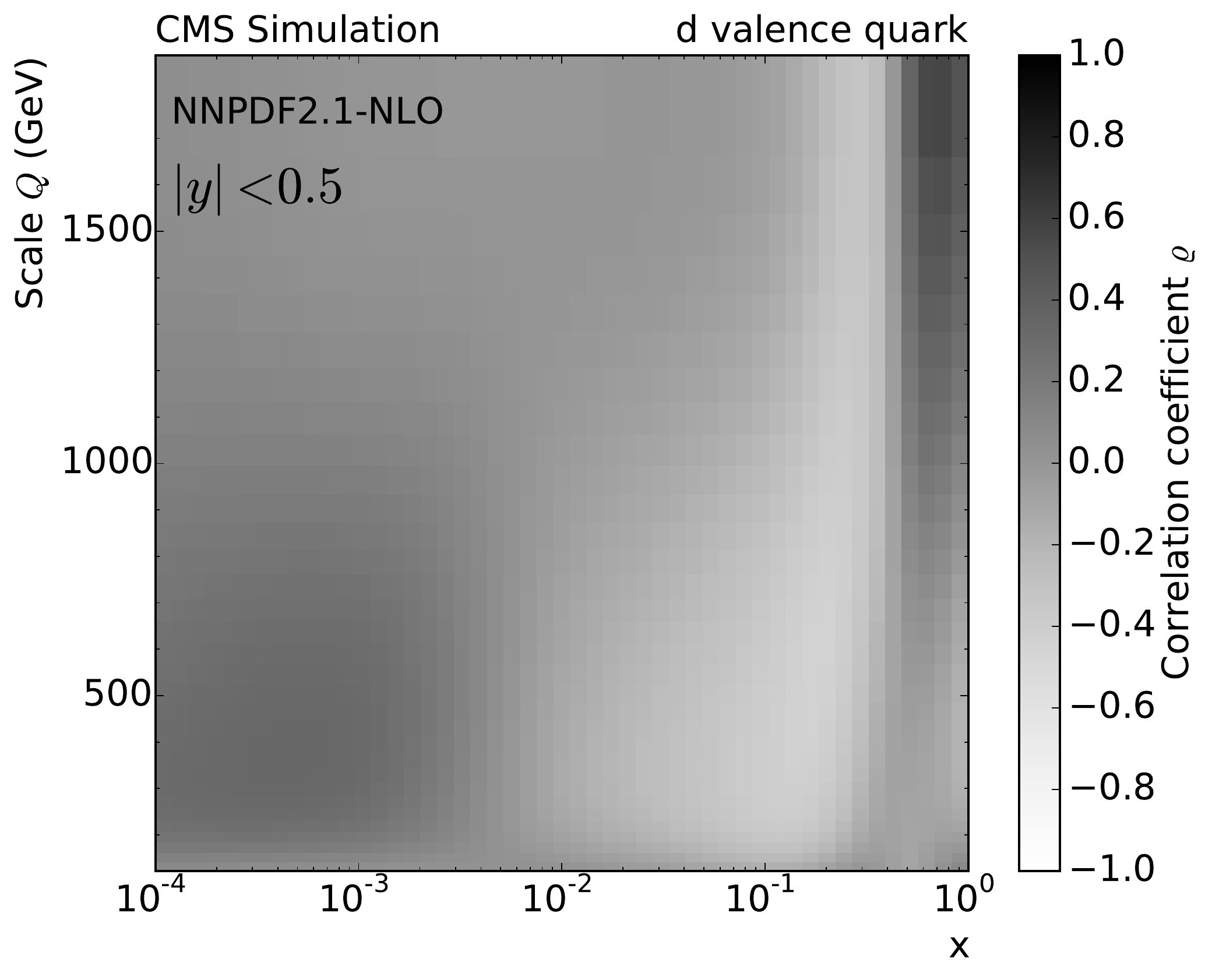}\hfill%
  \includegraphics[width=0.45\textwidth]{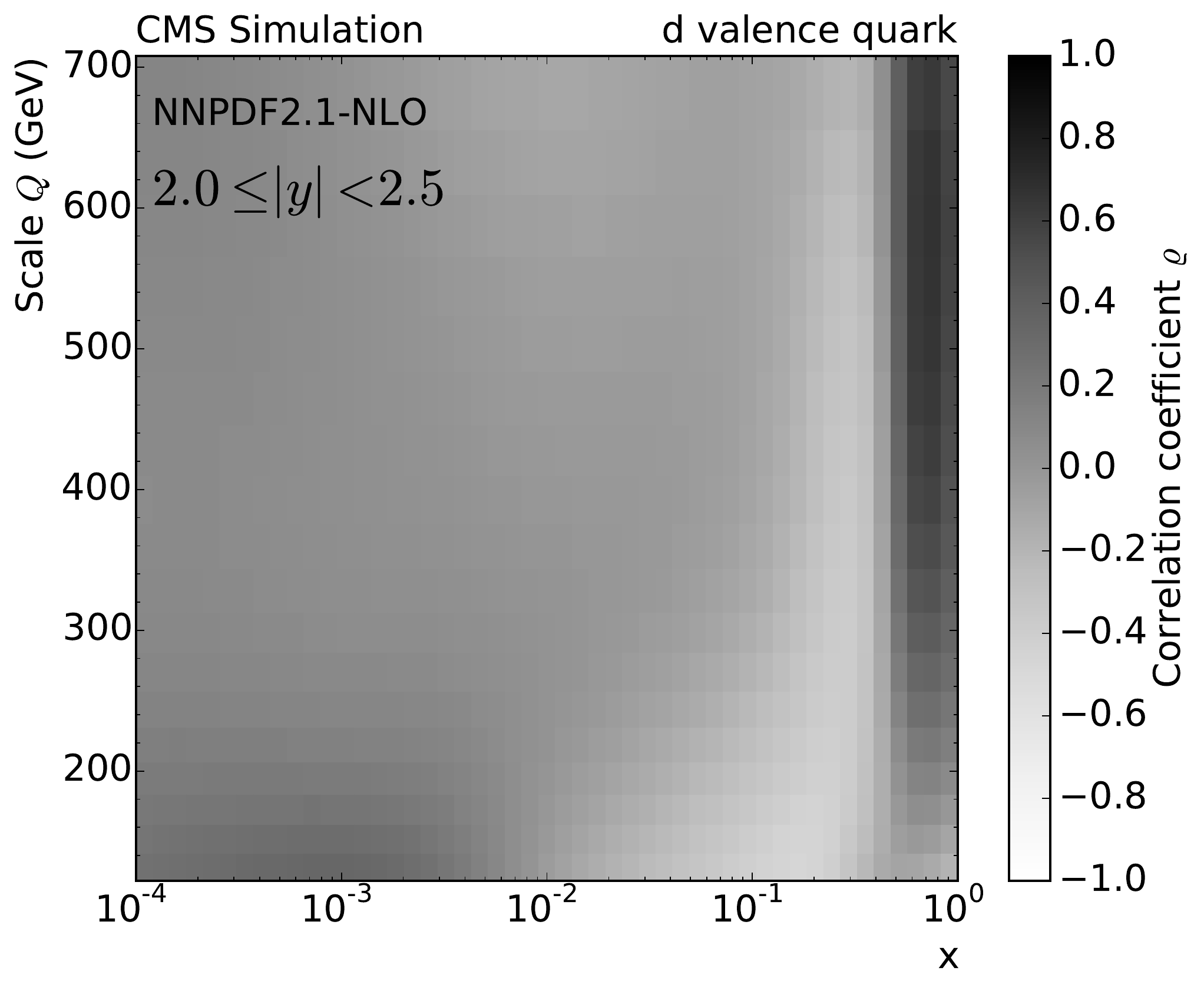}
  \caption{The correlation coefficient between the inclusive jet cross
    section and the gluon (top row), the u valence quark (middle row),
    and the d valence quark PDFs (bottom row), as a function of the
    momentum fraction $x$ of the proton and the energy scale $Q$ of
    the hard process. The correlation is shown for the central
    rapidity region $\yabs<0.5$ (left) and for $2.0<\yabs<2.5$
    (right).}
  \label{fig:correlation_pdf_xs_gqq}
\end{figure*}

The potential impact of the CMS inclusive jet data can be illustrated
by the correlation between the inclusive jet cross section
$\sigma_{\text{jet}}(Q)$ and the PDF $xf(x,Q^2)$ for any parton
flavour~$f$. The NNPDF Collaboration~\cite{Ball:2008by} provides PDF
sets in the form of an ensemble of replicas $i$, which sample
variations in the PDF parameter space within allowed uncertainties.
The correlation coefficient $\varrho_f(x,Q)$ between a cross section
and the PDF for flavour~$f$ at a point $(x,Q)$ can be computed by
evaluating means and standard deviations from an ensemble of $N$
replicas as
\ifthenelse{\boolean{cms@external}}{
\begin{multline}
  \varrho_f (x,Q) =
  \frac{N}{(N-1)} \times\\
  \frac{
    \langle \sigma_{\text{jet}}(Q)_i \cdot xf(x,Q^2)_i \rangle -
    \langle \sigma_{\text{jet}}(Q)_i \rangle \cdot
    \langle xf(x,Q^2)_i \rangle}
  {\Delta_{\sigma_{\text{jet}}(Q)} \Delta_{xf(x,Q^2)}}.
\end{multline}
}{
\begin{equation}
  \varrho_f (x,Q) =
  \frac{N}{(N-1)} \frac{%
    \langle \sigma_{\text{jet}}(Q)_i \cdot xf(x,Q^2)_i \rangle -
    \langle \sigma_{\text{jet}}(Q)_i \rangle \cdot
    \langle xf(x,Q^2)_i \rangle}
  {\Delta_{\sigma_{\text{jet}}(Q)} \Delta_{xf(x,Q^2)}}.
\end{equation}
}
Here, the angular brackets denote the averaging over the replica index
$i$, and $\Delta$ represents the evaluation of the corresponding
standard deviation for either the jet cross section,
$\Delta_{\sigma_{\text{jet}}(Q)}$, or a PDF, $\Delta_{xf(x,Q^2)}$.
Figure~\ref{fig:correlation_pdf_xs_gqq} presents the correlation
coefficient between the inclusive jet cross section and the gluon, u
valence quark, and d valence quark PDFs in the proton.

The correlation between the gluon PDF and the inclusive jet cross
section is largest at central rapidity for most jet \pt. In contrast,
the correlation between the valence quark distributions and the jet
cross section is rather small except for very high \pt such that some
impact can be expected at high $x$ from including these jet data in
PDF fits. In the forward region the correlation between the valence
quark distributions and the jet cross sections is more pronounced at
high $x$ and smaller jet \pt. Therefore, a significant reduction of
the PDF uncertainties is expected by including the CMS inclusive jet
cross section into fits of the proton structure.

\subsection{The fitting framework}
\label{section:fittingframework}

\subsubsection{The \texorpdfstring{\HERAFitter}{HERAFitter} setup}
\label{section:herafittersetup}

The impact of the CMS inclusive jet data on proton PDFs is
investigated by including the jet cross section measurement in a
combined fit at NLO with the HERA-I inclusive DIS cross
sections~\cite{Aaron:2009aa}, which were the basis for the
determination of the HERAPDF1.0 PDF set. The analysis is performed
within the \HERAFitter framework using the
Dokshitzer--Gribov--Lipatov--Altarelli--Parisi~\cite{Gribov:1972ri,Altarelli:1977zs,Dokshitzer:1977sg}
evolution scheme at NLO as implemented in the \QCDNUM
package~\cite{Botje:2010ay} and the generalised-mass variable-flavour
number Thorne--Roberts scheme~\cite{Thorne:1997ga,Thorne:2006qt}.

In contrast to the original HERAPDF fit, the results presented here
require the DIS data to fulfill $Q^2 > Q_\text{min}^2 = 7.5 \GeVsq$
instead of $3.5\GeVsq$. The amount of DIS data left out by the
increased $Q_\text{min}^2$ threshold is rather small and concerns a phase
space where a perturbative description is less reliable. A similar,
higher cutoff has been applied by the ATLAS
Collaboration~\cite{Aad:2012sb,Aad:2014qja}. As a crosscheck all fits
have been performed for a cutoff of $Q^2 > Q_\text{min}^2 = 3.5
\GeVsq$, and the results are consistent with the ones obtained using the
more stringent cutoff. Differences beyond the expected reduction of
uncertainties at low $x$ have not been observed.

The following PDFs are independent in the fit procedure:
$xu_v(x)$, $xd_v(x)$, $xg(x)$, and
$x\overline{U}(x)$, $x\overline{D}(x)$, where $x\overline{U}(x) =
x\overline{u}(x)$, and $x\overline{D}(x) = x\overline{d}(x) +
x\overline{s}(x)$.  Similar to Ref.~\cite{Abramowicz:1900rp}, a
parameterization with 13 free parameters is used.  At the starting
scale $Q_0$ of the QCD evolution, chosen to be $Q_0^2 = 1.9 \GeVsq$,
the PDFs are parameterized as follows:
\begin{equation}\begin{aligned}
  xg(x) &= A_g x^{B_g} (1-x)^{C_g} - A'_g x^{B'_g} (1-x)^{C'_g},\\
  xu_v(x) &= A_{u_{v}} x^{B_{u_{v}}} (1-x)^{C_{u_{v}}} (1 + E_{u_{v}}x^2),\\
  xd_v(x) &= A_{d_v} x^{B_{d_v}} (1-x)^{C_{d_{v}}},\\
  x\overline U(x) &= A_{\overline U} x^{B_{\overline U}}
  (1-x)^{C_{\overline U}}, \text{and}\\
  x\overline D(x) &= A_{\overline D} x^{B_{\overline D}}
  (1-x)^{C_{\overline D}}.
\end{aligned}\end{equation}

The normalisation parameters $A_g$, $A_{u_{v}}$, and $A_{d_{v}}$ are
constrained by QCD sum rules. Additional constraints $B_{\overline
  U}=B_{\overline D}$ and $A_{\overline U} = A_{\overline D}(1-f_s)$
are applied to ensure the same normalisation for the $\overline u$ and
$\overline d$ densities for $x \to 0$. The strangeness
fraction is set to $f_s = 0.31$, as obtained from neutrino-induced
dimuon production~\cite{Mason:2007zz}. The parameter $C'_g$ is fixed
to 25~\cite{Martin:2009iq,Thorne:2006qt} and the strong coupling constant
to $\alpsmz= 0.1176$.

\subsubsection{Definition of the goodness-of-fit estimator}
\label{sec:fitsetup}

The agreement between the $N$ data points $D_i$ and the theoretical
predictions $T_i$ is quantified via a least-squares method, where
\ifthenelse{\boolean{cms@external}}{
\begin{multline}
  \chi^2 = \sum_{ij}^N \left(D_i - T_i - \sum_k^K r_k \beta_{ik}\right)\\ \mathrm{C}_{ij}^{-1}
  \left(D_j - T_j - \sum_k^K r_k \beta_{jk} \right)
   + \sum_k^K r_k^2.
  \label{chi2_nuisance}
\end{multline}
}{
\begin{equation}
  \chi^2 = \sum_{ij}^N \left(D_i - T_i - \sum_k^K r_k \beta_{ik}\right) \mathrm{C}_{ij}^{-1}
  \left(D_j - T_j - \sum_k^K r_k \beta_{jk} \right) + \sum_k^K r_k^2.
  \label{chi2_nuisance}
\end{equation}
}
For fully correlated sources of uncertainty following a Gaussian
distribution with a zero mean and a root-mean-square of unity as
assumed here, this definition is equivalent to
Eq.~(\ref{chi2_square})~\cite{Stump:2001gu}. As a bonus, the
systematic shift of the nuisance parameter $r_k$ for each source in a
fit is determined. Numerous large shifts in either direction indicate
a problem as for example observed while fitting \alpsmz with this
technique and the old uncertainty correlation prescription.

In the following, the covariance matrix is defined as $\mathrm{C} =
\cov_{\text{stat}} + \cov_{\text{uncor}}$, while
the JES, unfolding, and luminosity determination are treated as fully
correlated systematic uncertainties $\beta_{ik}$ with nuisance
parameters $r_k$. Including also the NP uncertainties, treated via the
offset method in Section~\ref{sec:alphas}, in the form of one nuisance
parameter in total $K$ such sources are defined.  Of course, PDF
uncertainties emerge as results of the fits performed here, in
contrast to serving as inputs, as they do in the fits of \alpsmz
presented in Section~\ref{sec:alphas}.

All the fully correlated sources are assumed to be multiplicative to
avoid the statistical bias that arises from uncertainty estimations
taken from data~\cite{Lyons:1989gh,D'Agostini:2003nk,Ball:2009qv}.  As
a consequence, the covariance matrix of the remaining sources has to
be re-evaluated in each iteration step. To inhibit the compensation of
large systematic shifts by increasing simultaneously the theoretical
prediction and the statistical uncertainties, the systematic shifts of
the theory are taken into account before the rescaling of the
statistical uncertainty. Otherwise alternative minima in \chisq can
appear that are associated with large theoretical predictions and
correspondingly large shifts in the nuisance parameters. These
alternative minima are clearly undesirable~\cite{HERAFitter:2013hf}.

\subsubsection{Treatment of CMS data uncertainties}
\label{section:cmsdatauncertainties}

The JES is the dominant source of experimental systematic uncertainty
in jet cross sections. As described in
Section~\ref{sec:measurementjec}, the \pt- and $\eta$-dependent JES
uncertainties are split into 16 uncorrelated sources that are fully
correlated in \pt and $\eta$. Following the modified recommendation
for the correlations versus rapidity of the single-particle response
source as given in Section~\ref{sec:measurementjec}, it is necessary
to split this source into five parts for the purpose of using the
uncertainties published in Ref.~\cite{Chatrchyan:2012bja} within the
\chisq fits. The complete set of uncertainty sources is shown in
Table~\ref{tab:cmsjets2011:nuisance}.

By employing the technique of nuisance parameters, the impact of each
systematic source of uncertainty on the fit result can be examined
separately. For an adequate estimation of the size and the
correlations of all uncertainties, the majority of all systematic
sources should be shifted by less than one standard deviation from the
default in the fitting procedure.
Table~\ref{tab:cmsjets2011:nuisance} demonstrates that this is the
case for the CMS inclusive jet data.

\begin{table*}[tbp]
  \centering
  \topcaption{The 19 independent sources of systematic uncertainty considered in the
    CMS inclusive jet measurement. Out of these, 16 are related to the JES
    and are listed first. In order to implement the improved correlation
    treatment as described in Section~\ref{sec:measurementjec}, the
    single-particle response source JEC2, see also
    Appendix~\ref{sec:jessources}, has been split up into five
    sources: JEC2a--JEC2e. The shift from the default value in each source of systematic
    uncertainty is determined by nuisance parameters in the fit and is
    presented in units of standard deviations.}
  \label{tab:cmsjets2011:nuisance}
\begin{tabular}{llr}
    \multicolumn{2}{l}{\multirow{2}{*}{Systematic source}} & \multicolumn{1}{l}{\multirow{2}{*}{Shift in standard}}\\
    && \multicolumn{1}{l}{deviations}\rbthm\\\hline
    JEC0  & absolute jet energy scale & $ 0.09$\rbtrr\\
    JEC1  & MC extrapolation & $0.00$\rbtrr\\
    JEC2a & single-particle response barrel & $ 1.31$\rbtrr\\
    JEC2b & single-particle response endcap & $-1.46$\rbtrr\\
    JEC2c & single-particle decorrelation $\yabs<0.5$ & $0.20$\rbtrr\\
    JEC2d & single-particle decorrelation $0.5\leq\yabs<1.0$ & $ 0.19$\rbtrr\\
    JEC2e & single-particle decorrelation $1.0\leq\yabs<1.5$ & $ 0.92$\rbtrr\\
    JEC3  & jet flavor correction & $ 0.04$\rbtrr\\
    JEC4  & time-dependent detector effects & $-0.15$\rbtrr\\
    JEC5  & jet \pt resolution in endcap 1 & $ 0.76$\rbtrr\\
    JEC6  & jet \pt resolution in endcap 2 & $-0.42$\rbtrr\\
    JEC7  & jet \pt resolution in HF & $ 0.01$\rbtrr\\
    JEC8  & correction for final-state radiation & $0.03$\rbtrr\\
    JEC9  & statistical uncertainty of $\eta$-dependent correction for endcap & $-0.42$\rbtrr\\
    JEC10 & statistical uncertainty of $\eta$-dependent correction for HF & $ 0.00$\rbtrr\\
    JEC11 & data-MC difference in $\eta$-dependent pileup correction & $ 0.91$\rbtrr\\
    JEC12 & residual out-of-time pileup correction for prescaled triggers & $-0.17$\rbtrr\\
    JEC13 & offset dependence in pileup correction & $-0.03$\rbtrr\\
    JEC14 & MC pileup bias correction  & $ 0.39$\rbtrr\\
    JEC15 & jet rate dependent pileup correction & $ 0.29$\rbtrr\\
    \hline
    \multicolumn{2}{l}{Unfolding}     & $-0.26$\rbtrr\\
    \multicolumn{2}{l}{Luminosity}    & $-0.07$\rbtrr\\
    \multicolumn{2}{l}{NP correction} & $ 0.60$\rbtrr\\
  \end{tabular}
\end{table*}

In contrast, with the original assumption of full correlation within
the 16 JES systematic sources across all \yabs bins, shifts beyond two
standard deviations were apparent and led to a re-examination of this
issue and the improved correlation treatment of the JES uncertainties
as described previously in Section~\ref{sec:measurementjec}.

\subsection{Determination of PDF uncertainties according to the
  HERAPDF prescription}
\label{section:herapdf_pdf_uncertainties}

The uncertainty in the PDFs is subdivided into experimental, model,
and parameterization uncertainties that are studied separately. In the
default setup of the \HERAFitter framework, experimental uncertainties
are evaluated following a Hessian method~\cite{Stump:2001gu}, and
result from the propagated statistical and systematic uncertainties of
the input data.

For the model uncertainties, the offset method~\cite{Botje:2001fx} is
applied considering the following variations of model assumptions:

\begin{enumerate}
\item The strangeness fraction $f_s$, by default equal to $0.31$, is
  varied between $0.23$ and $0.38$.
\item The b-quark mass is varied by $\pm 0.25\GeV$ around the central
  value of $4.75\GeV$.
\item The c-quark mass, with the central value of $1.4\GeV$, is varied
  to $1.35\GeV$ and $1.65\GeV$. For the downwards variation the charm
  production threshold is avoided by changing the starting scale to
  $Q_0^2=1.8\GeVsq$ in this case.
\item The minimum $Q^2$ value for data used in the fit,
  $Q^2_\mathrm{min}$, is varied from $7.5\GeVsq$ to $5.0\GeVsq$ and
  $10\GeVsq$.
\end{enumerate}

The PDF parameterization uncertainty is estimated as described in
Ref.~\cite{Aaron:2009aa}. By employing the more general form of
parameterizations
\ifthenelse{\boolean{cms@external}}{
\begin{equation}\begin{split}
  xg(x) =& A_g x^{B_g} (1-x)^{C_g} (1  + D_g x + E_g x^2) \\
  &- A'_g x^{B'_g} (1-x)^{C'_g},\\
  xf(x) =& A_{f} x^{B_{f}} (1-x)^{C_{f}} (1 + D_{f}x + E_{f}x^2)
\end{split}\end{equation}
}{
\begin{equation}\begin{aligned}
  xg(x) &= A_g x^{B_g} (1-x)^{C_g} (1  + D_g x + E_g x^2) - A'_g x^{B'_g} (1-x)^{C'_g},\\
  xf(x) &= A_{f} x^{B_{f}} (1-x)^{C_{f}} (1 + D_{f}x + E_{f}x^2)
\end{aligned}\end{equation}
}
for gluons and the nongluon flavours, respectively, it is tested
whether the successive inclusion of additional fit parameters leads to
a variation in the shape of the fitted results.  Furthermore, the
starting scale $Q_0$ is changed to $Q^2_0 = 1.5\GeVsq$ and
$2.5\GeVsq$.  The maximal deviations of the resulting PDFs from those
obtained in the central fit define the parameterization
uncertainty. The experimental, model, and parameterization
uncertainties are added in quadrature to give the final PDF
uncertainty according to the HERAPDF prescription~\cite{Aaron:2009aa}.

\begin{figure*}[tbp]
  \centering
  \includegraphics[width=0.48\textwidth]{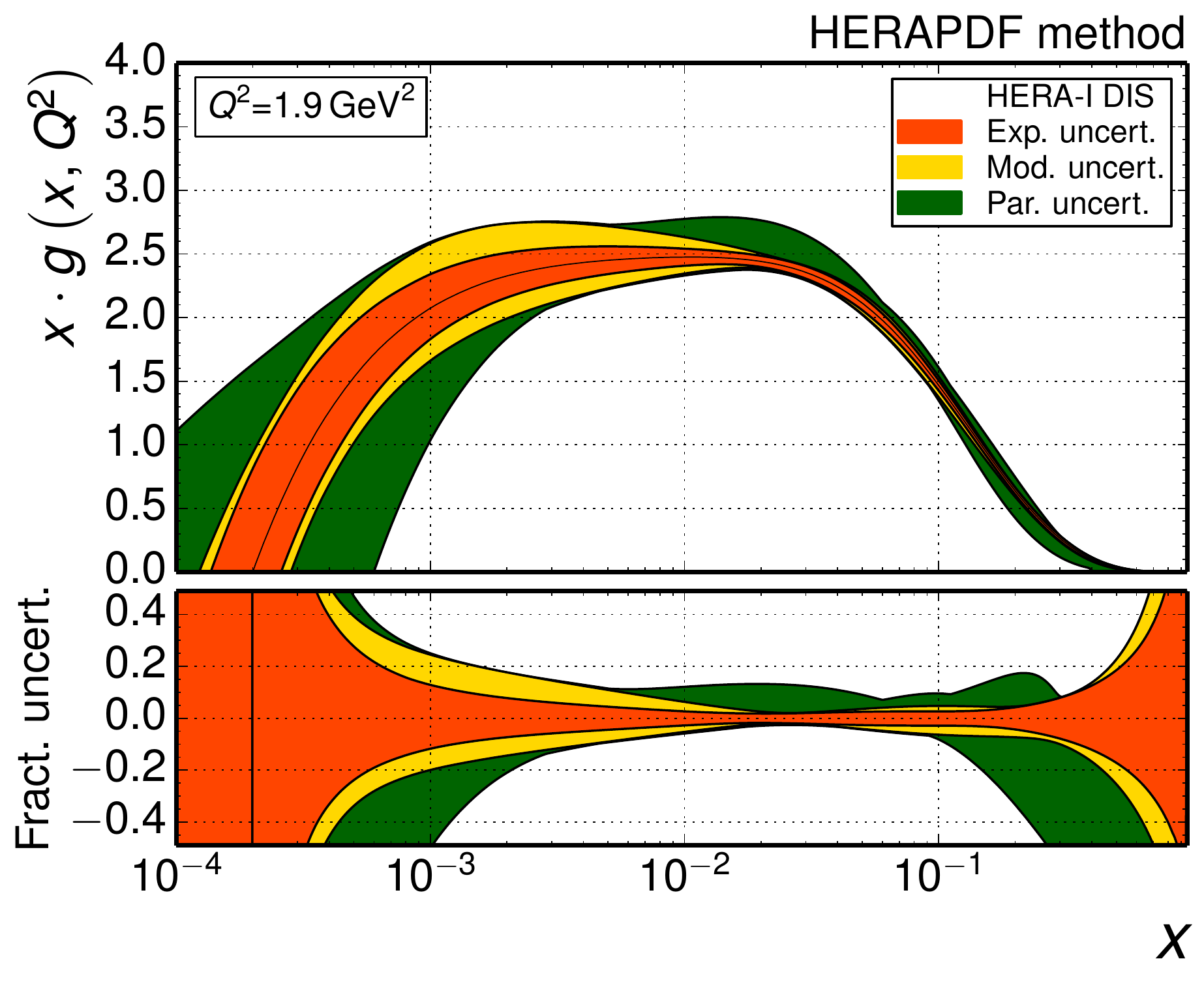}\hfill%
  \includegraphics[width=0.48\textwidth]{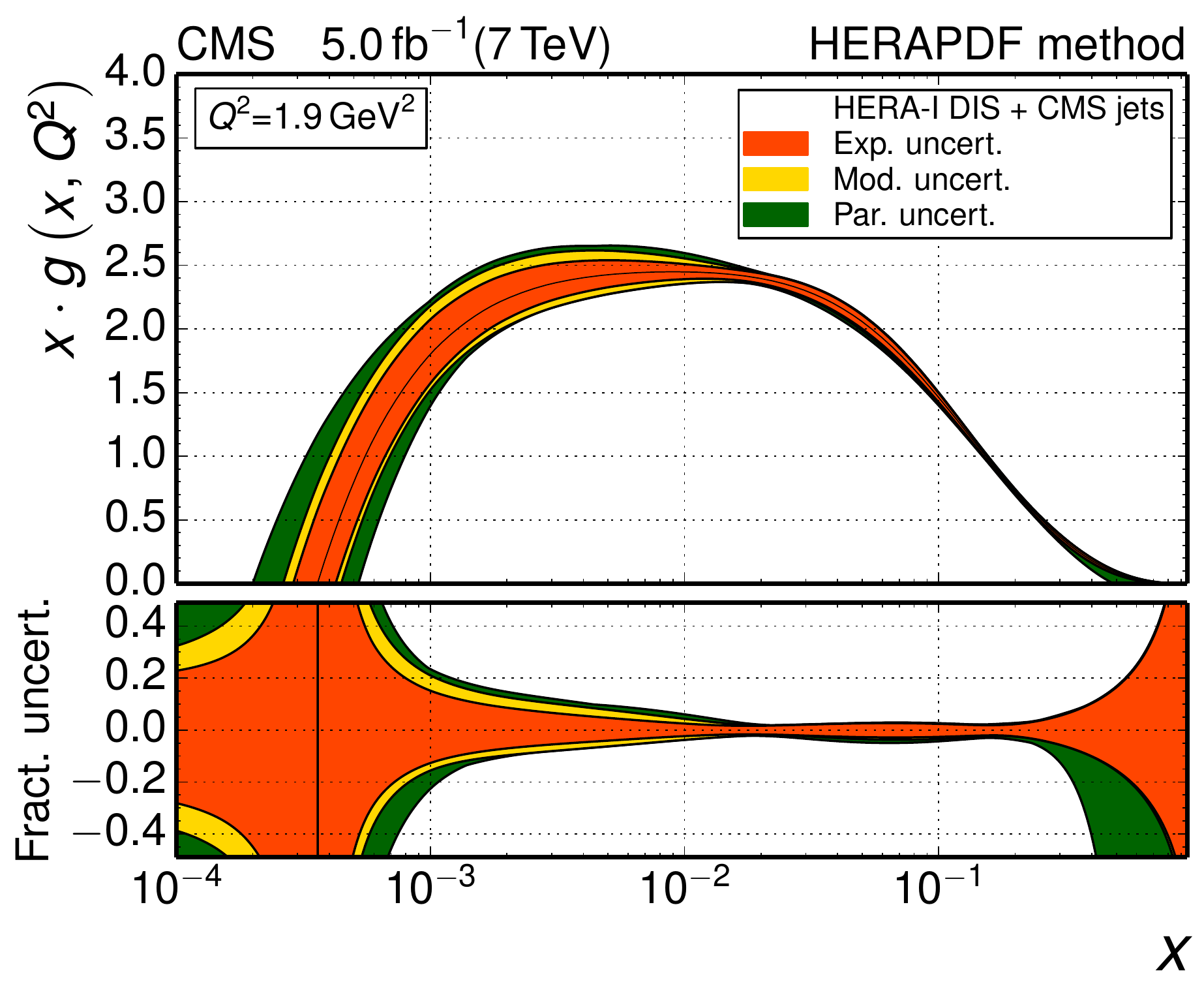}
  \includegraphics[width=0.48\textwidth]{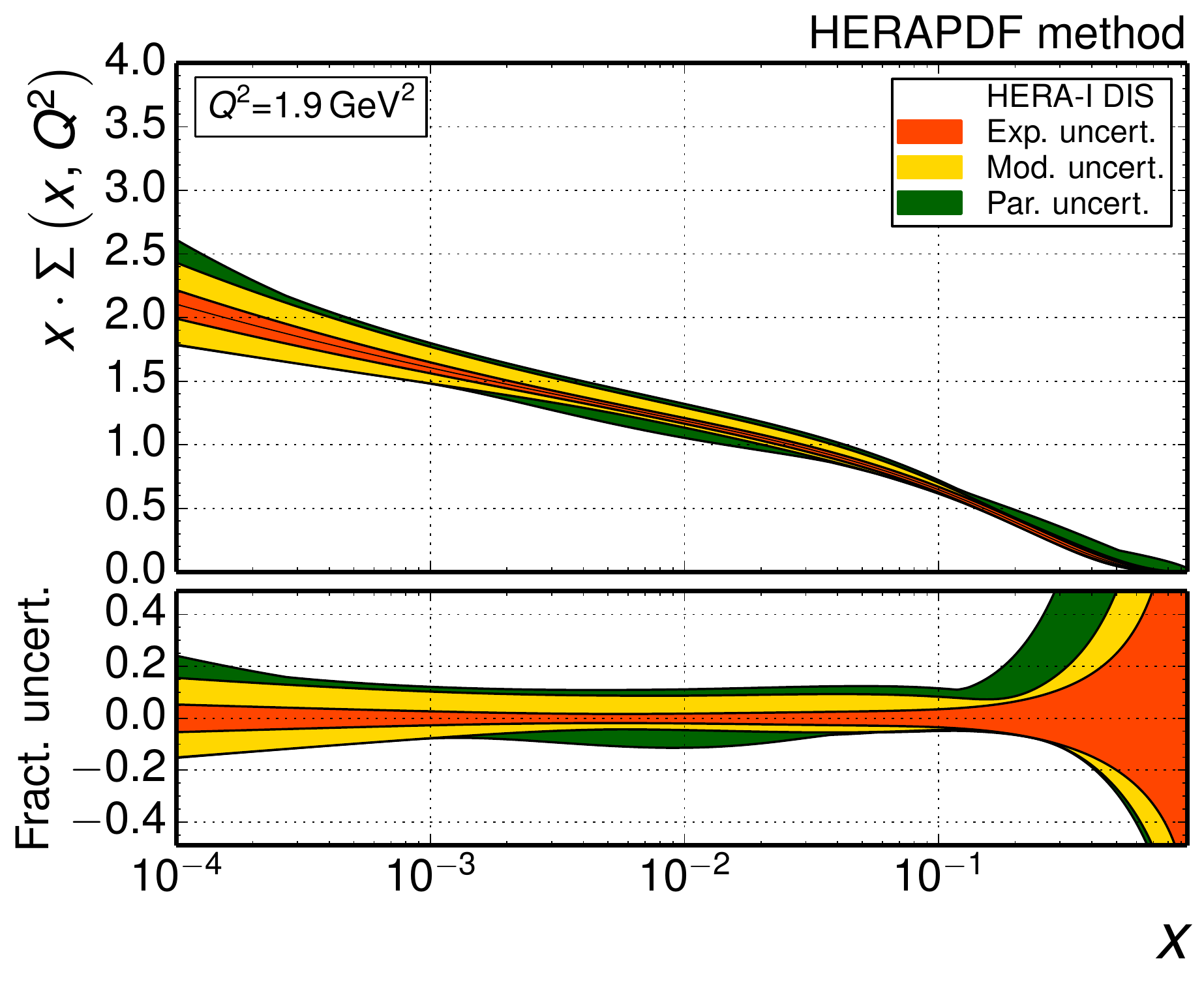}\hfill%
  \includegraphics[width=0.48\textwidth]{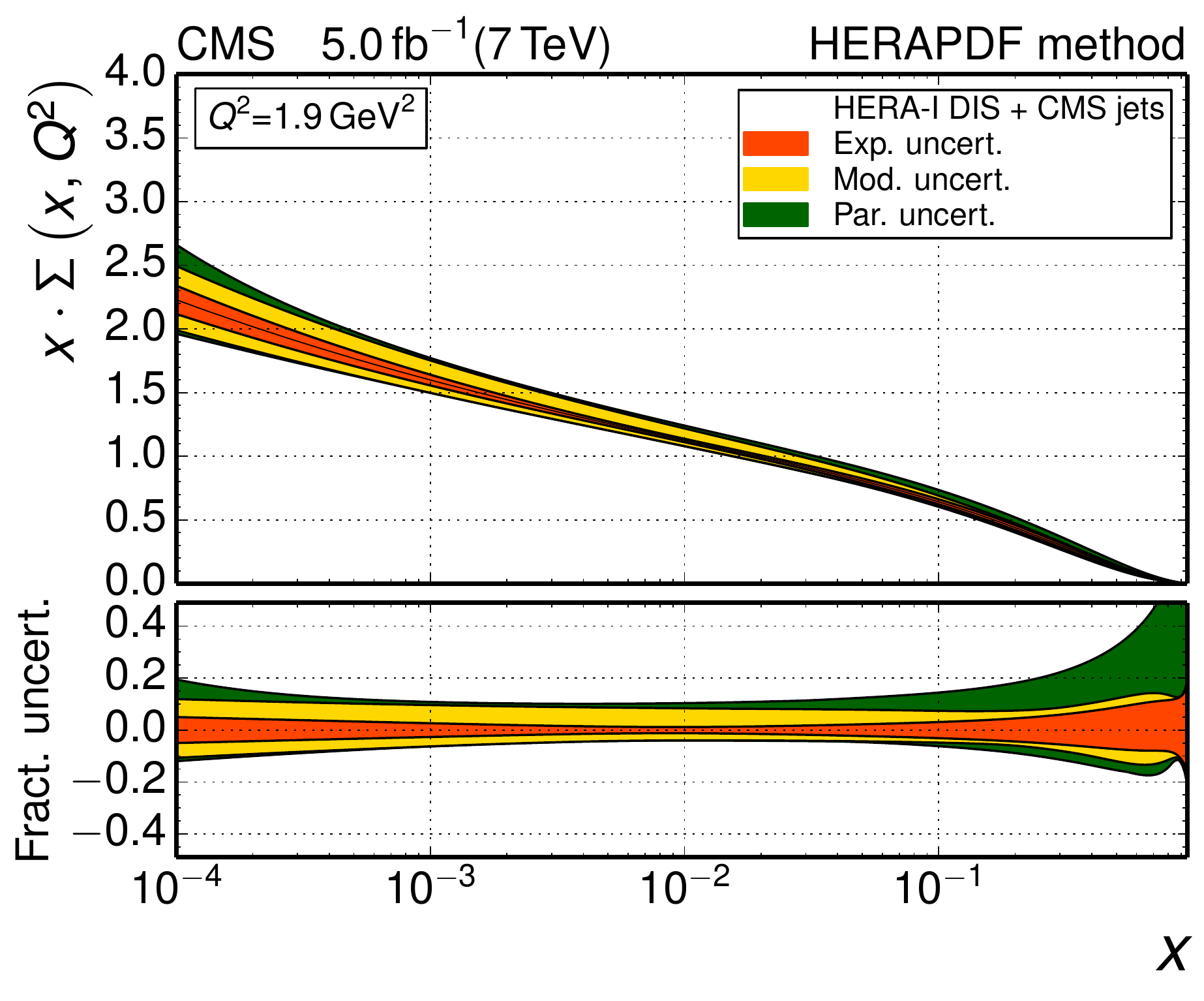}
  \caption{The gluon (top) and sea quark (bottom) PDFs as a function
    of $x$ as derived from HERA-I inclusive DIS data alone (left) and
    in combination with CMS inclusive jet data (right). The PDFs are
    shown at the starting scale $Q^2 = 1.9 \GeVsq$. The experimental
    (inner band), model (middle band), and parameterization
    uncertainties (outer band) are successively added quadratically to
    give the total uncertainty.}
  \label{fit:cmsjets2011:gsea:fitscale:qcut75}
\end{figure*}

\begin{figure*}[tbp]
  \centering
  \includegraphics[width=0.48\textwidth]{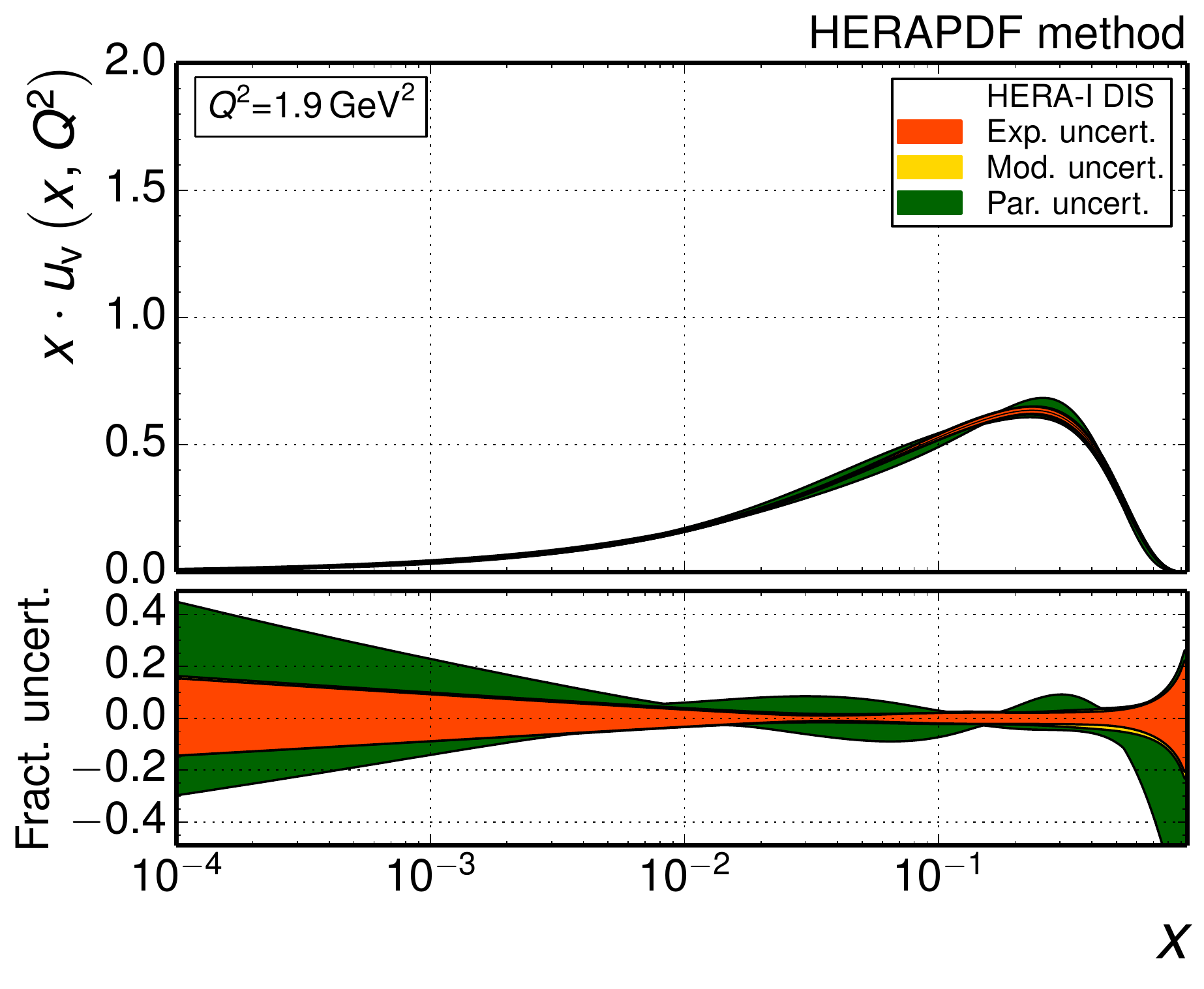}\hfill%
  \includegraphics[width=0.48\textwidth]{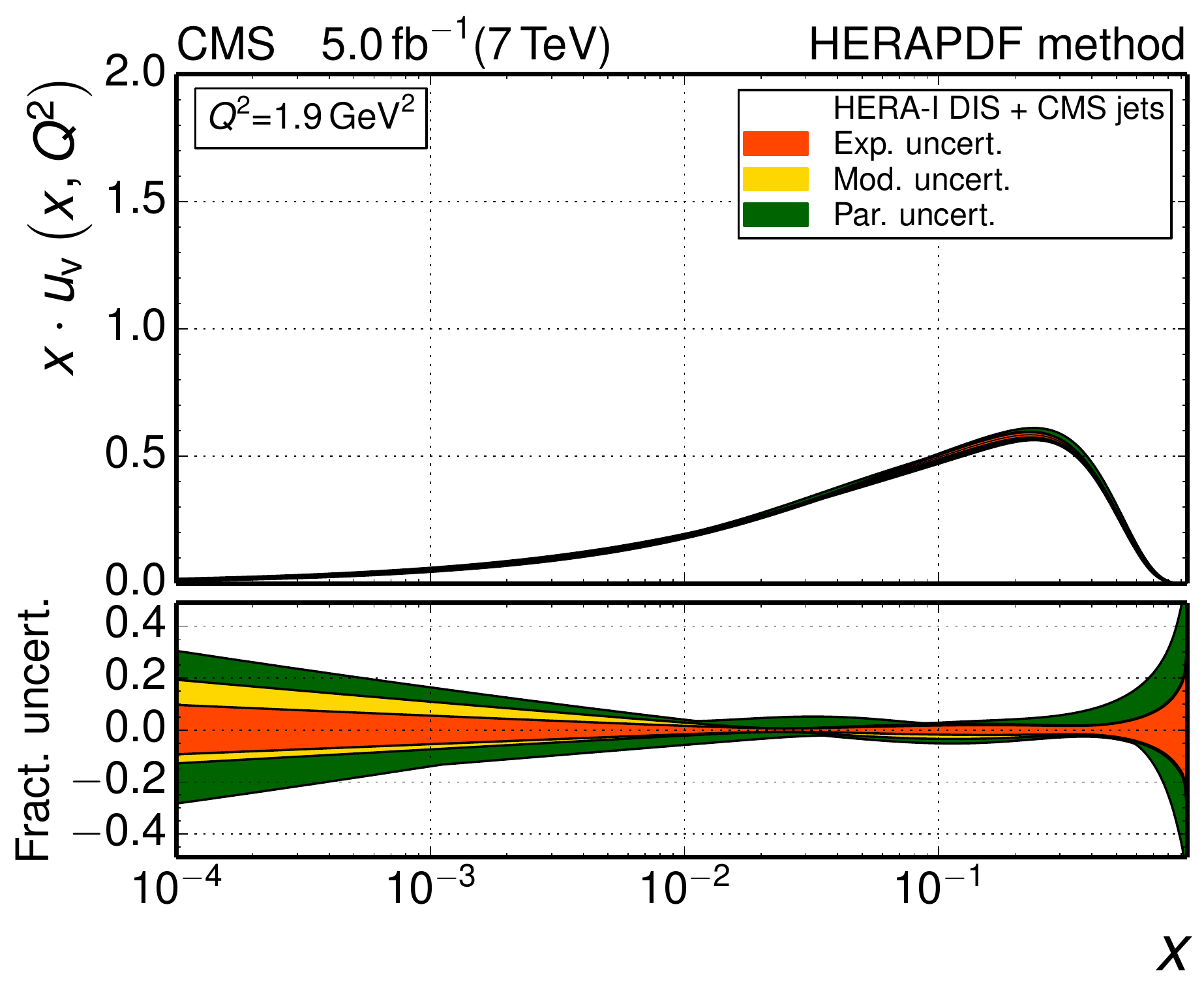}
  \includegraphics[width=0.48\textwidth]{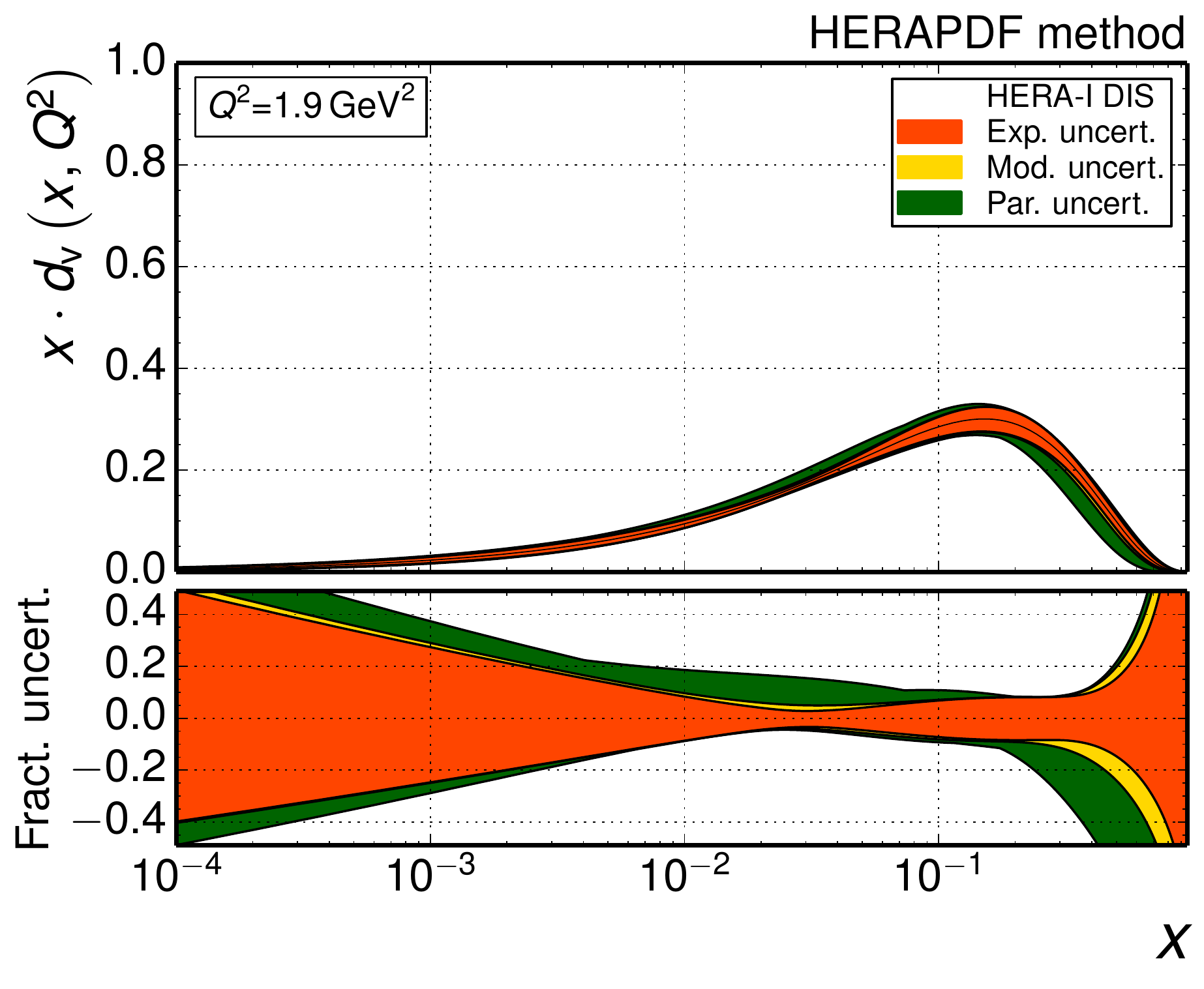}\hfill%
  \includegraphics[width=0.48\textwidth]{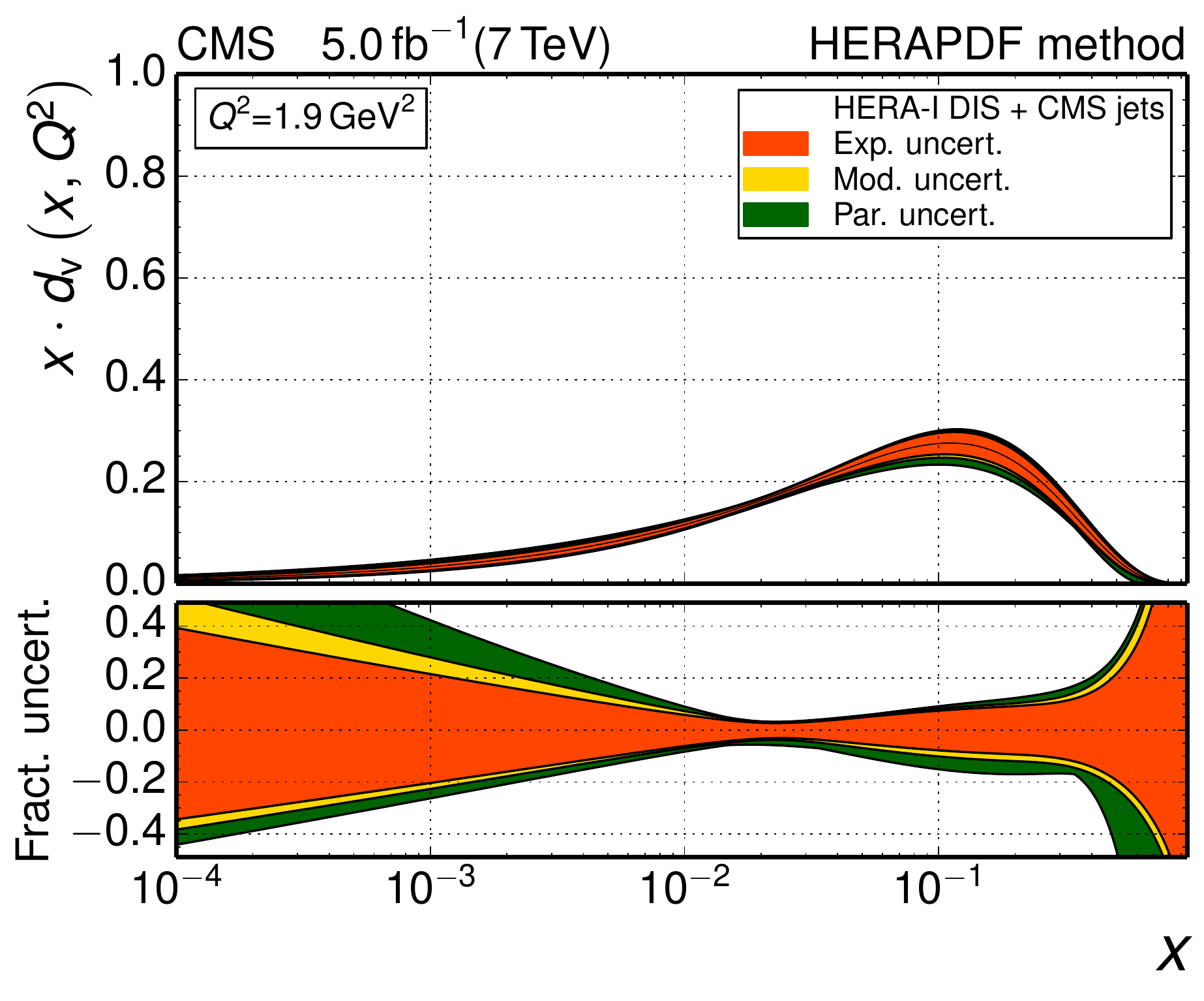}
  \caption{The u valence quark (top) and d valence quark (bottom) PDFs
    as a function of $x$ as derived from HERA-I inclusive DIS data
    alone (left) and in combination with CMS inclusive jet data
    (right). The PDFs are shown at the starting scale $Q^2 = 1.9
    \GeVsq$. The experimental (inner band), model (middle band), and
    parameterization uncertainties (outer band) are successively added
    quadratically to give the total uncertainty.}
  \label{fit:cmsjets2011:uvdv:fitscale:qcut75}
\end{figure*}

\begin{figure*}[tbp]
  \centering
  \includegraphics[width=0.48\textwidth]{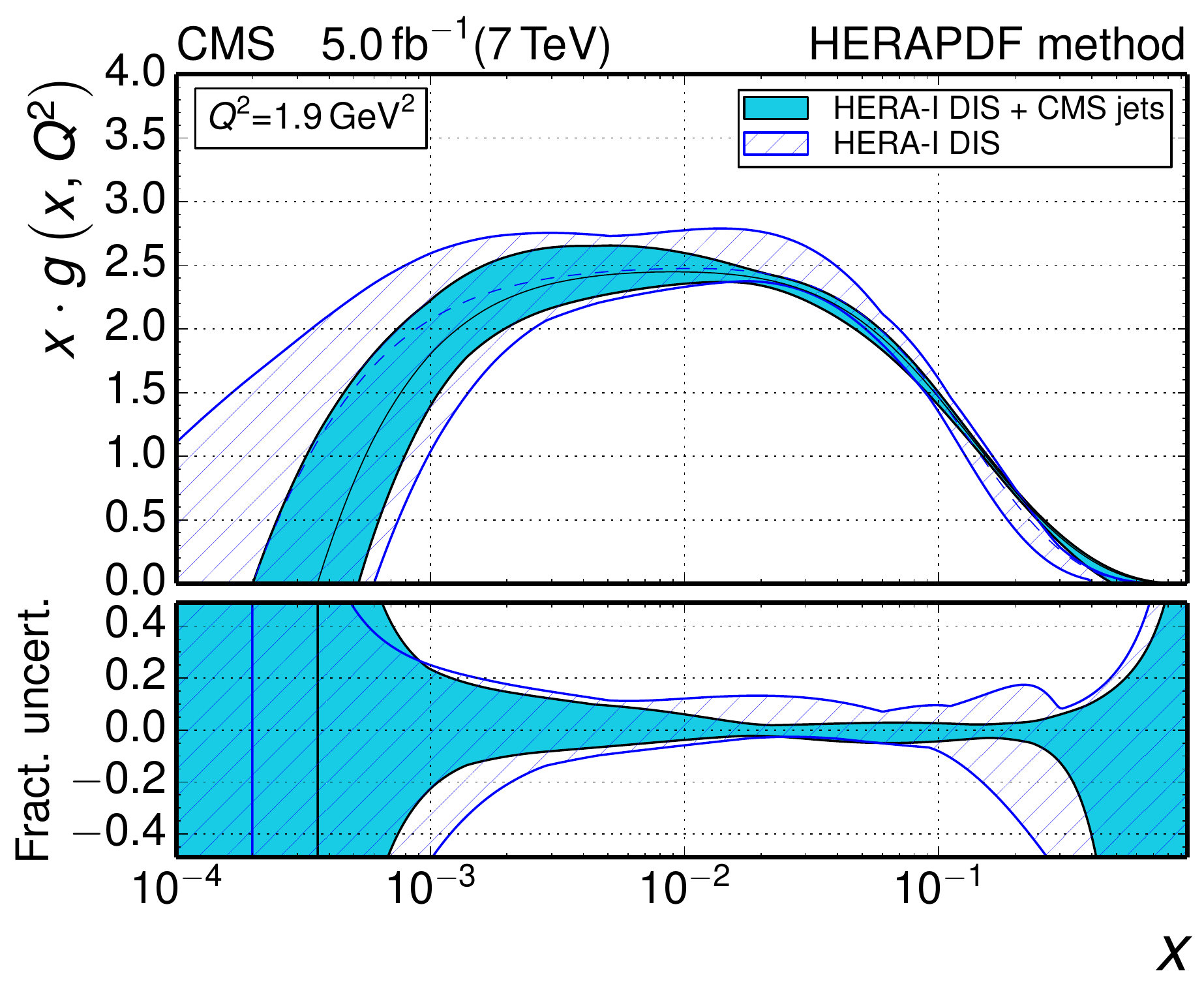}\hfill%
  \includegraphics[width=0.48\textwidth]{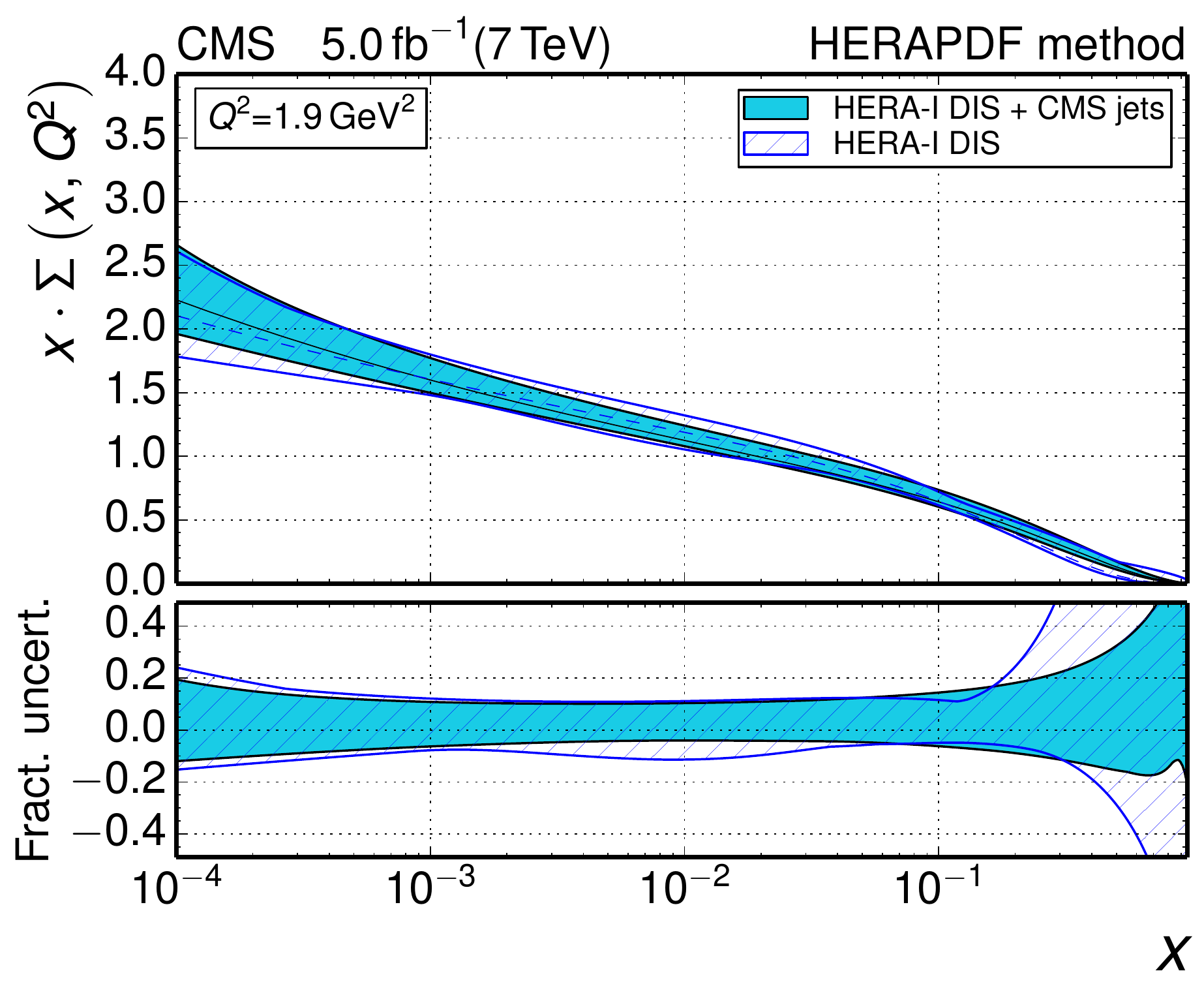}
  \includegraphics[width=0.48\textwidth]{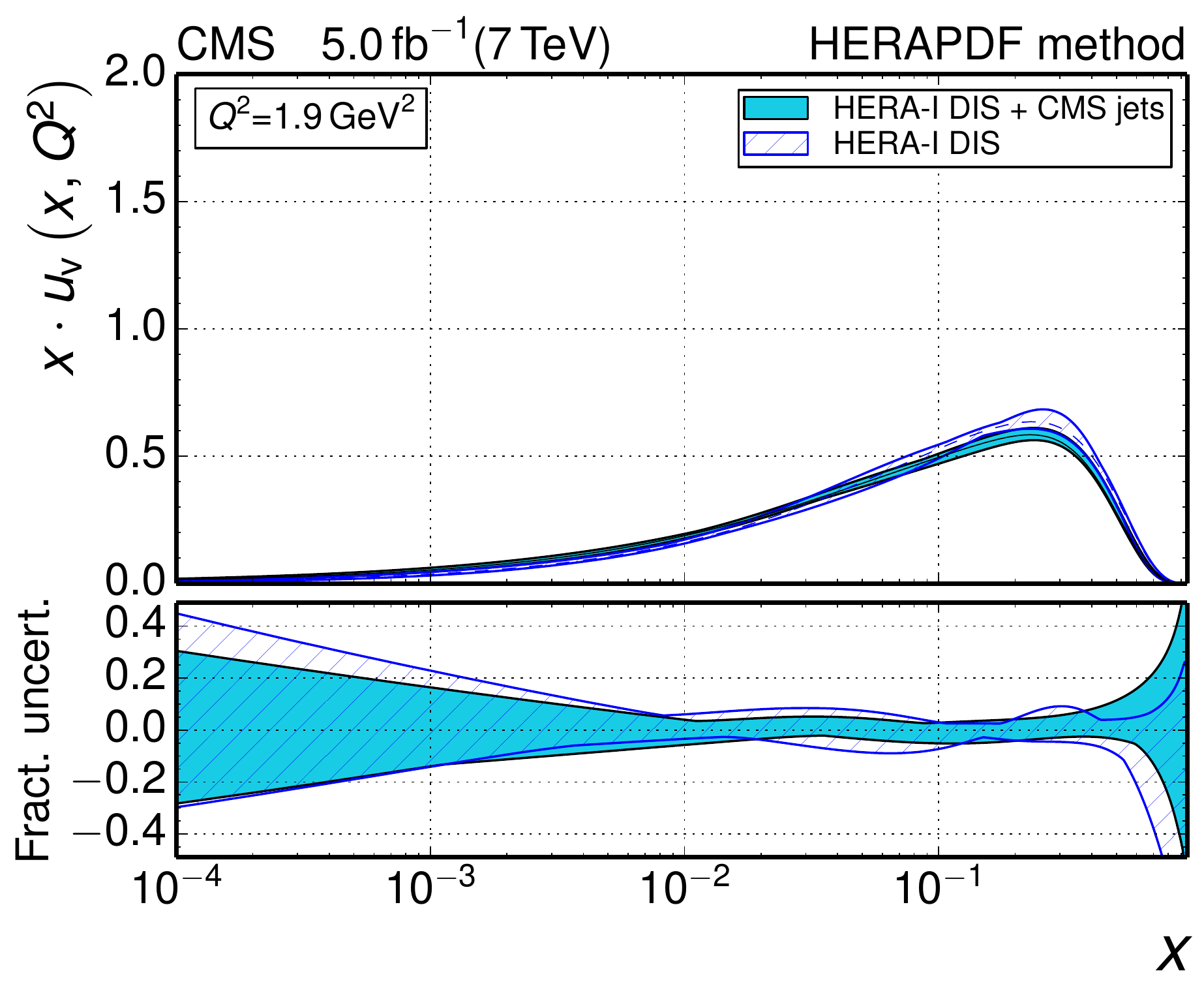}\hfill%
  \includegraphics[width=0.48\textwidth]{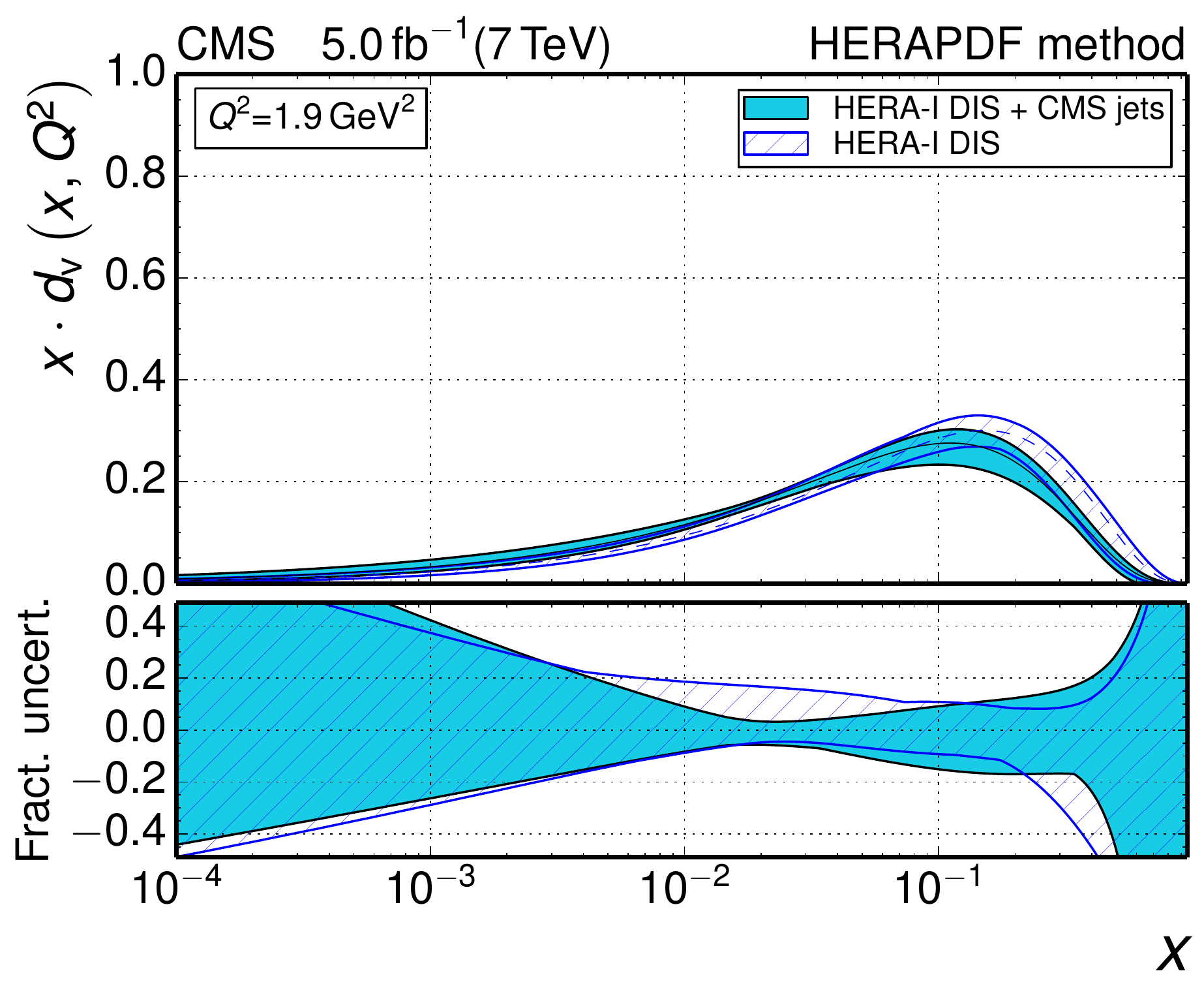}
  \caption{The gluon (top left), sea quark (top right), u valence
    quark (bottom left), and d valence quark (bottom right) PDFs as a
    function of $x$ as derived from HERA-I inclusive DIS data alone
    (dashed line) and in combination with CMS inclusive jet data (full
    line). The PDFs are determined employing the HERAPDF method with
    a $Q^2_\mathrm{min} = 7.5\GeVsq$ selection criterion.
    The PDFs are shown at the starting scale $Q^2 = 1.9\GeVsq$. Only the total
    uncertainty in the PDFs is shown (hatched and solid bands).}
  \label{fit:cmsjets2011:hera:directcomparison:1_4:all}
\end{figure*}

Using this fitting setup, the partial \chisq values per number of data
points, \ndata, are reported in Table~\ref{tab:fit:results} for each
of the neutral current (NC) and charged current (CC) data sets in the
HERA-I DIS fit and for the combined fit including the CMS inclusive
jet data. The achieved fit qualities demonstrate the compatibility of
all data within the presented PDF fitting framework. The resulting
PDFs with breakdown of the uncertainties for the gluon, the sea, u
valence, and d valence quarks with and without CMS inclusive jet data
are arranged next to each other in
Figs.~\ref{fit:cmsjets2011:gsea:fitscale:qcut75}
and~\ref{fit:cmsjets2011:uvdv:fitscale:qcut75}.
Figure~\ref{fit:cmsjets2011:hera:directcomparison:1_4:all} provides
direct comparisons of the two fit results with total uncertainties.
The parameterization and model uncertainties of the gluon distribution
are significantly reduced for almost the whole $x$ range from
$10^{-4}$ up to 0.5. When DIS data below $Q^2_\mathrm{min} = 7.5
\GeVsq$ are included in the fit, the effect is much reduced for the
low $x$ region $x < 0.01$, but remains important for medium to high
$x$. Also, for the u valence, d valence, and sea quark distributions
some reduction in their uncertainty is visible at high $x$ ($x \gtrsim
0.1$).

At the same time, some structure can be seen, particularly in the
parameterization uncertainties that might point to a still
insufficient flexibility in the parameterizations. Therefore, a
comparison is presented in the next
Section~\ref{section:mcddr_pdf_uncertainties}, using the MC method
with the regularisation based on data, which is also implemented within the
\HERAFitter framework.

\begin{table*}[tbp]
  \centering
  \topcaption{Partial \chisq values, \chipsq, for each data set in the HERA-I DIS (middle
    section) or in the combined fit including CMS inclusive jet data (right
    section). Here, \ndata is the number of data points available for the
    determination of the 13 parameters. The bottom two lines show the
    total \chisq and \chisqndof. The difference between the sum of all
    \chipsq and the total \chisq for the combined fit is attributed to the
    nuisance parameters.}
  \label{tab:fit:results}
\begin{tabular}{lrrcrc}
    \multicolumn{2}{c}{} &
    \multicolumn{2}{c}{HERA-I data} &
    \multicolumn{2}{c}{HERA-I \& CMS data}\rbthm\\\cline{3-6}
    data set &
    \multicolumn{1}{c}{\ndata} &
    \multicolumn{1}{c}{\chipsq} &
    \multicolumn{1}{c}{\chipsqndata} &
    \multicolumn{1}{c}{\chipsq} &
    \multicolumn{1}{c}{\chipsqndata}\rbthm\\\hline
    NC HERA-I H1-ZEUS combined $\Pem\Pp$ & 145 & 109 & 0.75 & 109 & 0.75 \rbtrr\\
    NC HERA-I H1-ZEUS combined $\Pep\Pp$ & 337 & 309 & 0.91 & 311 & 0.92 \rbtrr\\
    CC HERA-I H1-ZEUS combined $\Pem\Pp$ &  34 &  20 & 0.59 &  22 & 0.65 \rbtrr\\
    CC HERA-I H1-ZEUS combined $\Pep\Pp$ &  34 &  29 & 0.85 &  35 & 1.03 \rbtrr\\
    CMS inclusive jets              & 133 & --- &  --- & 102 & 0.77 \rbtrr\\\hline
    data set(s) & \ndof &
    \multicolumn{1}{c}{\chisq} &
    \multicolumn{1}{c}{\chisqndof} &
    \multicolumn{1}{c}{\chisq} &
    \multicolumn{1}{c}{\chisqndof}\rbthm\\\hline
    HERA-I data                       & 537 & 468 & 0.87 &    --- &  --- \rbtrr\\
    HERA-I \& CMS data                & 670 &    --- &  --- & 591 & 0.88 \rbtrr\\
  \end{tabular}
\end{table*}

\subsection{Determination of PDF uncertainties using the MC method
  with regularisation}
\label{section:mcddr_pdf_uncertainties}

{\tolerance=800
To study more flexible PDF parameterizations, a MC method based on
varying the input data within their correlated uncertainties is
employed in combination with a data-based regularisation technique.
This method was first used by the NNPDF Collaboration and uses a more
flexible parameterization to describe the $x$ dependence of the
PDFs~\cite{Ball:2008by}. To avoid the fitting of statistical
fluctuations present in the input data (over-fitting) a data-based
stopping criterion is introduced.  The data set is split randomly into
a ``fit'' and a ``control'' sample. The \chisq minimisation is
performed with the ``fit'' sample while simultaneously the \chisq of
the ``control'' sample is calculated using the current PDF parameters.
It is observed that the \chisq of the ``control'' sample at first
decreases and then starts to increase again because of
over-fitting. At this point, the fit is stopped. This regularisation
technique is used in combination with a MC method to estimate the
central value and the uncertainties of the fitted PDFs. Before a fit,
several hundred replica sets are created by allowing the central
values of the measured cross section to fluctuate within their
statistical and systematic uncertainties while taking into account all
correlations.  For each replica, a fit to NLO QCD is performed, which
yields an optimum value and uncertainty for each parameter. The
collection of all replica fits can then provide an ensemble average
and root-mean-square.  Moreover, the variations to derive the model
dependence of the HERAPDF prescription do not lead to any further
increase of the uncertainty.\par}

\begin{figure*}[tbp]
  \centering
  \includegraphics[width=0.48\textwidth]{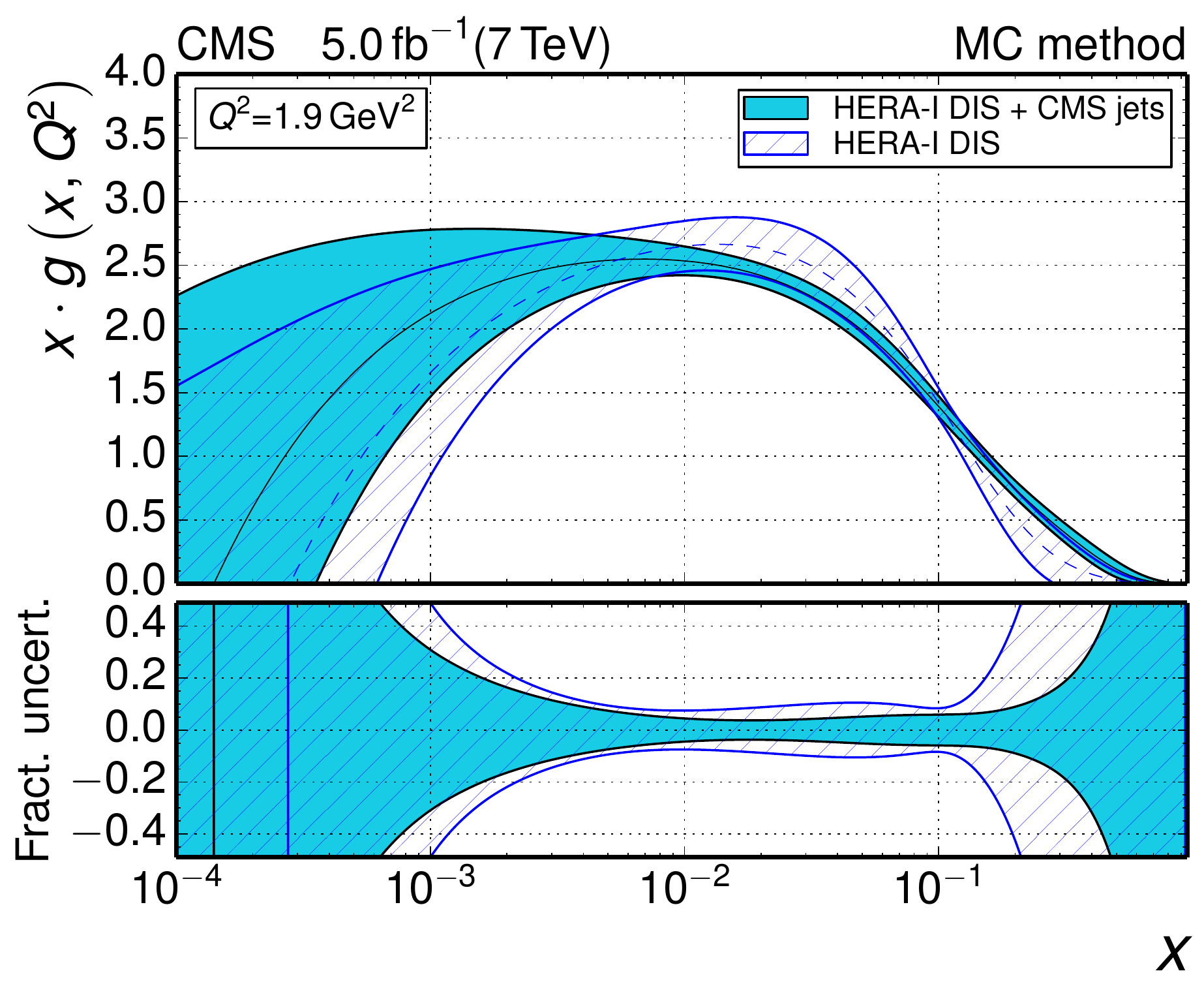}\hfill
  \includegraphics[width=0.48\textwidth]{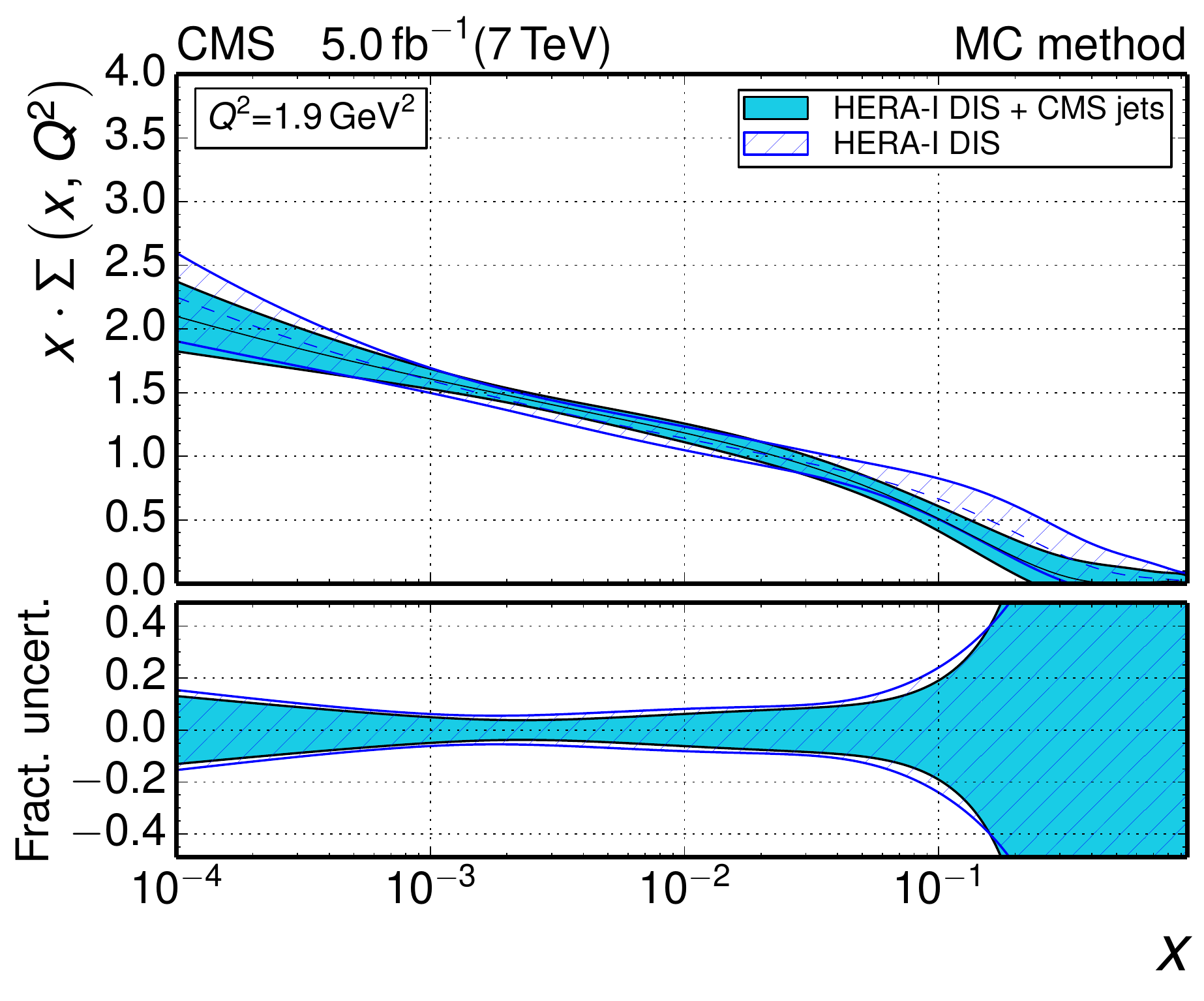}\hfill
  \includegraphics[width=0.48\textwidth]{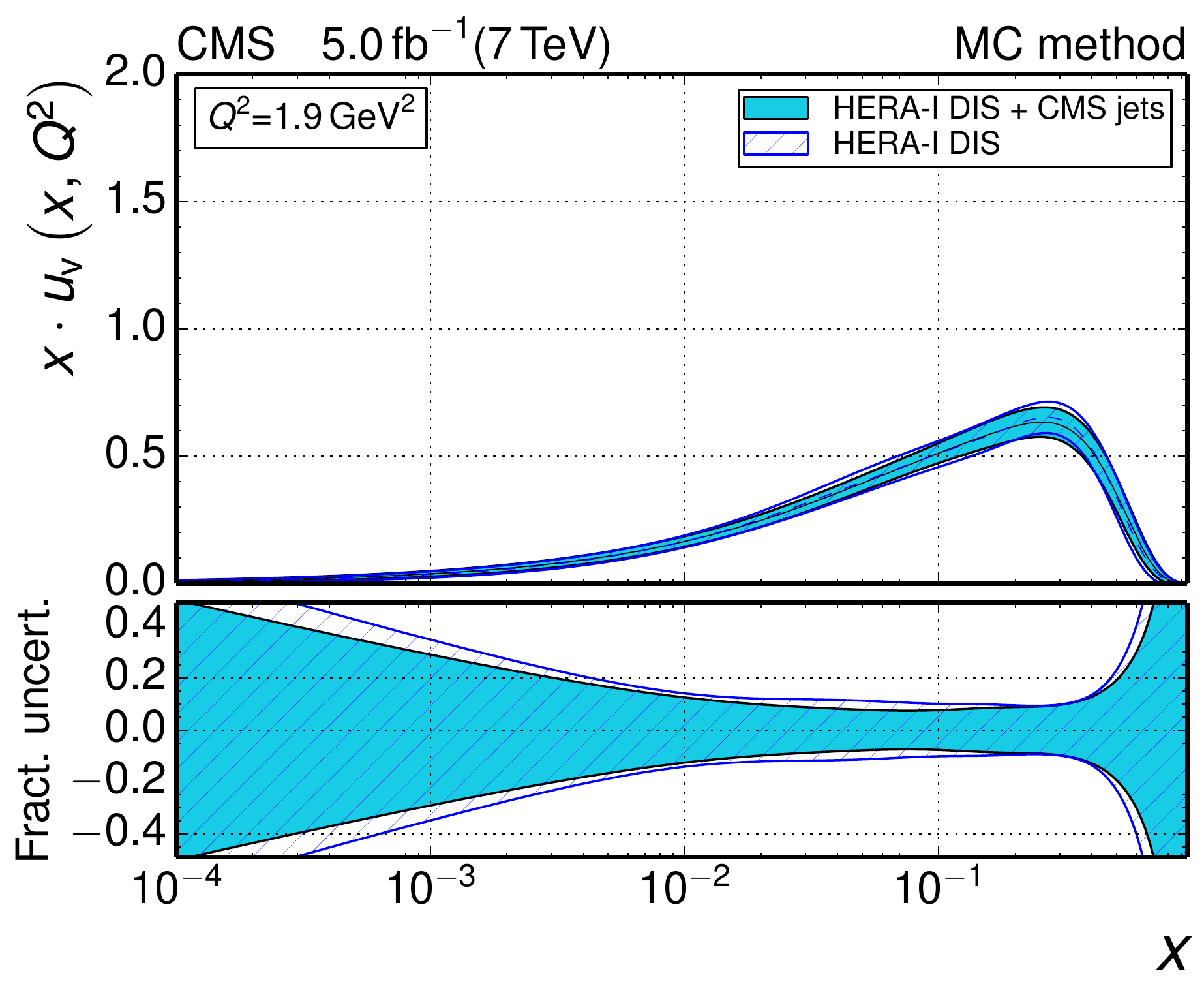}\hfill
  \includegraphics[width=0.48\textwidth]{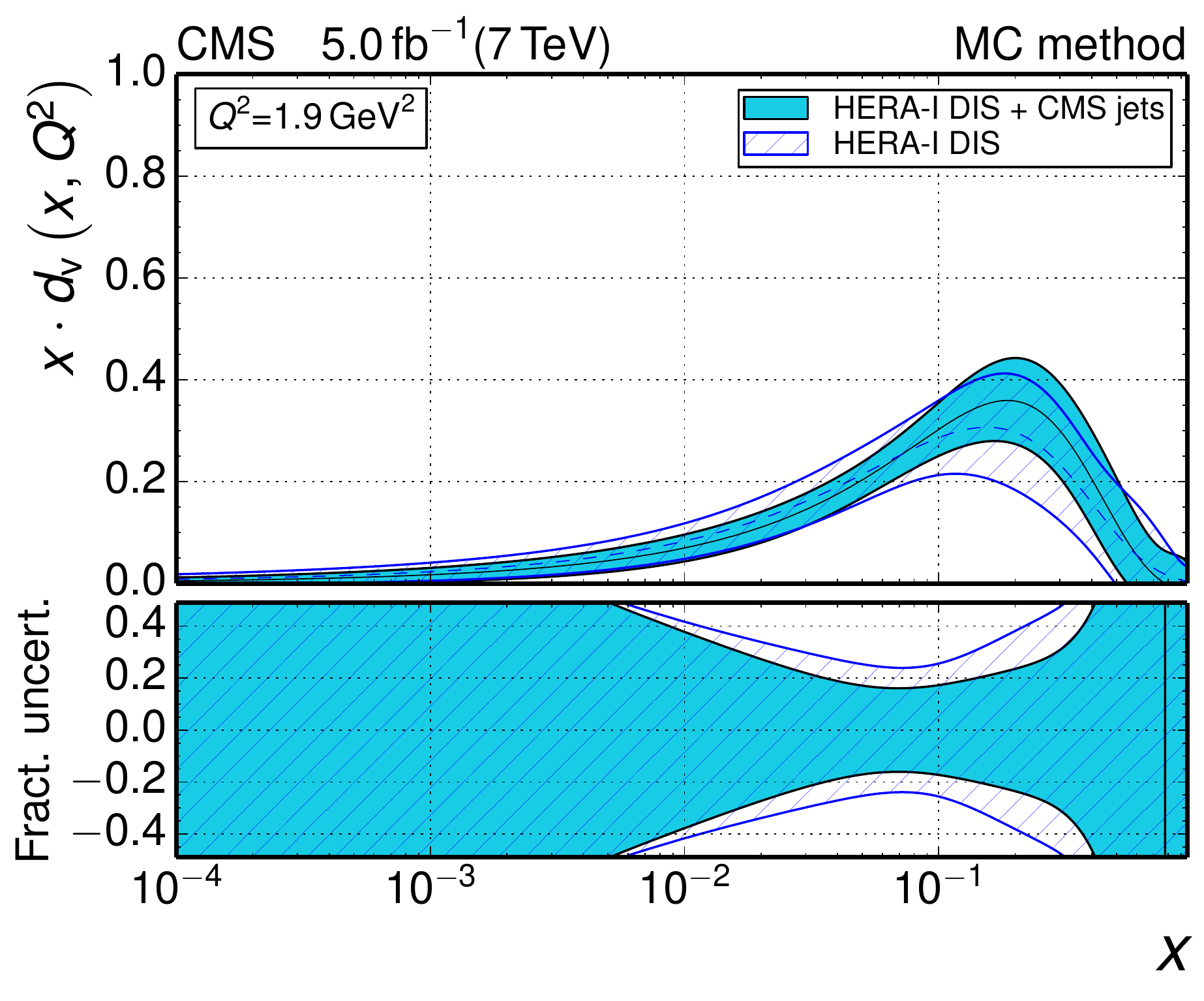}\hfill
  \caption{The gluon (top left), sea quark (top right), u valence
    quark (bottom left), and d valence quark (bottom right) PDFs as a
    function of $x$ as derived from HERA-I inclusive DIS data alone
    (dashed line) and in combination with CMS inclusive jet data (full
    line). The PDFs are determined employing the MC method with
    data-derived regularisation.
    The PDFs are shown at the starting scale $Q^2 = 1.9\GeVsq$. Only the total
    uncertainty in the PDFs is shown (hatched and solid bands).}
  \label{fit:cmsjets2011:ddr:directcomparison:1_4:all}
\end{figure*}

Similarly to Fig.~\ref{fit:cmsjets2011:hera:directcomparison:1_4:all}
for the HERAPDF method, a direct comparison of the two fit results with
total uncertainties is shown in
Fig.~\ref{fit:cmsjets2011:ddr:directcomparison:1_4:all} for the MC
method. The total uncertainty derived with the MC method is almost
always larger than with the HERAPDF technique, and in the case of
the gluon at low $x$, it is much larger. In both cases a significant reduction of
the uncertainty in the gluon PDF is observed, notably in the $x$
range from $10^{-2}$ up to 0.5. Both methods also lead to a decrease
in the gluon PDF between $10^{-2}$ and $10^{-1}$ and an increase for
larger $x$. Although this change is more pronounced when applying the
MC method, within the respective uncertainties both results are
compatible. For the sea quark only small differences in shape are
observed, but, in contrast to the HERAPDF method that exhibits reduced
uncertainties for $x > 0.2$, this is not visible when using the MC
method. Both methods agree on a very modest reduction in uncertainty
at high $x > 0.05$ in the u valence quark PDF and a somehwat larger
improvement for the d valence quark PDF, which is expected from the
correlations, studied in Fig.~\ref{fig:correlation_pdf_xs_gqq}, where
the quark distributions are constrained via the \cPq\cPq~contribution
to jet production at high \yabs and \pt. Changes in shape of the d
valence quark PDF go into opposite directions for the two methods, but
are compatible within uncertainties.

\begin{figure*}[tbp]
  \centering
  \includegraphics[width=0.48\textwidth]{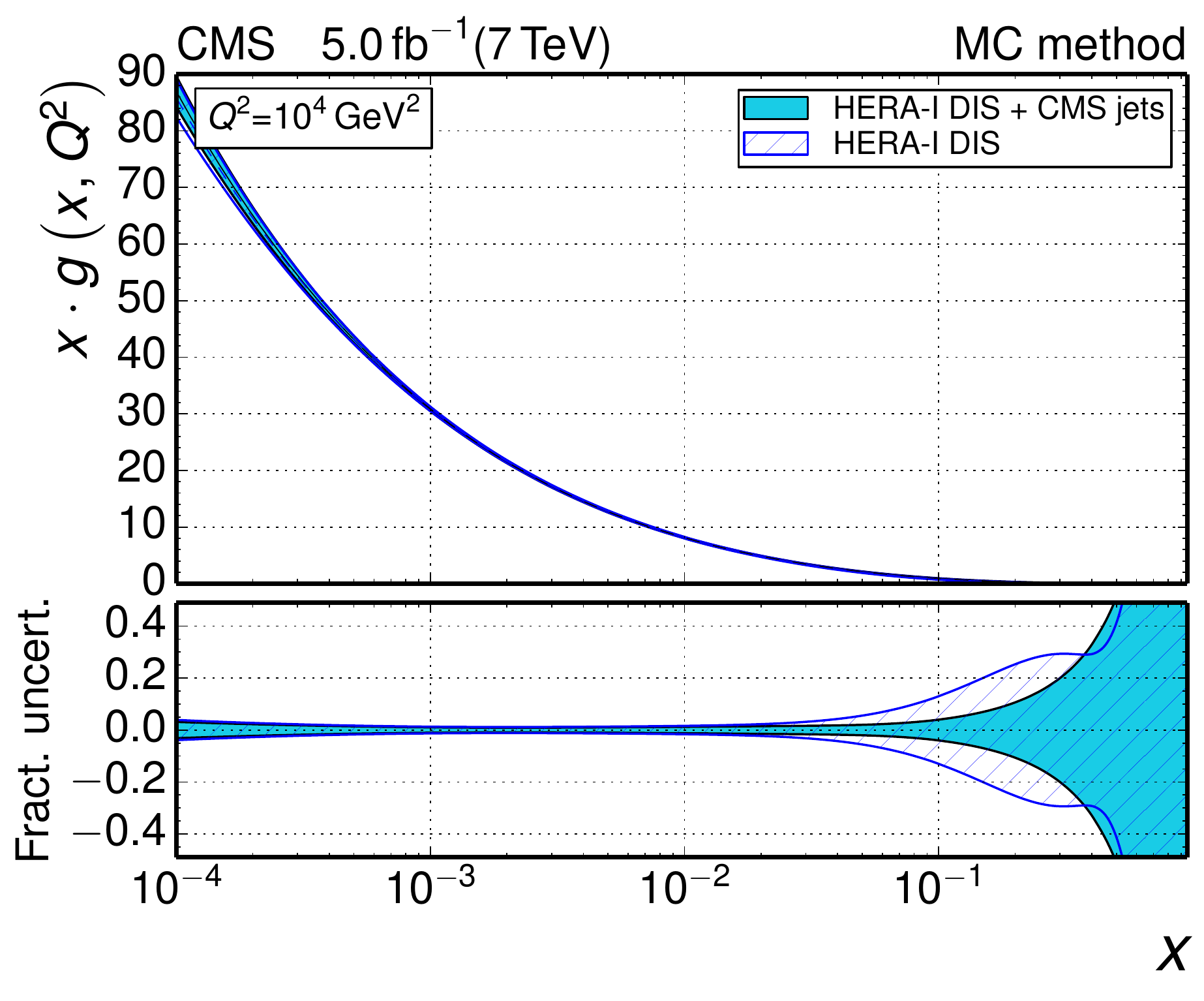}\hfill
  \includegraphics[width=0.48\textwidth]{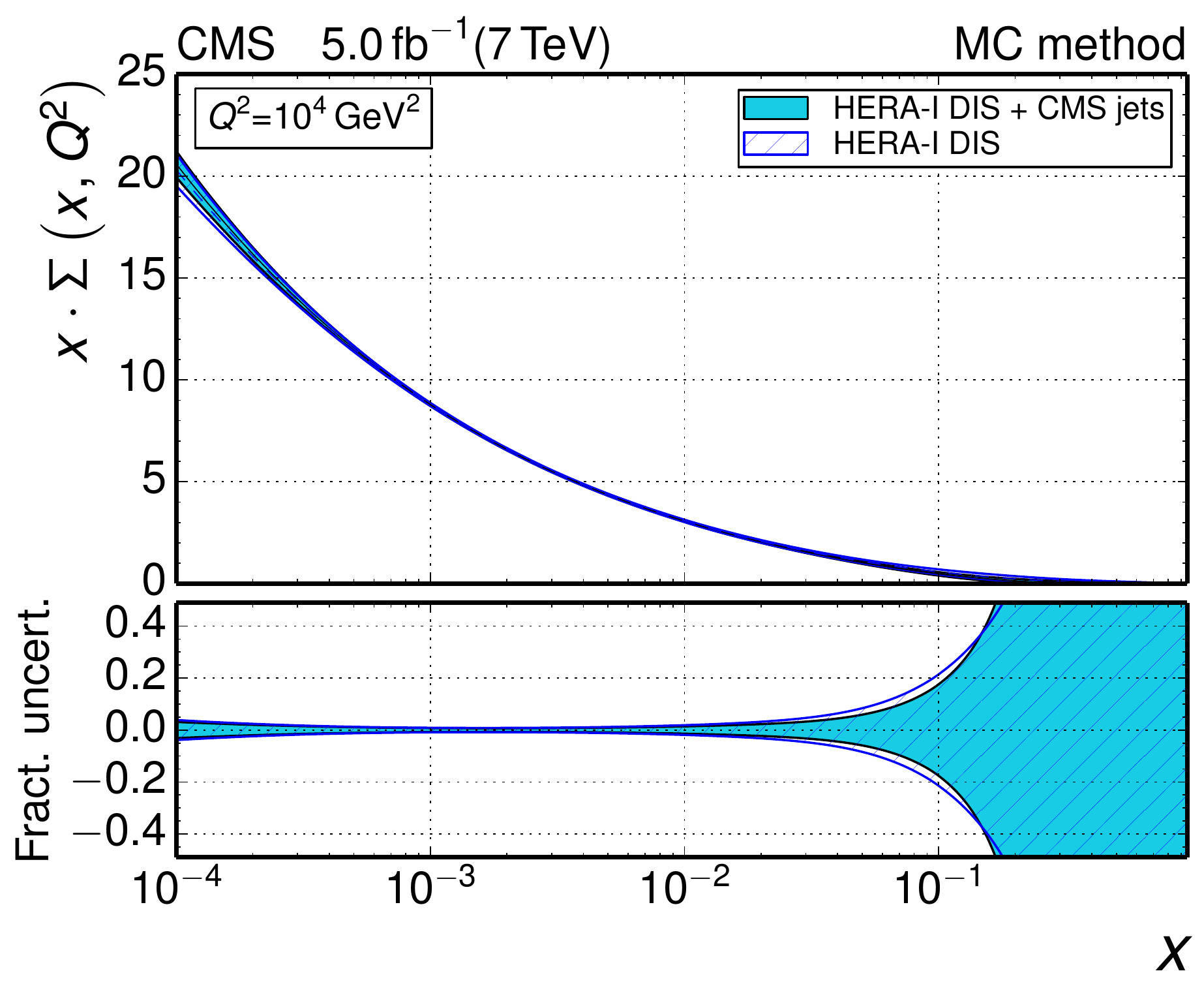}\hfill
  \includegraphics[width=0.48\textwidth]{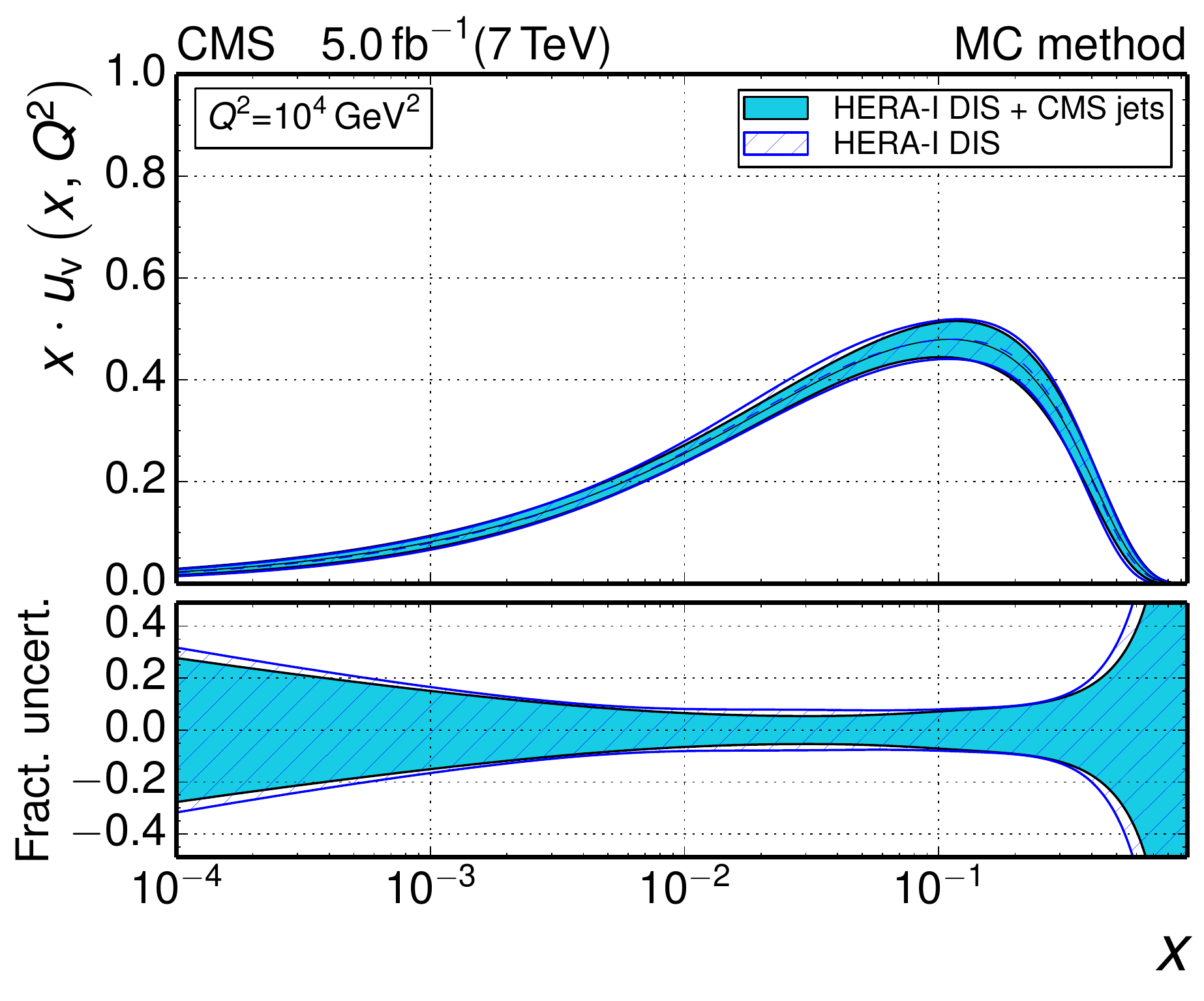}\hfill
  \includegraphics[width=0.48\textwidth]{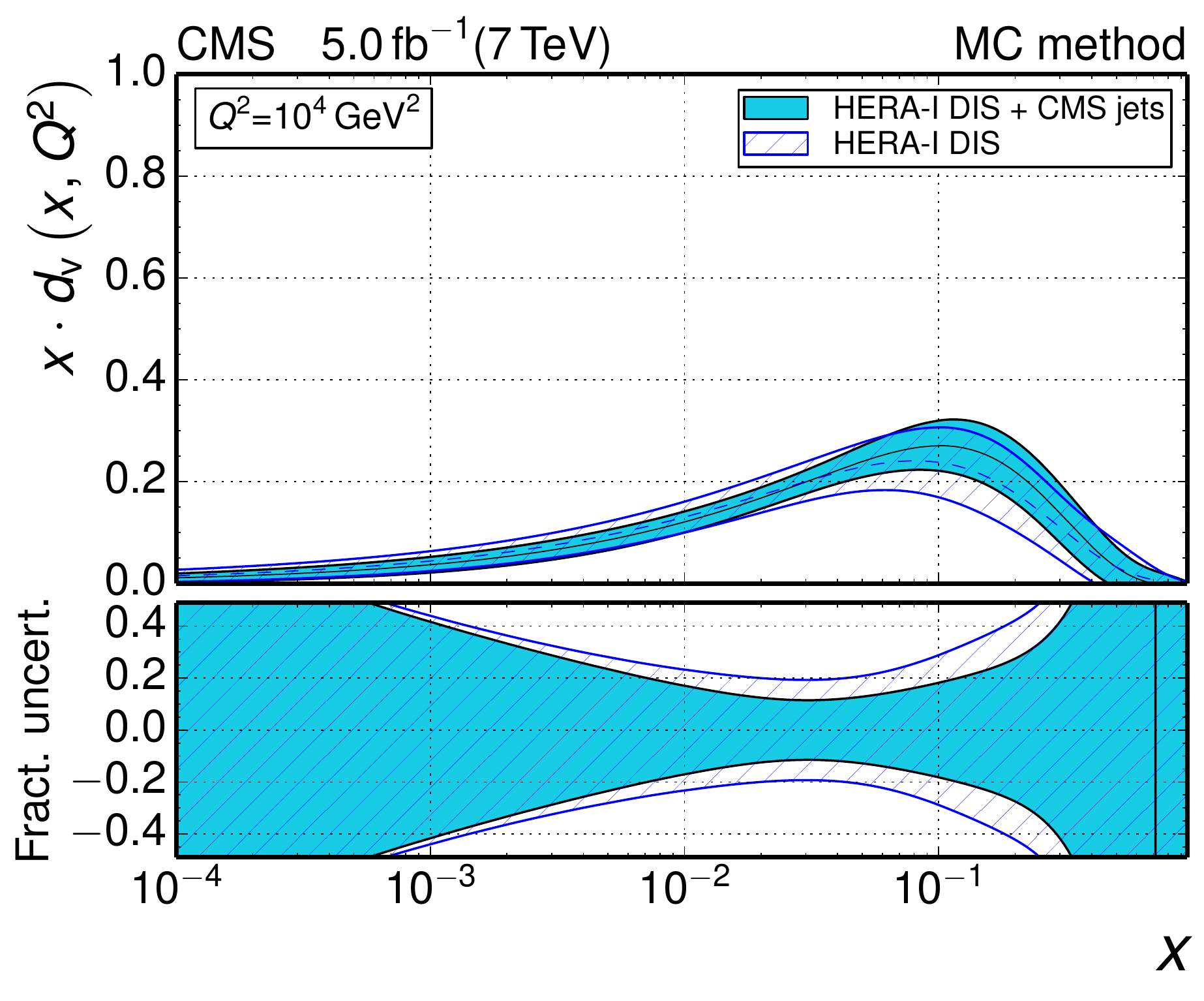}\hfill
  \caption{The gluon (top left), sea quark (top right), u valence
    quark (bottom left), and d valence quark (bottom right) PDFs as a
    function of $x$ as derived from HERA-I inclusive DIS data alone
    (dashed line) and in combination with CMS inclusive jet data (full
    line). The PDFs are determined employing the MC method with
    data-derived regularisation.
    The PDFs are evolved to $Q^2 = 10^4 \GeVsq$. Only the total
    uncertainty in the PDFs is shown (hatched and solid bands).}
  \label{fit:cmsjets2011:ddr:directcomparison:10_4:all}
\end{figure*}

\begin{figure}[tbp]
  \centering
  \includegraphics[width=0.48\textwidth]{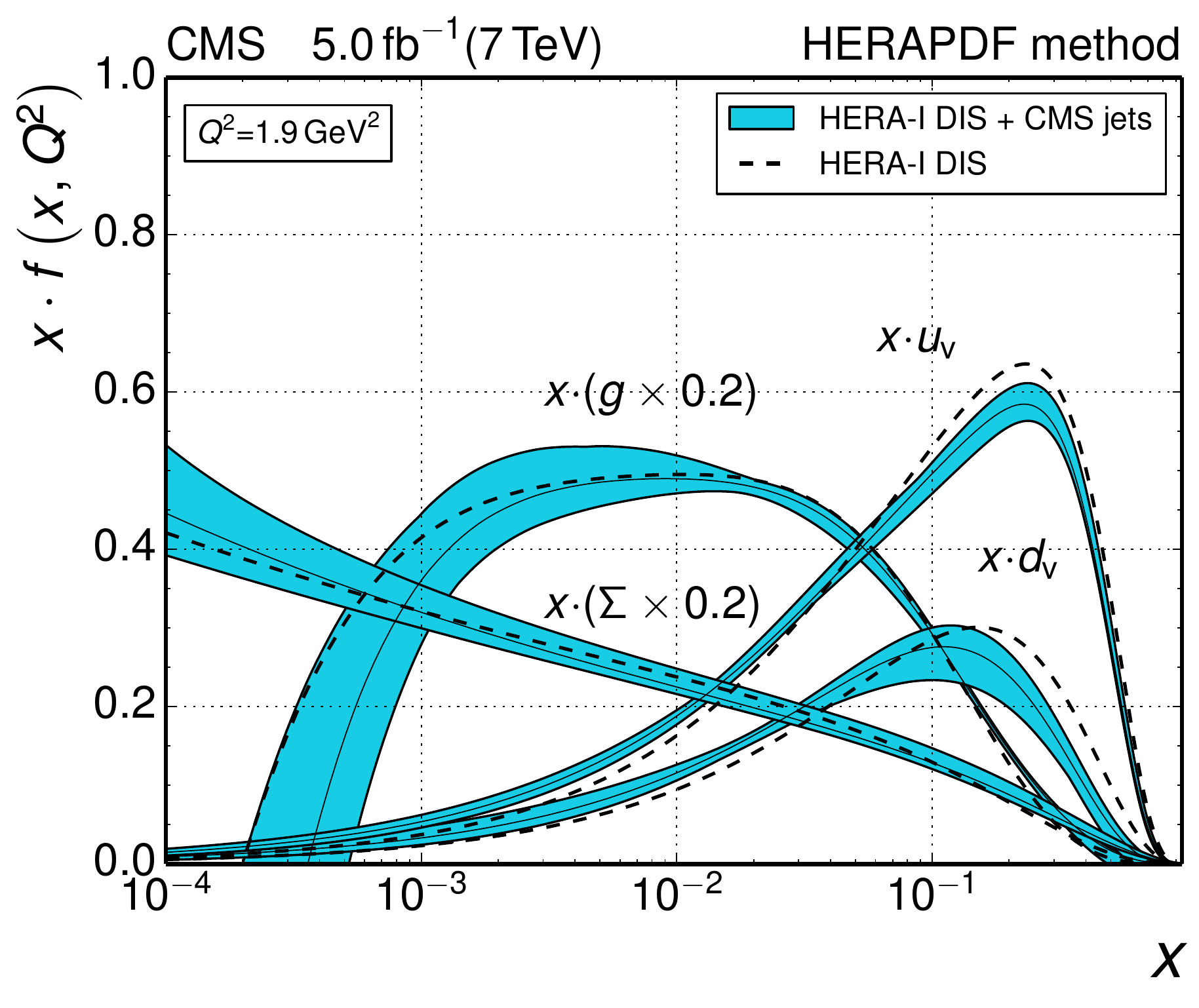}
  \includegraphics[width=0.48\textwidth]{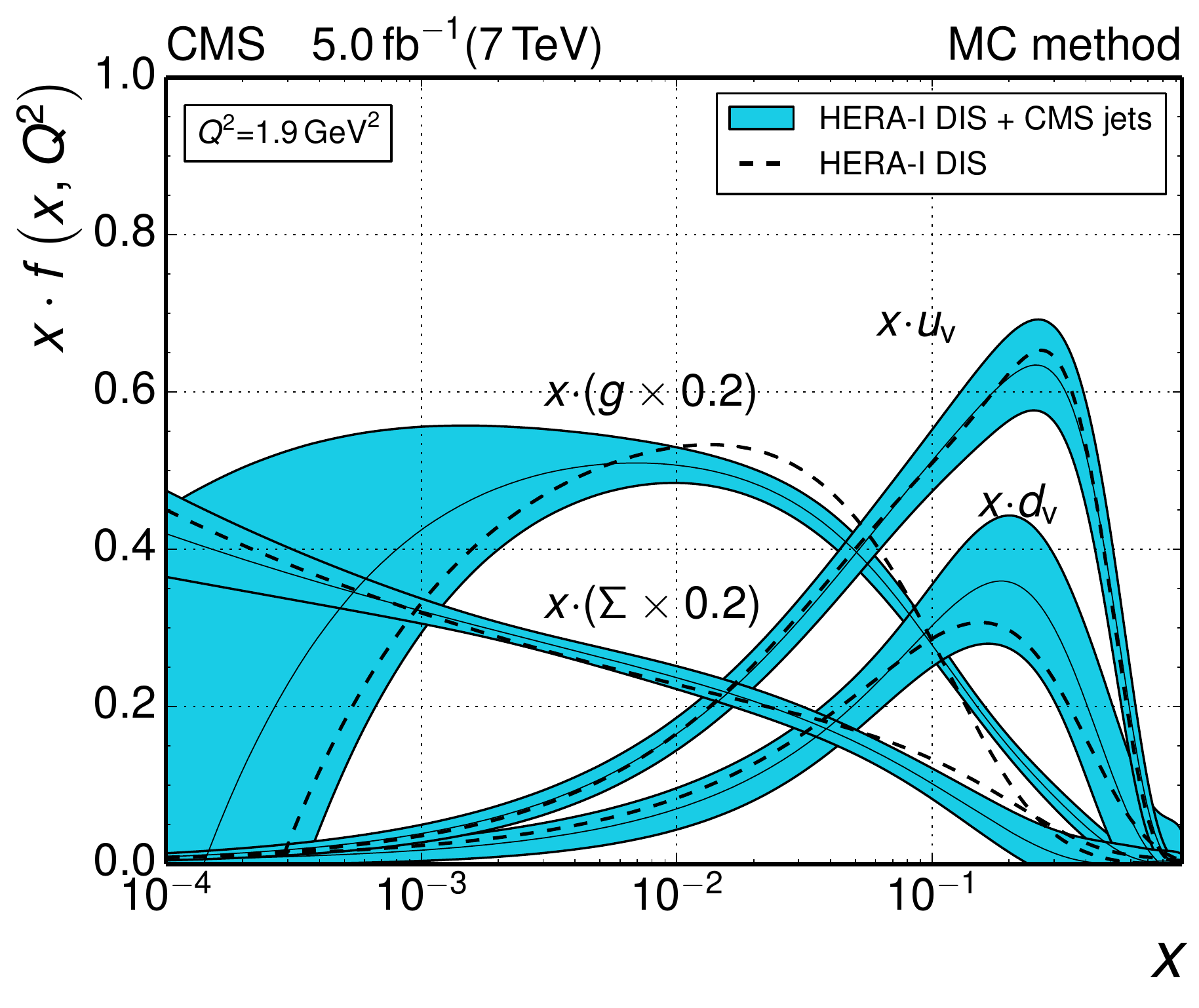}
  \caption{Overview of the gluon, sea, u valence, and d valence PDFs
    before (dashed line) and after (full line) including the CMS
    inclusive jet data into the fit. The plots show the PDF fit
    outcome from the HERAPDF method (\cmsLeft) and from the MC method with
    data-derived regularisation (\cmsRight). The PDFs are shown at the
    starting scale $Q^2 = 1.9 \GeVsq$. The total uncertainty including
    the CMS inclusive jet data is shown as a band around the central
    fit result.}
  \label{fit:cmsjets2011:overview}
\end{figure}

All preceding figures presented the PDFs at the starting scale of the
evolution of $Q^2 = 1.9 \GeVsq$. For illustration,
Fig.~\ref{fit:cmsjets2011:ddr:directcomparison:10_4:all} displays the
PDFs derived with the regularised MC method after evolution to a scale
of $Q^2 = 10^4 \GeVsq$. Finally, Fig.~\ref{fit:cmsjets2011:overview}
shows an overview of the gluon, sea, u valence, and d valence
distributions at the starting scale of $Q^2 = 1.9 \GeVsq$ for both
techniques, the HERAPDF and the regularised MC method.

\subsection{Combined fit of PDFs and the strong coupling constant}
\label{sec:combinedfits}

Inclusive DIS data alone are not sufficient to disentangle effects on
cross section predictions from changes in the gluon distribution or
\alpsmz simultaneously. Therefore \alpsmz was always fixed to 0.1176
in the original HERAPDF1.0 derivation.  When the CMS inclusive jet
data are added, this constraint can be dropped and \alpsmz and its
uncertainty (without $Q$ scale variations) is determined to $\alpsmz =
0.1192\,^{+0.0023}_{-0.0019}\,\text{(all except scale)}$.  Repeating
the fit with the regularised MC method gives $\alpsmz =
0.1188\pm0.0041\,\text{(all except scale)}$.

{\tolerance=1000
Since a direct correspondence among the different components of the
uncertainty can not easily be established, only the quadratic sum of
experimental, PDF, and NP uncertainties are presented, which is
equivalent to the total uncertainty without scale uncertainty. For
example, the HERA-I DIS data contribute to the experimental
uncertainty in the combined fits, but contribute only to the PDF
uncertainty in separate \alpsmz fits. The HERAPDF prescription for PDF
fits tends to small uncertainties, while the uncertainties of the MC
method with data-derived regularisation are twice as large. For
comparison, the corresponding uncertainty in \alpsmz using more
precisely determined PDFs from global fits as in
Section~\ref{sec:alphas} gives a result between the two: $\alpsmz =
0.1185\pm0.0034\,\text{(all except scale)}$.\par}

The evaluation of scale uncertainties is an open issue, which is
ignored in all global PDF fits given in Table~\ref{tab:pdfsets}. The
impact is investigated in
Refs.~\cite{Martin:2009iq,Gao:2012he,Ball:2012wy,Watt:2013oha}, where
scale definitions and $K$-factors are varied. Lacking a recommended
procedure for the scale uncertainties in combined fits of PDFs and
\alpsmz, two evaluations are reported here for the HERAPDF method. In
the first one, the combined fit of PDFs and \alpsmz is repeated for
each variation of the scale factors from the default choice of
$\mur=\muf=\pt$ for the same six combinations as explained in
Section~\ref{section:fit_proc}. The scale for the HERA DIS data
is not changed. The maximal
observed upward and downward changes of \alpsmz with respect to the
default scale factors are then taken as scale uncertainty,
irrespective of changes in the PDFs: $\Delta\alpsmz
=\,^{+0.0022}_{-0.0009}\,\mathrm{(scale)}$.

The second procedure is analogous to the method employed to determine
\alpsmz in Section~\ref{sec:alphas}.  The best PDFs are derived for a
series of fixed values of \alpsmz as done for the global PDF sets.
Using this series of PDFs with varying values of \alpsmz, the
combination of PDF and \alpsmz that best fits the HERA-I DIS and CMS
inclusive jet data is found. The \alpsmz values determined both ways
are consistent with each other. The fits are now repeated for the same
scale factor variations, and the maximal observed upward and downward
changes of \alpsmz with respect to the default scale factors are taken
as scale uncertainty: $\Delta\alpsmz
=\,^{+0.0024}_{-0.0039}\,\mathrm{(scale)}$.

In contrast to the scale uncertainty of the first procedure, there is
less freedom for compensating effects between different gluon
distributions and \alpsmz values in the second procedure, and the
latter procedure leads to a larger scale uncertainty as expected. In
overall size the uncertainty is similar to the final results on
\alpsmz reported in the last section: $\Delta\alpsmz
=\,^{+0.0053}_{-0.0024}\,\mathrm{(scale)}$.

\section{Summary}
\label{sec:summary}

An extensive QCD study has been performed based on the CMS inclusive
jet data in Ref.~\cite{Chatrchyan:2012bja}. Fits dedicated to
determine \alpsmz have been performed involving QCD predictions at NLO
complemented with electroweak and nonperturbative (NP)
corrections. Employing global parton distribution functions (PDFs),
where the gluon is constrained through data from various experiments,
the strong coupling constant has been determined to be
\ifthenelse{\boolean{cms@external}}{
\begin{multline*}
  \alpsmz = 0.1185 \pm 0.0019\,(\text{exp}) \pm
  0.0028\,(\mathrm{PDF})\\ \pm 0.0004\,(\mathrm{NP})
  \,^{+0.0053}_{-0.0024}\,(\text{scale}),
\end{multline*}
}{
\begin{equation*}
  \alpsmz = 0.1185 \pm 0.0019\,(\text{exp}) \pm
  0.0028\,(\mathrm{PDF}) \pm 0.0004\,(\mathrm{NP})
  \,^{+0.0053}_{-0.0024}\,(\text{scale}),
\end{equation*}
}
which is consistent with previous results.

It was found that the published correlations of the experimental
uncertainties adequately reflect the detector characteristics and
reliable fits of standard model parameters could be performed within
each rapidity region.  However, when combining several rapidity
regions, it was discovered that the assumption of full correlation in
rapidity $y$ had to be revised for one source of uncertainty in the
jet energy scale, which suggested a modified correlation treatment
that is described and applied in this work.

To check the running of the strong coupling, all fits have also been
carried out separately for six bins in inclusive jet \pt, where the
scale $Q$ of \alpsq is identified with \pt. The observed behaviour of
\alpsq is consistent with the energy scale dependence predicted by the
renormalization group equation of QCD, and extends the H1, ZEUS, and
D0 results to the \TeV region.

The impact of the inclusive jet measurement on the PDFs of the proton
is investigated in detail using the \HERAFitter tool.  When the CMS
inclusive jet data are used together with the HERA-I DIS measurements,
the uncertainty in the gluon distribution is significantly reduced for
fractional parton momenta $x \gtrsim 0.01$.  Also, a modest
improvement in uncertainty in the u and d valence quark distributions
is observed.

The inclusion of the CMS inclusive jet data also allows a combined fit
of \alpsmz and of the PDFs, which is not possible with the HERA-I
inclusive DIS data alone. The result is consistent with the reported
values of \alpsmz obtained from fits employing global PDFs.

\begin{acknowledgments}
\hyphenation{Bundes-ministerium Forschungs-gemeinschaft Forschungs-zentren} We congratulate our colleagues in the CERN accelerator departments for the excellent performance of the LHC and thank the technical and administrative staffs at CERN and at other CMS institutes for their contributions to the success of the CMS effort. In addition, we gratefully acknowledge the computing centres and personnel of the Worldwide LHC Computing Grid for delivering so effectively the computing infrastructure essential to our analyses. Finally, we acknowledge the enduring support for the construction and operation of the LHC and the CMS detector provided by the following funding agencies: the Austrian Federal Ministry of Science, Research and Economy and the Austrian Science Fund; the Belgian Fonds de la Recherche Scientifique, and Fonds voor Wetenschappelijk Onderzoek; the Brazilian Funding Agencies (CNPq, CAPES, FAPERJ, and FAPESP); the Bulgarian Ministry of Education and Science; CERN; the Chinese Academy of Sciences, Ministry of Science and Technology, and National Natural Science Foundation of China; the Colombian Funding Agency (COLCIENCIAS); the Croatian Ministry of Science, Education and Sport, and the Croatian Science Foundation; the Research Promotion Foundation, Cyprus; the Ministry of Education and Research, Estonian Research Council via IUT23-4 and IUT23-6 and European Regional Development Fund, Estonia; the Academy of Finland, Finnish Ministry of Education and Culture, and Helsinki Institute of Physics; the Institut National de Physique Nucl\'eaire et de Physique des Particules~/~CNRS, and Commissariat \`a l'\'Energie Atomique et aux \'Energies Alternatives~/~CEA, France; the Bundesministerium f\"ur Bildung und Forschung, Deutsche Forschungsgemeinschaft, and Helmholtz-Gemeinschaft Deutscher Forschungszentren, Germany; the General Secretariat for Research and Technology, Greece; the National Scientific Research Foundation, and National Innovation Office, Hungary; the Department of Atomic Energy and the Department of Science and Technology, India; the Institute for Studies in Theoretical Physics and Mathematics, Iran; the Science Foundation, Ireland; the Istituto Nazionale di Fisica Nucleare, Italy; the Korean Ministry of Education, Science and Technology and the World Class University program of NRF, Republic of Korea; the Lithuanian Academy of Sciences; the Ministry of Education, and University of Malaya (Malaysia); the Mexican Funding Agencies (CINVESTAV, CONACYT, SEP, and UASLP-FAI); the Ministry of Business, Innovation and Employment, New Zealand; the Pakistan Atomic Energy Commission; the Ministry of Science and Higher Education and the National Science Centre, Poland; the Funda\c{c}\~ao para a Ci\^encia e a Tecnologia, Portugal; JINR, Dubna; the Ministry of Education and Science of the Russian Federation, the Federal Agency of Atomic Energy of the Russian Federation, Russian Academy of Sciences, and the Russian Foundation for Basic Research; the Ministry of Education, Science and Technological Development of Serbia; the Secretar\'{\i}a de Estado de Investigaci\'on, Desarrollo e Innovaci\'on and Programa Consolider-Ingenio 2010, Spain; the Swiss Funding Agencies (ETH Board, ETH Zurich, PSI, SNF, UniZH, Canton Zurich, and SER); the Ministry of Science and Technology, Taipei; the Thailand Center of Excellence in Physics, the Institute for the Promotion of Teaching Science and Technology of Thailand, Special Task Force for Activating Research and the National Science and Technology Development Agency of Thailand; the Scientific and Technical Research Council of Turkey, and Turkish Atomic Energy Authority; the National Academy of Sciences of Ukraine, and State Fund for Fundamental Researches, Ukraine; the Science and Technology Facilities Council, UK; the US Department of Energy, and the US National Science Foundation.

Individuals have received support from the Marie-Curie programme and the European Research Council and EPLANET (European Union); the Leventis Foundation; the A. P. Sloan Foundation; the Alexander von Humboldt Foundation; the Belgian Federal Science Policy Office; the Fonds pour la Formation \`a la Recherche dans l'Industrie et dans l'Agriculture (FRIA-Belgium); the Agentschap voor Innovatie door Wetenschap en Technologie (IWT-Belgium); the Ministry of Education, Youth and Sports (MEYS) of the Czech Republic; the Council of Science and Industrial Research, India; the HOMING PLUS programme of Foundation for Polish Science, cofinanced from European Union, Regional Development Fund; the Compagnia di San Paolo (Torino); the Consorzio per la Fisica (Trieste); MIUR project 20108T4XTM (Italy); the Thalis and Aristeia programmes cofinanced by EU-ESF and the Greek NSRF; and the National Priorities Research Program by Qatar National Research Fund.
\end{acknowledgments}
\bibliography{auto_generated}
\clearpage
\appendix

\section{Sources of uncertainty in the calibration of jet energies in
  CMS}
\label{sec:jessources}

In the following, the full list of uncertainty sources of the jet
energy calibration procedure that were originally considered by CMS
and that were used in Ref.~\cite{Chatrchyan:2012bja} is presented
including a short description. It is recommended to apply the
procedure with updated correlations for the JEC2 source, as described
in Section~\ref{sec:measurementjec}. A general description of the jet
energy calibration procedure of CMS is given in
Ref.~\cite{Chatrchyan:2011ds}.

When simulations were employed, the following event generators have
been used: \PYTHIA version~6.4.22~\cite{Sjostrand:2006za} tune Z2 and
\HERWIGPP version~2.4.2~\cite{Bahr:2008pv} with the default tune of
version~2.3.

\begin{description}
\item[JEC0:] Absolute uncertainty.\\
  Using data with photon+jet and $Z$+jet events an absolute
  calibration of jet energies is performed in the jet \pt range of
  30--600\GeV. Uncertainties in the determination of electromagnetic
  energies in the ECAL, of the muon momenta from $Z\to\mu\mu$
  decays, and of the corrections for initial- and final-state (ISR and
  FSR) radiation are propagated together with the statistical
  uncertainty to give the absolute JES uncertainty.

\item[JEC1:] High- and low-\pt extrapolation uncertainty.\\
  Where an absolute calibration with data is not possible, events are
  produced with the event generators \PYTHIAS and \HERWIGPP and are
  subsequently processed through the CMS detector simulation based on
  \GEANTfour~\cite{Agostinelli:2002hh}. Differences in particular in
  modelling the fragmentation process and the underlying event lead to
  an extrapolation uncertainty relative to the directly calibrated jet
  \pt range of $30$--$600$\GeV.

\item[JEC2:] High-\pt extrapolation uncertainty.\\
  This source accounts for a $\pm 3$\% variation in the
  single-particle response that is propagated to jets using a
  parameterized fast simulation of the CMS
  detector~\cite{Abdullin:2011zz}.

\item[JEC3:] Jet flavour related uncertainty.\\
  Differences in detector response to light, charm, and bottom quark
  as well as gluon jets relative to the mixture predicted by QCD for
  the measured processes are evaluated on the basis of simulations
  with \PYTHIAS and \HERWIGPP.

\item[JEC4:] Uncertainty caused by time dependent detector effects.\\
  This source considers residual time-dependent variations in the
  detector conditions such as the endcap ECAL crystal transparency.

\item[JEC5--JEC10:] $\eta$-dependent uncertainties coming from the
  dijet balance method:

  \begin{description}
  \item[JEC5--JEC7:] Caused by the jet energy resolution. These three
    sources are assumed to be fully correlated for the endcap with
    upstream tracking detectors (JEC5), the endcap without upstream
    tracking detectors (JEC6), and the HF calorimeter (JEC7).

  \item[JEC8:] $\eta$-dependent uncertainty caused by corrections for
    final-state radiation. The uncertainty is correlated from one
    region to the other and increases towards HF\@.

  \item[JEC9--JEC10:] Statistical uncertainty in the determination of
    $\eta$-dependent corrections. These are two separate sources for
    the endcap without upstream tracking detectors (JEC9), and the HF
    calorimeter (JEC10).
  \end{description}

\item[JEC11--JEC15:] Uncertainties for the pileup corrections:

  \begin{description}
  \item[JEC11:] parameterizes differences between data and MC events
    versus $\eta$ in zero-bias data.
  \item[JEC12:] estimates residual out-of-time pileup for prescaled
    triggers, if MC events are reweighted to unprescaled data.
  \item[JEC13:] covers an offset dependence on jet \pt (due to, e.g.\
    zero-suppression effects), when the correction is calibrated for
    jets in the \pt range of 20--30\GeV.
  \item[JEC14:] accounts for differences in measured offset from
    zero-bias MC events and from generator-level information in a QCD
    sample.
  \item[JEC15:] covers observed jet rate variations versus the average
    number of reconstructed primary vertices in the 2011 single-jet
    triggers after applying L1 corrections.
  \end{description}

\end{description}

\section{Comparison to theoretical predictions by
  \texorpdfstring{\POWHEG}{POWHEG} +
  \texorpdfstring{\PYTHIAS}{PYTHIA6}}
\label{theory_data}

\begin{figure*}[pt]
  \centering
  \includegraphics[width=0.4\textwidth]{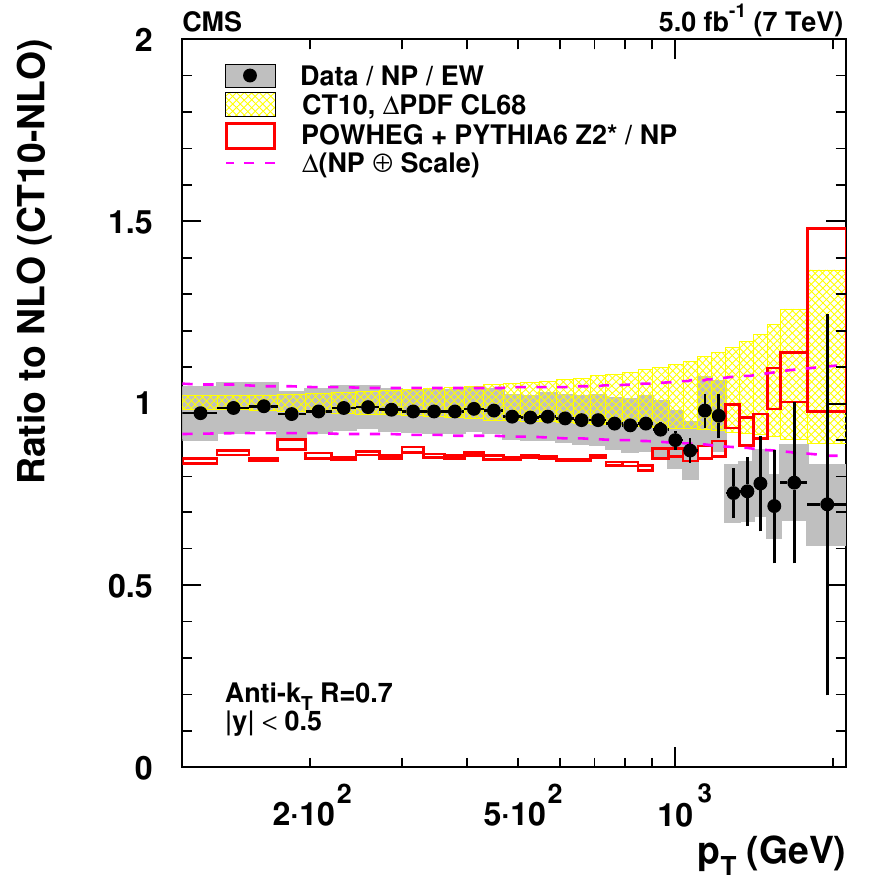}%
  \includegraphics[width=0.4\textwidth]{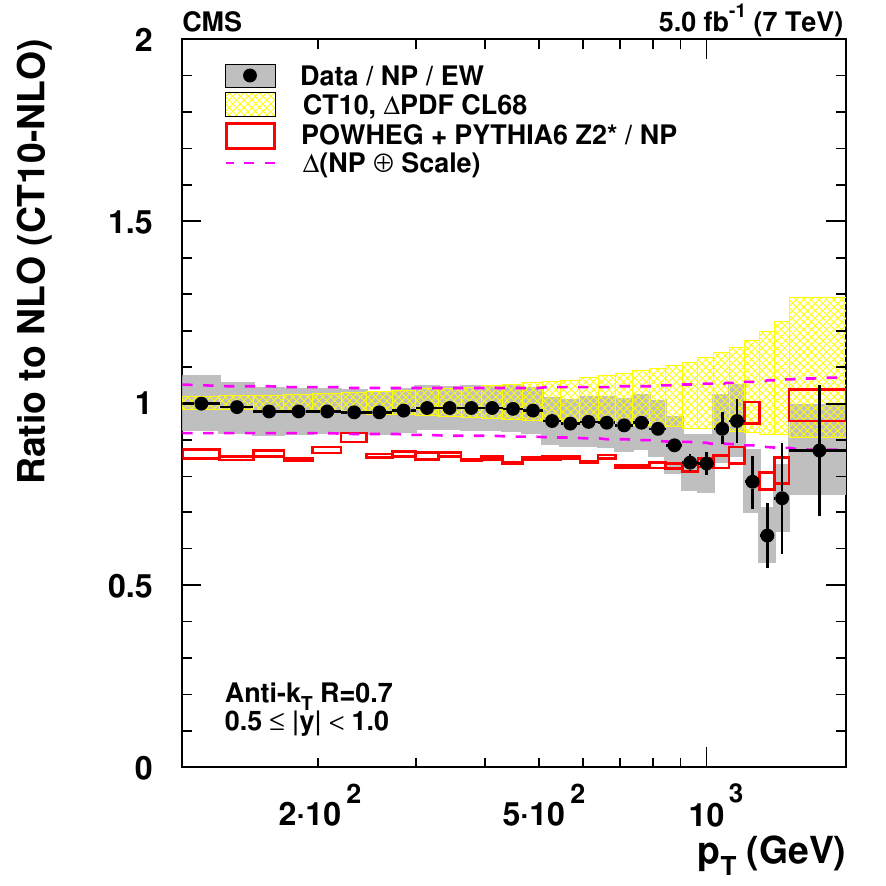}
  \includegraphics[width=0.4\textwidth]{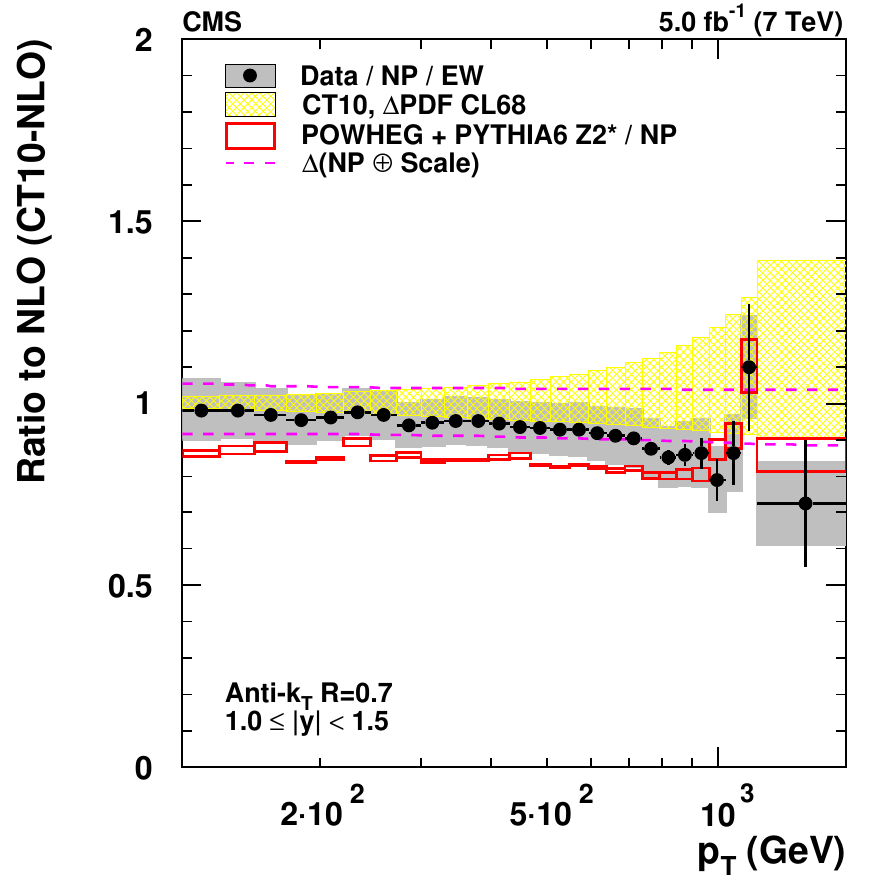}%
  \includegraphics[width=0.4\textwidth]{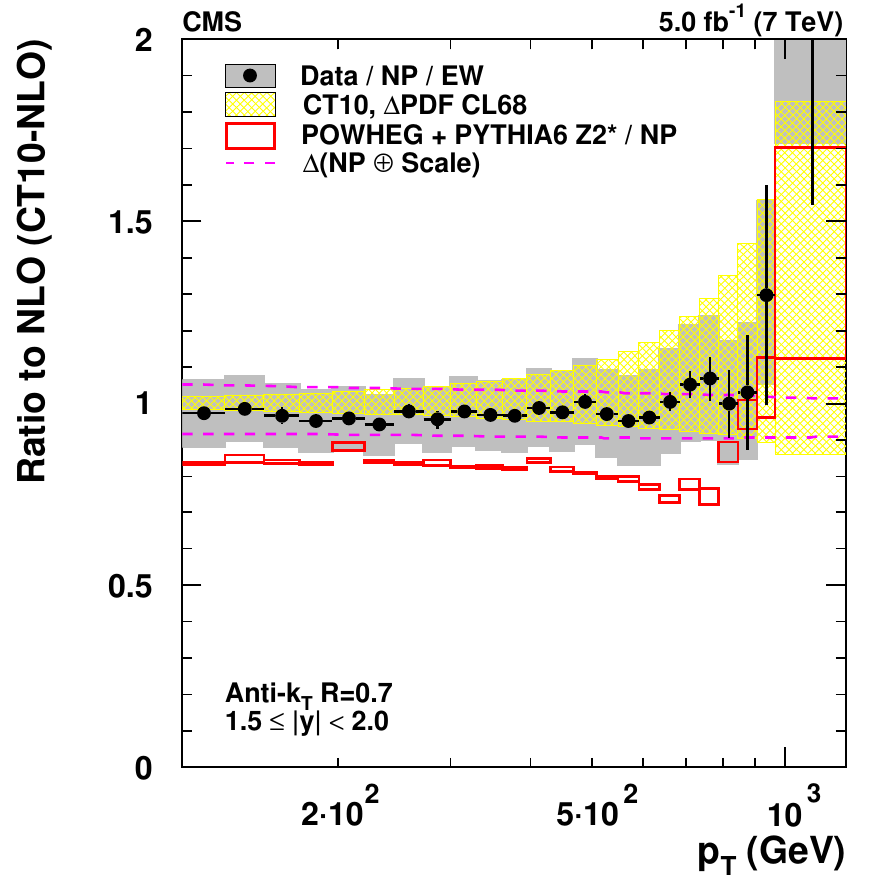}
  \includegraphics[width=0.4\textwidth]{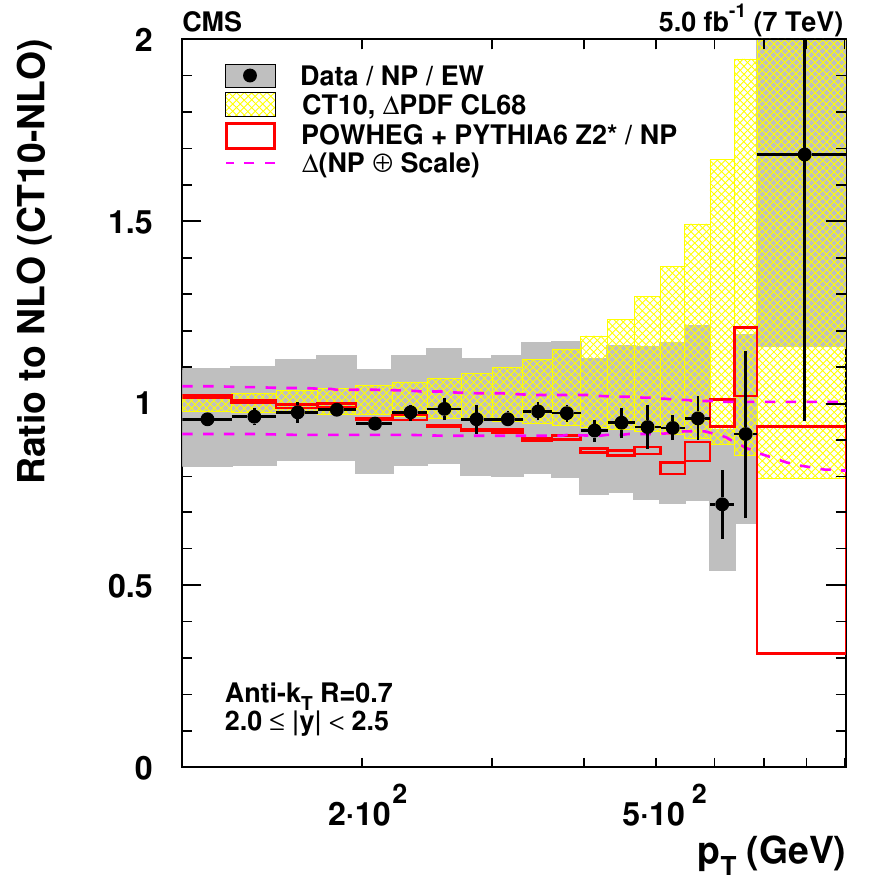}
  \caption{Ratio of data to pQCD at NLO with the CT10-NLO PDF set
    multiplied by electroweak and NP corrections for the five bins in
    rapidity together with bands representing the CT10 PDF uncertainty
    (hatched), and the quadratically added scale and NP uncertainty
    (dashed lines). In addition, the ratio of the prediction by
    \POWHEG + \PYTHIAS tune Z2* at particle level is shown with boxes
    indicating the statistical uncertainty. The error bars and the
    grey boxes correspond to the statistical and systematic
    uncertainty in the data.}
  \label{fig:DataTheory_comp4}
\end{figure*}

Figure~\ref{fig:DataTheory_comp4} presents ratios of data over theory
predictions at NLO using the CT10-NLO PDF set multiplied by
electroweak and NP corrections including PDF uncertainties.

\cleardoublepage \section{The CMS Collaboration \label{app:collab}}\begin{sloppypar}\hyphenpenalty=5000\widowpenalty=500\clubpenalty=5000\textbf{Yerevan Physics Institute,  Yerevan,  Armenia}\\*[0pt]
V.~Khachatryan, A.M.~Sirunyan, A.~Tumasyan
\vskip\cmsinstskip
\textbf{Institut f\"{u}r Hochenergiephysik der OeAW,  Wien,  Austria}\\*[0pt]
W.~Adam, T.~Bergauer, M.~Dragicevic, J.~Er\"{o}, M.~Friedl, R.~Fr\"{u}hwirth\cmsAuthorMark{1}, V.M.~Ghete, C.~Hartl, N.~H\"{o}rmann, J.~Hrubec, M.~Jeitler\cmsAuthorMark{1}, W.~Kiesenhofer, V.~Kn\"{u}nz, M.~Krammer\cmsAuthorMark{1}, I.~Kr\"{a}tschmer, D.~Liko, I.~Mikulec, D.~Rabady\cmsAuthorMark{2}, B.~Rahbaran, H.~Rohringer, R.~Sch\"{o}fbeck, J.~Strauss, W.~Treberer-Treberspurg, W.~Waltenberger, C.-E.~Wulz\cmsAuthorMark{1}
% ...
\vskip\cmsinstskip
% ...
1:~~Also at Vienna University of Technology, Vienna, Austria\\
2:~~Also at CERN, European Organization for Nuclear Research, Geneva, Switzerland\\
% ...

\end{sloppypar}
\end{document}